%% file: roulettes.tex
\documentclass[aps,prd,onecolumn,10pt,notitlepage,amsmath,amssymb,nofootinbib]{revtex4-2}
\usepackage[T1]{fontenc}
\usepackage{graphicx,mathtools,mathrsfs,bm}
\usepackage{microtype,booktabs,xcolor,placeins,etoolbox,xurl}
\usepackage{needspace}
\usepackage[colorlinks=true,linkcolor=blue,citecolor=blue,urlcolor=blue]{hyperref}
\usepackage{roulette-headings}
\makeatletter
\let\roulette@tocsub\l@subsection
\let\roulette@tocsubsub\l@subsubsection
\renewcommand{\l@subsection}[2]{\ifnum\c@tocdepth>0\relax\roulette@tocsub{#1}{#2}\fi}
\renewcommand{\l@subsubsection}[2]{\ifnum\c@tocdepth>1\relax\roulette@tocsubsub{#1}{#2}\fi}
\makeatother

\newcommand{\dd}{\mathrm d}
\newcommand{\calM}{\mathcal M}
\newcommand{\calR}{\mathcal R}
\newcommand{\calF}{\mathcal F}

\makeatletter

\renewenvironment{description}{\list{}{\labelwidth\z@\leftmargin\z@
  \itemindent 1em}}{\endlist}
\makeatother
\allowdisplaybreaks[2]
\ifdefined\pdfsuppressptexinfo
\fi

\begin{document}
\title{Roulettes: A weak lensing formalism for strong lensing\\
III. Nonlinear propagation and extended images}
\author{Chris Clarkson}
\email{chris.clarkson@qmul.ac.uk}
\affiliation{Centre for Theoretical Physics and Astronomy, School of Physical and Chemical Sciences,\\
Queen Mary University of London, 327 Mile End Road, London E1 4NS, United Kingdom}
\affiliation{Department of Physics and Astronomy, University of the Western Cape, Cape Town 7535, South Africa}
\hypersetup{pdftitle={Roulettes: A weak lensing formalism for strong lensing. III. Nonlinear
propagation and extended images},pdfauthor={Chris Clarkson}}
\renewcommand{\dbltopfraction}{0.95}
\renewcommand{\textfraction}{0.05}
\renewcommand{\topfraction}{0.95}
\renewcommand{\bottomfraction}{0.90}
\renewcommand{\floatpagefraction}{0.65}
\setcounter{topnumber}{3}
\setcounter{bottomnumber}{2}
\setcounter{totalnumber}{4}
\begin{abstract}
The stretching and bending of a strongly lensed image reflect how gravity
acts across a beam of light. The nonlinear geodesic deviation equation
is the fundamental equation governing this process: it describes how
spacetime curvature changes the separation and relative directions of
the rays. Roulettes express the resulting distortion as elementary
patterns, extending convergence and shear to the shapes of extended
images. We derive their propagation from this equation in general
spacetimes, giving a complete local description of gravitational lensing
degree by degree. We show how these patterns build strongly distorted
images for a variety of lenses, including galaxy clusters, rotating
black holes and gravitational waves. The examples reveal how the
curvature encountered by a beam generates its image distortions, and
how nonlinear propagation changes those distortions as the beam travels.
Combining expansions about several rays reconstructs extended images
and multiple images beyond the reach of a single expansion, with
explicit error bounds for the analytic lens models.

\end{abstract}
\maketitle
\raggedbottom
\setcounter{secnumdepth}{3}
\setcounter{tocdepth}{1}

\begingroup\hypersetup{linkcolor=black}\tableofcontents\endgroup
\begingroup\let\boldsymbol\bm

\input{sections/01_introduction}
\input{sections/02_roulette_amplitudes}
\input{sections/03_covariant_propagation}
\input{sections/04_propagation_examples}
\input{sections/05_extended_images}
\input{sections/06_discussion}

\section*{Acknowledgements}
CC is supported by STFC (UK) grant ST/X000931/1. CC acknowledges complimentary access through
OpenAI's ChatGPT for Academic Researchers programme and access through
Anthropic's Claude Team Plan for Scientists. GPT-6 Astra assisted
with the calculations. The author also used OpenAI ChatGPT and Codex, and
Anthropic Claude, for algebraic consistency checks, numerical
implementation and testing, figure preparation, literature comparisons,
and manuscript drafting and editing. The author takes full responsibility
for the calculations, interpretation and final text.
\addtocontents{toc}{\protect\setcounter{tocdepth}{0}}
\appendix

\input{appendices/a_connector_geometry}
\input{appendices/b_curvature_and_optics}
\input{appendices/c_calculating_maps}
\input{appendices/d_lens_models}
\input{appendices/e_wavelets}

\setlength{\bibsep}{-0.5pt}
\bibliographystyle{apsrev4-2}
\bibliography{references}
\end{document}

%% file: sections/01_introduction.tex
\section{Introduction}\label{sec:introduction}
A small lensed image is approximately a stretched and magnified copy of its source; convergence
and shear describe that distortion. Across a long arc the stretching changes: the image bends
and its width varies, carrying information about the gravitational field beyond a single local
Jacobian. Flexion describes the first departure from a constant stretch
\citep{goldbergbacon,bacon,hessian}, while roulettes extend this description to higher degrees
by expressing the local lens equation as a sum of elementary distortion patterns
\citep{clarksonI,clarksonII,normann}. The expansion is in angular offset about a reference ray,
so strong magnification can be retained while the variation across the image is resolved into
successive patterns. The quadratic Hessian map and the original roulette construction were
developed in Refs.~\citep{hessian,clarksonI,clarksonII}, which we refer to as Papers~0, I and
II. Here we derive how spacetime curvature generates the roulette amplitudes and combine
expansions about several rays to reconstruct extended images.

Geodesic deviation relates the separation of neighbouring rays to spacetime curvature. For an
infinitesimal bundle, its linear form gives the Jacobi map and the Sachs optical equations
\citep{sachs1961,bartelmann,grasso}; Jacobi fields also underlie direct spacetime descriptions
of image distortion \citep{frittelli2001}. Across a finite bundle, variations in curvature and
relative ray direction contribute beyond linear order in the separation and its rate of change.
The nonlinear geodesic deviation equation (NLGDE) describes this evolution. Higher-order and
generalised deviation equations provide several approximations to it
\citep{hodgkinson,bazanski1977,bazanski1977dynamics,chicone2002,kerner2001,puetzfeld2016,waldsteinbrown2022},
while Vines and Kim give exact formulations using the geodesic connecting the two rays
\citep{vines,kim}. The world function associated with that connecting geodesic also appears in
action and Gaussian-beam formulations \citep{uzun2020,uzun2025}. We use the connector geometry
to follow the distortion across a finite beam, with the observed directions and source surface
specified. Fleury et al.~\citep{fleury2021arbitrary} formulate lensing through thin deflectors
linked by infinitesimal-beam propagation, whereas we follow the variation across the finite
beam.

Kim gives an all-order construction of the relative-acceleration equation, with explicit
expressions through degree ten \citep{kim}. We derive the equation in a form suited to the
optical hierarchy, keeping the curvature derivatives, relative-direction terms and curvature
products explicit. In doing so, we correct the slot order in a mixed cubic term of Kim's
printed expansion and four coefficients in Vines's quartic expansion, three of which Kim had
already identified. Independent calculations from the geodesic equations test the corrected
relative-acceleration expansion through quartic degree (appendix~\ref{arch-evidence}). We also
obtain the terms containing one curvature tensor and its ordered derivatives at arbitrary
degree, and the terms cubic in curvature at fifth degree. These expressions determine how the
gravitational field generates the roulette amplitudes; the angular patterns themselves remain
those of Papers~I and II.

Papers~I and II retained the leading transverse gradients of the tidal field, an approximation
we denote by TG. \citet{normann2025review} discuss the missing terms in their review of joint
weak- and strong-lensing reconstruction. Extending the transverse-gradient calculation involves
two physical effects. First, the distortion acquired earlier along the ray changes the beam on
which the subsequent tidal field acts. Retaining this nonlinear coupling, denoted by C, gives
the TG+C hierarchy. The effect is present in standard multi-plane lensing
\citep{schneider1992}: the first lens changes the positions at which the rays encounter the
second, and hence their subsequent deflections. For a reduced single-plane lens equation, TG,
TG+C and the complete Taylor expansion coincide. Second, the full spacetime equation includes
curvature variation along the ray, the changing relative directions of neighbouring rays and
products of curvature arising from finite separation. The quadratic along-ray and
relative-direction terms were included in Paper~0. Here we incorporate these contributions into
the propagation hierarchy and carry the resulting maps from the observer's sky to a physical
source surface.

Strongly magnified images can form while the additional spacetime terms remain small. In the
weak-field limit of the isothermal and Schwarzschild families studied here, their relative
contribution scales approximately in proportion to the Einstein angle in radians. A weak-field
estimate explains this behaviour through the impact scale relative to the distance to the
nearer endpoint, and the depth of the gravitational potential. The compact Schwarzschild and
gravitational-wave examples explore situations in which these terms become appreciable. There,
completing the propagation at fixed cubic degree recovers image positions and shapes that the
restricted hierarchy misplaces.

These additional terms change the amplitudes of patterns already present at a given angular
degree. Increasing the degree of an incomplete expansion therefore leaves errors in its
lower-degree coefficients. We demonstrate this in a spacetime with covariantly constant
curvature, where curvature gradients vanish but nonlinear beam evolution remains: feeding the
full quadratic distortion back into the propagation reverses the sign of one cubic coefficient.
This involves the quadratic distortion generated by the full equation, beyond the
transverse-gradient coupling denoted by C above. Once the coefficients have been calculated
correctly, the remaining error comes from truncating the angular expansion, and determines how
far it can follow an image.

A second difficulty arises when an image extends beyond the region accurately described by one
local expansion. Increasing the degree improves the map within its domain of convergence, but
nearby lens structure can restrict that domain. A truncated map can also produce spurious
images when used outside the region where its error is controlled
\citep{clarksonII,schaathun2023}. We therefore expand about several reference rays and assign
each part of the image to one of the resulting local maps. For the analytic thin lenses we
obtain explicit position-error bounds over these regions; for images lensed by a rotating black
hole we assess the reconstruction against directly traced rays. The high-degree coefficients
are calculated from analytic lens equations or by differentiating the geodesic flow.

Roulettes connect the shapes of resolved images to the Riemann curvature sampled by their rays.
Curved-arc, fold and cusp measurements constrain local derivatives of the lens map
\citep{birrer2021curved,wagner2016local}, while reduced flexion and mappings between multiple
images constrain combinations that survive lensing degeneracies
\citep{schneiderer2008,tessore2017}. Higher roulette amplitudes carry further information:
\citet{normann2026beyond} show that higher-order amplitudes improve parameter recovery in their
noisy synthetic cluster examples. Finite-beam moments and morphology statistics describe image
shape
\citep{fleury,fleury2019distortions,mecke1994,kratochvil2012,munshi2021,schroederturk2013};
roulettes retain the correspondence between image and source positions. We calculate how their
amplitudes depend on the Riemann tensor, its derivatives and the distortions accumulated along
the light path. Combining these local maps extends that relation across an image, connecting
its resolved structure to the gravitational field.

Section~\ref{sec:review} introduces the roulette amplitudes through a singular isothermal sphere (SIS) arc.
Section~\ref{sec:theory} derives their covariant propagation and the conversion to observed
angles and a source surface. Section~\ref{sec:spacetime} examines nonlinear coupling and the
remaining spacetime contributions using two-plane lenses, static spacetimes and a
gravitational-wave burst. Section~\ref{sec:lenses} then combines local expansions to
reconstruct fold pairs, images near compact structure, cluster fields and images lensed by a
rotating black hole. Figure~\ref{fig:model-gallery} previews these geometries: the source
cross, white circle and colour boundaries show how the same features stretch and bend into
different images. The appendices give the detailed derivations and model inputs.

\begingroup\expandafter\def\csname fps@figure\endcsname{p}\begin{figure}\centering
\includegraphics[width=\textwidth,height=0.755\textheight,keepaspectratio]{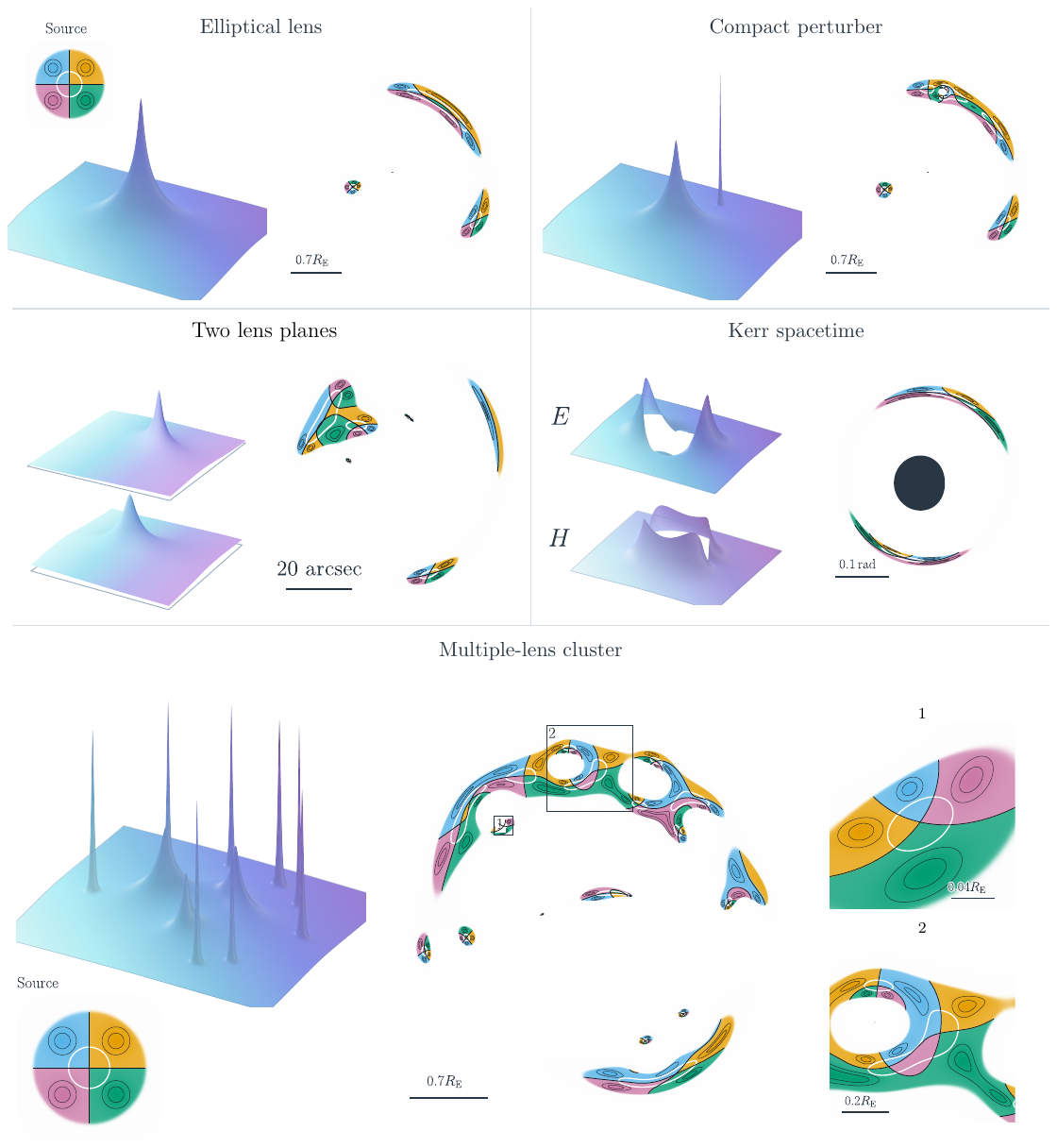}
\caption{A common four-quadrant source forms arcs and multiple images across
five lens geometries. The source is shown at top left and repeated
beneath the cluster lens. Its black cross separates the four colour
regions; a centred white circle and two contrast contours in each
quadrant mark features that can be followed through the lens map.
The single-plane thin-lens examples use radius \(0.12R_{\rm E}\),
the two-plane case uses \(0.12\theta_0=2.4\,\mathrm{arcsec}\), and
Kerr uses \(0.045D\). Numbered cluster boxes locate the enlargements
at right, each with its own scale bar.
\protect\\[0.35em]
The thin-lens surfaces show projected density on a logarithmic height
scale. The two planes share a height normalisation; their spacing is
schematic. For Kerr, the surfaces show
\(\mathsf M^2\sqrt{E_{ab}E^{ab}}\) and
\(\mathsf M^2\sqrt{H_{ab}H^{ab}}\), measured by locally nonrotating
(ZAMO) observers at \(a/\mathsf M=0.9\). They share a linear height
scale in a meridional Kerr--Schild section, with schematic vertical
separation and the horizon interior omitted. These Weyl magnitudes and
the thin-lens densities are different physical quantities, so their
heights cannot be compared numerically. The dark disc in the Kerr image
is the observer's shadow.
\protect\\[0.35em]
The two-plane panel uses the stronger pair analysed in
figures~\ref{fig:two-plane} and \ref{fig:two-plane-coupling}.}
\label{fig:model-gallery}
\end{figure}\endgroup

%% file: sections/02_roulette_amplitudes.tex
\section{How roulettes build an image}
\label{sec:review}\label{sec:local}
We introduce the roulette amplitudes through a thin-lens example, before
calculating them from geodesic deviation in section~\ref{sec:theory}.

\subsection{From an angle on the sky to a position at the source}
The usual lens equation sends an observed direction
\(\boldsymbol\theta\) to a source position \(\boldsymbol\beta\).
We choose one reference ray and describe neighbouring directions by the
two angular components \(\zeta^A\), where \(A=1,2\). Their angular
radius is \(\rho=(\delta_{AB}\zeta^A\zeta^B)^{1/2}\). Screen indices are raised and lowered
with \(\delta_{AB}\); contractions pair one upper and one lower index.

In small-angle coordinates,
\(\boldsymbol\zeta=\boldsymbol\theta-\boldsymbol\theta_*\). With
background source distance \(\chi_S\), define
\[
 \xi_A(\boldsymbol\zeta)=\chi_S
 [\beta_A(\boldsymbol\theta_*+\boldsymbol\zeta)-\beta_A(\boldsymbol\theta_*)].
\]
This is a displacement from the central source position; that position
must be added back before evaluating the source brightness. The exact sky coordinates in
section~\ref{sec:endpoint-choice}
generalise this angular offset.

The source displacement, expressed in the chosen angular coordinates, has the expansion
\begin{equation}
\begin{aligned}{}
\xi_A(\boldsymbol\zeta)
 &=\calM_{AB}\zeta^B+\frac12\calM_{ABC}\zeta^B\zeta^C\\
 &\quad+\frac16\calM_{ABCD}\zeta^B\zeta^C\zeta^D+O(\rho^4).
 \label{eq:map-review}
\end{aligned}
\end{equation}
The tensors \(\calM_{AB_1\cdots B_m}\) are the angular derivatives
\(\partial_{B_1}\cdots\partial_{B_m}\xi_A|_0\), where
\(\partial_B=\partial/\partial\zeta^B\). They depend on the observer
coordinates, endpoint and output screen. These are the observer-to-source maps
\(\mathcal M^{\searrow}\) of Papers~I and II; we suppress the arrow.
Degree \(m\) counts powers of the angular offset, and \(m_{\max}\)
is the largest degree retained. Section~\ref{sec:theory} calculates these
derivatives by covariant propagation and endpoint composition.

For the flat single-plane lens equation
\(\boldsymbol\beta=\boldsymbol\theta-\boldsymbol\nabla_\theta\psi\),
with \(\psi\) the scaled lensing potential, this reduces to
\(\calM_{AB}=\chi_S(\delta_{AB}-\psi_{,AB})\) and
\(\calM_{ABC}=-\chi_S\psi_{,ABC}\): the usual shear and flexion
expansion \citep{bartelmann,bacon}. Section~\ref{sec:weak-flexion-limit} derives this limit from the
covariant propagation equation.
A degree-\(m\) coefficient has one output index and \(m\) input indices. Its input indices can
be interchanged because they label
commuting Taylor derivatives in the chosen angular coordinates. In the
thin-lens formulas, \(\nabla_A\) denotes an ordinary derivative with
respect to the stated angular or lens-plane coordinates. Spacetime
nablas in section~\ref{sec:theory} are covariant derivatives.

At linear degree \(\mathcal M_{AB}=\chi_S\mathcal A_{AB}\), where
\(\mathcal A\) is the full dimensionless amplification matrix of
Paper~II. Allowing also for an antisymmetric part,
\begin{equation}
\begin{aligned}{}
\calM_{AB}
 &=\chi_S\begin{pmatrix}
 1-\kappa-\gamma_1&-\gamma_2+\varpi\\
 -\gamma_2-\varpi&1-\kappa+\gamma_1
 \end{pmatrix}.
 \label{eq:jacobi-review}
\end{aligned}
\end{equation}
The map includes the unlensed contribution \(\xi_A=\chi_S\zeta_A\).
The antisymmetric coefficient has sign \(\epsilon_{12}=+1\): positive
\(\varpi\) adds a clockwise tangential displacement. Physical cosmological
distances are treated in appendix~\ref{app:cosmology}.

These effects can be separated directly:
\begin{equation}
\begin{aligned}{}
\calM_{AB}&=\calM_{(AB)}+\epsilon_{AB}\overline{\calM},\\
 \overline{\calM}&=\frac{\calM_{12}-\calM_{21}}2=\chi_S\varpi,\\
 \epsilon_{12}&=+1 .
 \label{eq:linear-split-review}
\end{aligned}
\end{equation}
Parentheses denote the average over index permutations. The symmetric
part supplies convergence and shear; the barred part supplies the extra
tangential freedom. Its relation to a finite rotation angle also depends
on the symmetric map (section~\ref{sec:weak-two-lens}).

\begin{figure}[!htb]
  \centering
  \includegraphics[width=\textwidth,height=.72\textheight,keepaspectratio]{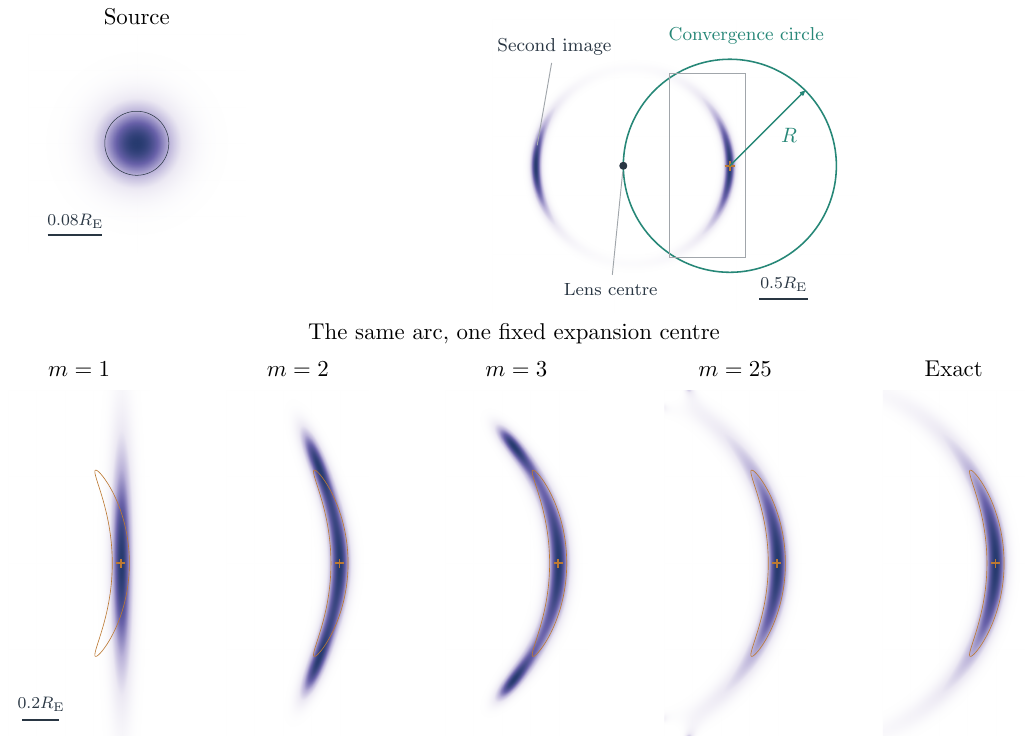}
  \caption{Successive degrees recover an SIS arc within one Taylor domain.
Above: the Gaussian source and exact image. The cross is the expansion
centre at \(R=1.10R_{\rm E}\), the dot the lens centre, and the teal
circle the convergence boundary of radius \(R\). Below: degrees one,
two, three and 25, then the exact map, with common spatial and colour
scales. Orange contours mark exact half-maximum intensity. The source
width is \(0.04R_{\rm E}\). The second image lies outside the domain.}
  \label{fig:worked}
\end{figure}

To form an image, we add the roulette displacements to find the source
position associated with each image pixel, then evaluate the brightness
there. This order matters because brightness is generally a nonlinear
function of position. We call analytic reference
maps \emph{exact}, and numerical evaluations without angular truncation
\emph{direct rays}. Figure~\ref{fig:worked} shows the construction for
an SIS arc.

\subsection{Degree, spin and the elementary patterns}\label{sec:degree-spin-review}
Roulettes separate each Taylor degree into patterns with different angular
symmetries. We can see the patterns by sending a circle of directions
through the forward map. Write
\(\zeta^A=\rho n^A(\vartheta)\), where
\(n_A=(\cos\vartheta,\sin\vartheta)\) points radially outwards and
\(t_A=(-\sin\vartheta,\cos\vartheta)\) points around the circle.
The scalar \(\vartheta\) is the azimuth of the local offset;
\(\boldsymbol\theta\) denotes the observed image position. In the polar
notation of Paper~II, \((r,\theta,\hat\zeta_A,e_A^\theta)\) correspond to
\((\rho,\vartheta,n_A,t_A)\) here. We reserve \(r,\theta\) for spacetime
coordinates when discussing Schwarzschild and Kerr rays. The contribution at degree \(m\)
is
\begin{equation}
\begin{aligned}{}
P_A^{(m)}(\vartheta)&=\calM_{AB_1\cdots B_m}n^{B_1}\cdots n^{B_m},&
 \delta\xi_A^{(m)}&=\frac{\rho^m}{m!}P_A^{(m)}(\vartheta).
 \label{eq:pattern}
\end{aligned}
\end{equation}
The vector \(P_A^{(m)}\) is a displacement arrow for each position
around the circle. Its radial component changes the radius there; its
tangential component shifts the point around the circle.

The degree \(m\) sets the power of angular distance from the expansion centre: doubling
\(\rho\) multiplies a quadratic
contribution by four. Spin describes how the coefficient pair transforms
when the screen axes are rotated: a rotation of the axes through an angle
\(\varphi\) rotates that pair through \(s\varphi\). This axis rotation
is distinct from varying the offset azimuth \(\vartheta\) around a fixed
input circle. Shear therefore has degree one and spin
two. We keep the amplitudes \(\alpha_s^m,\beta_s^m\) of Papers~I
and II, and write their cosine and sine combinations as
\begin{equation}
\begin{aligned}{}
\sigma_s^m&=\alpha_s^m\cos s\vartheta+\beta_s^m\sin s\vartheta,&
 \omega_s^m&=-\alpha_s^m\sin s\vartheta+\beta_s^m\cos s\vartheta .
 \label{eq:polarisation}
\end{aligned}
\end{equation}
For \(s>0\), \(\omega_s^m=(\mathrm d\sigma_s^m/\mathrm d\vartheta)/s\);
it supplies the tangential component in \eqref{eq:potential-spins}.
The pair \(\alpha_s^m,\beta_s^m\) gives the two orientations of one
pattern, just as \(\gamma_1,\gamma_2\) give the two components of
shear. Appendix~\ref{app:amplitude-dictionary} gives the equivalent complex
potential formula.

For a single scalar-potential lens, a homogeneous potential contribution
has the form \(\rho^{m+1}f_m(\vartheta)/(m+1)\). Its gradient is
\[
 \rho^m\left[f_m n_A+\frac{f'_m}{m+1}t_A\right].
\]
The radial component is \(H_m=n^AP_A^{(m)}\) and the tangential
component is \(K_m=t^AP_A^{(m)}\); hence \(K_m=H'_m/(m+1)\).
Expanding \(H_m\) in Fourier modes gives
\begin{equation}
\begin{aligned}{}
P_A^{(m)}&=\sum_{\substack{0\leq s\leq m+1\\m+s\ {\rm odd}}}
 \left(\sigma_s^m n_A+\frac{s}{m+1}\omega_s^m t_A\right).
 \label{eq:potential-spins}
\end{aligned}
\end{equation}
Only the displayed spins occur: \(s=0,2\) at degree one,
\(s=1,3\) at degree two, and \(s=0,2,4\) at degree three.
The spin-zero sine coefficient is absent.

A general linear map also has the independent antisymmetric part of
\eqref{eq:linear-split-review}, labelled by \(\varpi\). Including it,
the degree-one amplitudes of the full map are
\begin{equation}
\begin{aligned}{}
\alpha_0^1&=\chi_S(1-\kappa),&
 \alpha_2^1&=-\chi_S\gamma_1,&
 \beta_2^1&=-\chi_S\gamma_2,&
 \bar\alpha_0^1&=\chi_S\varpi .
 \label{eq:full-linear-amplitudes}
\end{aligned}
\end{equation}
The bar denotes the independent tangential component in
\eqref{eq:linear-split-review}; we construct its higher-degree versions
in section~\ref{sec:barred-review}. Throughout this section the amplitudes
have dimensions of length and include the unlensed linear map.
Appendix~\ref{app:amplitude-dictionary} gives the conversion to the
dimensionless, lensing-only amplitudes used for potential derivatives in
Papers~I and II. The two quadratic spins are the familiar flexion sectors
\citep{bacon,hessian}.

For a scalar-potential lens, define the angular complex derivative
\(\partial=\partial/\partial\theta_1+i\partial/\partial\theta_2\)
and shear \(\gamma=\gamma_1+i\gamma_2\). The first flexion
\(\mathcal F=\partial\kappa\) has spin one and describes the
leading lopsided distortion; the second flexion
\(\mathcal G=\partial\gamma\) has spin three and describes the
threefold distortion \citep{goldbergbacon,bacon}. In our forward-map
normalisation, \(\alpha_1^2+i\beta_1^2=-3\chi_S\mathcal F/2\)
and \(\alpha_3^2+i\beta_3^2=-\chi_S\mathcal G/2\).
These are precisely the spin-one and spin-three quadratic patterns
obtained above; both come from the third angular derivatives of \(\psi\).\footnote{With
\(\mathcal G=\partial\gamma\), the radial derivative expression for
\(\mu_3^2\) in Paper~II's equation~(120) equals \(\mathcal G/2\) on the
positive \(X\)-axis. Its final identification with \(\mathcal G\) differs
by a factor of two; the roulette coefficient itself agrees.}
The unindexed \(\mathcal F\) here is flexion; the indexed
\(\mathcal F_n^a\) defined in section~\ref{sec:nlgde-result} denotes a
spacetime relative-acceleration term.

\subsection{A worked singular isothermal sphere arc}\label{sec:sis-worked}
\begin{figure}[!htb]
  \centering
\includegraphics[width=\textwidth,height=.72\textheight,keepaspectratio]{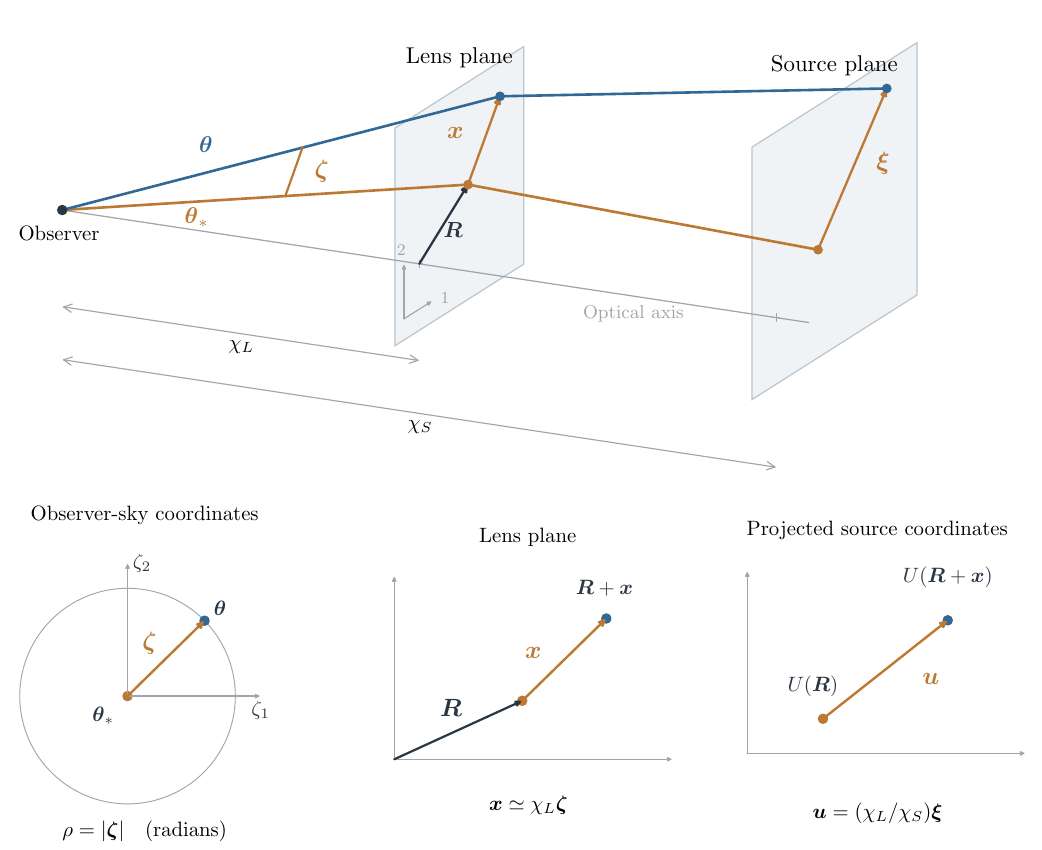}
  \caption{Observed angular offsets and thin-lens displacements.
The reference and neighbouring rays meet the lens at \(\boldsymbol R\) and
\(\boldsymbol R+\boldsymbol x\); \(\boldsymbol\xi\) joins their source endpoints. The lower panels show
the same positions in the sky coordinates, lens plane and projected source
coordinates, with \(\boldsymbol x\simeq\chi_L\boldsymbol\zeta\) and
\(\boldsymbol u=(\chi_L/\chi_S)\boldsymbol\xi\). The enlarged sky panel shows the angular radius
\(\rho\) in radians. The line of sight is
shortened; the transverse axes are common, with \(\boldsymbol R\) oblique to them.}
  \label{fig:coordinates}
\end{figure}

For the SIS we can calculate the map by ordinary differentiation.
We align the first transverse axis with the reference ray's lens-plane
position; figure~\ref{fig:coordinates} shows a general orientation.
Let \(R_A=(R,0)\) be this position,
and let \(x_A=\chi_L\zeta_A\) be a nearby displacement, with
\(\chi_L\) the observer--lens distance. To compare all transverse displacements in the same length units, we
project source positions back onto the lens plane and denote them by \(U_A\). Their local
difference is \(u_A=U_A(\boldsymbol R+\boldsymbol x)-U_A(\boldsymbol R)=(\chi_L/\chi_S)\xi_A\).
Figure~\ref{fig:coordinates} summarises these coordinates.

For an SIS with Einstein radius \(b=R_{\rm E}\), the lens equation is
\begin{equation}
\begin{aligned}{}
U_A(\boldsymbol r)&=r_A-b\frac{r_A}{|\boldsymbol r|},& R_A&=(R,0),\qquad R>b .
 \label{eq:SIS-map}
\end{aligned}
\end{equation}
Here \(r_A\) is an absolute lens-plane position measured from the mass.
Expanding at \(R_A\) through cubic degree gives
\begin{equation}
\begin{aligned}{}
u_1&=x_1+\frac{b}{2R^2}x_2^2-\frac{b}{R^3}x_1x_2^2
                 +O\!\left(b\frac{|\boldsymbol x|^4}{R^4}\right),\\
 u_2&=\left(1-\frac bR\right)x_2+\frac b{R^2}x_1x_2
       -\frac b{R^3}x_1^2x_2+\frac b{2R^3}x_2^3
                 +O\!\left(b\frac{|\boldsymbol x|^4}{R^4}\right).
 \label{eq:SIS-worked}
\end{aligned}
\end{equation}
The small factor \(1-b/R\) produces tangential elongation: a source
width \(\sigma\) occupies an image width \(\sigma/(1-b/R)\) in
the linear approximation. The quadratic terms then bend the image and
allow its width to vary. The cubic terms refine this shape. Figure~\ref{fig:worked}
uses \(R=1.10b\), close enough to the Einstein radius that the cubic
approximation still leaves a visible discrepancy.

The degree-25 map
recovers more of the arc, but its accuracy deteriorates near the circle
\(|\boldsymbol x|=R\). This is the radius of convergence of the local
SIS series: the circle reaches the singular lens centre. In the angular
coordinates used here the domain is \(\rho<R/\chi_L\). The second
image lies outside it, so extending that reconstruction requires another
centre. Section~\ref{sec:image-centres} explains the distinction between
image multiplicity and convergence, and appendix~\ref{app:bounds} gives
the analytic recurrence and remainder bounds.

To find the amplitudes we separate each degree into its radial and
tangential patterns. For example, the radial quadratic contribution is
\begin{equation}
\begin{aligned}{}
n^AP_A^{(2)}
 &=\frac{3b\chi_S\chi_L}{R^2}\cos\vartheta\sin^2\vartheta
 =\frac{3b\chi_S\chi_L}{4R^2}(\cos\vartheta-\cos3\vartheta).
 \label{eq:SIS-radial}
\end{aligned}
\end{equation}
Reading off the cosine coefficients, together with the linear terms,
we obtain
\begin{equation}
\begin{aligned}{}
\frac{\alpha_0^1}{\chi_S}&=1-\frac b{2R},&
 \frac{\alpha_2^1}{\chi_S}&=\frac b{2R},&\beta_2^1&=0,\\
 \frac{\alpha_1^2}{\chi_S}&=\frac{3b\chi_L}{4R^2},&
 \frac{\alpha_3^2}{\chi_S}&=-\frac{3b\chi_L}{4R^2},&
 \beta_1^2&=\beta_3^2=0 .
 \label{eq:SIS-amplitudes}
\end{aligned}
\end{equation}
The factors of \(\chi_L\) convert length derivatives to angular
derivatives. Substitution in \eqref{eq:potential-spins} recovers
\eqref{eq:SIS-worked} through quadratic degree;
appendix~\ref{app:practical-amplitudes} lists the cubic amplitudes.

We now place a Gaussian source at \(\boldsymbol U(\boldsymbol R)\), with width \(\sigma\).
The source and its predicted image are
\begin{equation}
\begin{aligned}{}
I_s(\boldsymbol U)&=\exp\!\left[-\frac{|\boldsymbol U-\boldsymbol U(\boldsymbol R)|^2}{2\sigma^2}\right],\\
 I^{[m_{\max}]}(\boldsymbol x)&=\exp\!\left[-\frac{|\boldsymbol u^{[m_{\max}]}(\boldsymbol
x)|^2}{2\sigma^2}\right].
 \label{eq:SIS-image}
\end{aligned}
\end{equation}
The superscript \([m_{\max}]\) denotes the truncated polynomial.
We add its map contributions before evaluating the source. Surface
brightness is conserved; magnification changes the area of the image.

\subsection{Barred roulettes: the extra tangential freedom}
\label{sec:barred-review}
The radial arrows around an input circle determine the unbarred
amplitudes. For a potential lens they also fix the tangential arrows
through \eqref{eq:potential-spins}. A general map has additional
tangential freedom: its output index need not share the symmetry of its
Taylor input indices. Misaligned shears on two lens planes provide an
example (section~\ref{sec:weak-two-lens}). Separating the fully symmetric
part gives

\begin{equation}
\begin{aligned}{}
\calM_{AB_1\cdots B_m}
 &=\calM_{(AB_1\cdots B_m)}
   +\epsilon_{A(B_1}\overline{\calM}_{B_2\cdots B_m)},\\
 \overline{\calM}_{B_2\cdots B_m}
 &=\frac{m}{m+1}\epsilon^{AB}\calM_{AB B_2\cdots B_m} .
 \label{eq:split}
\end{aligned}
\end{equation}
The first tensor is symmetric in all indices, as a scalar-potential
derivative must be. The barred tensor is symmetric in its own indices
and supplies the additional components of a general map. Contracting
the first line with \(\epsilon^{AB_1}\) removes the fully symmetric
part; the averaged terms in the second part give
\((m+1)\overline{\calM}/m\), yielding the inverse in the second line.
For \(m=1\) this is exactly \eqref{eq:linear-split-review}.

\begin{figure}[!htb]
  \centering
\includegraphics[width=\textwidth,height=.72\textheight,keepaspectratio]{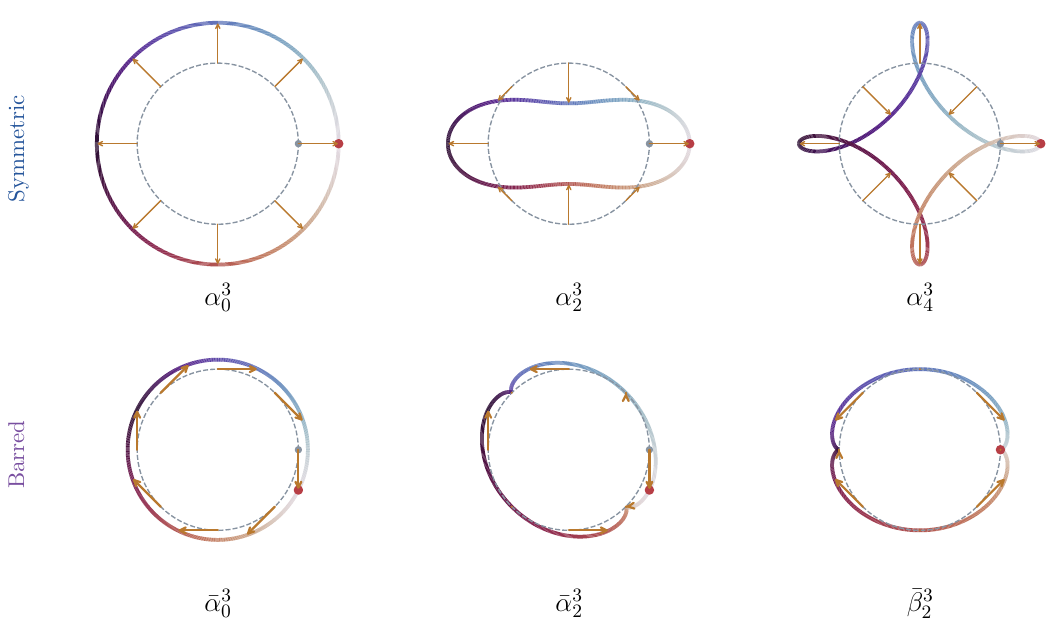}
  \caption{Symmetric and barred cubic roulettes displace an input circle in
different ways. Colours identify the same initial direction on the dashed
circle and on the displaced curve; orange arrows join corresponding points.
The upper row shows symmetric modes and the lower row independent barred
modes. In the barred spin-zero example (lower left), the displacement is
uniform along the local tangent: the endpoints form an enlarged, rotated
circle. Each panel plots \(n_A+0.5P_A^{(3)}\), with the indicated
coefficient set to unity, showing one basis pattern at a time.}
  \label{fig:basis}
\end{figure}

We write
\(\bar\sigma_s^m=\bar\alpha_s^m\cos s\vartheta+\bar\beta_s^m\sin s\vartheta\).
The following expression combines the potential pattern in
\eqref{eq:potential-spins} with this independent tangential displacement.
It is the general reconstruction formula, including barred amplitudes
(Paper~II, equation (58)):
\begin{equation}
\begin{aligned}{}
P_A^{(m)}&=\sum_{\substack{0\leq s\leq m+1\\m+s\ {\rm odd}}}
 \left[\sigma_s^m n_A+
       \left(\frac{s}{m+1}\omega_s^m-\bar\sigma_s^m\right)t_A\right] .
 \label{eq:spins}
\end{aligned}
\end{equation}
The barred term points around the input circle, along \(t_A\). Barred
spins end at \(s=m-1\), two below the highest unbarred spin, and
\(\beta_0^m=\bar\beta_0^m=0\). Table~\ref{tab:spins} lists the
resulting spins and component counts, and figure~\ref{fig:basis} shows
the cubic patterns. Section~\ref{sec:curvature-orders} relates the
higher-degree families to their curvature sources.

\begin{table}[!htb]\centering\small
\begin{minipage}[c]{0.43\textwidth}
\caption{Spin content at each angular degree \(m\), whose map tensor
has rank \(m+1\). The \(2(m+1)\) real components divide into
\(m+2\) symmetric and \(m\) barred components. A scalar-potential
lens occupies the symmetric sector.}
\label{tab:spins}
\end{minipage}\hfill
\begin{minipage}[c]{0.54\textwidth}\centering
\begin{tabular}{cccc}\toprule
\(m\)&Symmetric spins&Barred spins&Components\\\midrule
1&\(0,2\)&\(0\)&\(3+1=4\)\\
2&\(1,3\)&\(1\)&\(4+2=6\)\\
3&\(0,2,4\)&\(0,2\)&\(5+3=8\)\\
4&\(1,3,5\)&\(1,3\)&\(6+4=10\)\\
5&\(0,2,4,6\)&\(0,2,4\)&\(7+5=12\)\\\bottomrule
\end{tabular}
\end{minipage}\end{table}

The amplitudes can be extracted without manipulating high-rank tensors.
For a calculated pattern, let \(H_m=n^AP_A^{(m)}\) and
\(K_m=t^AP_A^{(m)}\) be its radial and tangential components.
Equation~\eqref{eq:spins} gives
\begin{equation}
\begin{aligned}{}
H_m&=\sum_s\sigma_s^m,&
 \frac{\mathrm d_\vartheta H_m}{m+1}-K_m&=\sum_s\bar\sigma_s^m .
 \label{eq:fourier}
\end{aligned}
\end{equation}
Here \(\mathrm d_\vartheta\) differentiates with respect to the azimuth.
Fourier coefficients of the first function give \(\alpha_s^m,\beta_s^m\);
those of the second give their barred counterparts. The integrals and
normalisations are in appendix~\ref{app:practical-amplitudes}.

\subsection{Reconstructing an image}\label{sec:guide}\label{app:image-procedure}
Once the map and source brightness are specified, we reconstruct an image
by evaluating the source at the positions predicted by the local expansion.
For a thin lens the map is the usual
\(\boldsymbol\beta(\boldsymbol\theta)\). Starting from a spacetime
metric, we first specify the observer, angular coordinates and source endpoint;
section~\ref{sec:theory} explains how to calculate the corresponding map. At each chosen expansion direction
\(\boldsymbol\theta_j\), we calculate the map derivatives through degree \(m_{\max}\) and retain
the central source position
\(\boldsymbol\beta_j=\boldsymbol\beta(\boldsymbol\theta_j)\).
The latter fixes where the local displacement lands in the source.
For each image pixel we set
\(\boldsymbol\zeta_j=\boldsymbol\theta-\boldsymbol\theta_j\)
in these angular coordinates and calculate
\begin{equation}
\begin{aligned}{}
\beta_A^{[m_{\max}]}(\boldsymbol\theta)
 &=\beta_{jA}+\frac1{\chi_S}\sum_{m=1}^{m_{\max}}
       \frac{\calM_{AB_1\cdots B_m}(\boldsymbol\theta_j)}{m!}
       \zeta_j^{B_1}\cdots\zeta_j^{B_m},\\
 I^{[m_{\max}]}(\boldsymbol\theta)
 &=I_s\!\left(\boldsymbol\beta^{[m_{\max}]}(\boldsymbol\theta)\right).
 \label{eq:render-recipe}
\end{aligned}
\end{equation}
For length-valued outputs we evaluate the source in those coordinates.
In the SIS example, this means adding \(\boldsymbol U(\boldsymbol R)\)
to the polynomial displacement before evaluating the source profile.
Adding the roulette displacement patterns first and then evaluating the
brightness gives the same image as the tensor polynomial.
Our examples prescribe the source position, size and brightness pattern;
the expansion does not require a circular source. An observational fit
would jointly infer the lens and source, convolve the predicted brightness
with the PSF, and integrate over detector pixels.

We determine the useful region of each expansion by checking
source positions and brightness over the full intended image window,
including any false polynomial images. The analytic thin-lens bounds
in appendix~\ref{app:bounds} control position errors. Magnification
and critical curves also require accurate derivatives of the map.
Where one expansion is insufficient, we calculate derivatives
and the source position at additional centres, and assign pixels to
regions where the required accuracy is reached. Translating an existing
truncated polynomial cannot recover its missing derivatives.
Section~\ref{sec:changing-centres} constructs such regions for the thin
lenses; section~\ref{sec:kerr-coverage} illustrates several centres along
images lensed by a rotating black hole.

\subsection{How one map can give several images}
\label{sec:image-centres}
One local expansion can describe several images if its forward map
remains accurate over the region containing them. Near a fold caustic,
a quadratic map already gives an image pair. In locally rescaled dimensionless
coordinates, consider
\begin{equation}
\begin{aligned}{}
u_1&=x_1,&u_2&=x_2^2,&\det(\nabla_Bu_A)&=2x_2 .
 \label{eq:fold-normal-form}
\end{aligned}
\end{equation}
A source with \(u_2>0\) has two images at
\(x_2=\pm\sqrt{u_2}\). As it approaches \(u_2=0\), the images
meet on the \emph{critical curve} \(x_2=0\), where the linear map
vanishes in one direction; for \(u_2<0\) the pair is absent.
The source-plane image of the critical curve is the \emph{caustic}.
Throughout this change the forward map remains smooth, so one roulette
expansion can describe both images if it is accurate across the pair.
Figure~\ref{fig:fold} demonstrates this for an extended source.

The SIS in figure~\ref{fig:worked} shows a different limitation. Its
singular centre restricts the radius of convergence, leaving the second
image outside the series centred on the primary arc. Increasing the
degree improves the map within this domain; additional centres extend
the region covered, as proposed in Paper~II \citep{clarksonII} and
developed in section~\ref{sec:changing-centres}. Each candidate image must lie within a patch
where the map error
is controlled.

We call a map \emph{complete through degree} \(m_{\max}\) when every
Taylor coefficient through that degree is calculated from the specified
lens model, angular coordinates and endpoint. The neglected remainder then
sets its useful angular region. At a fold, the inverse image position
becomes multivalued while the forward map stays smooth; the singular
centre of the SIS can obstruct the forward expansion itself. Finite
Taylor accuracy requires sufficient differentiability, while convergence
of the infinite series requires analyticity. A ray tangent to the
chosen source surface can independently spoil the endpoint map
(section~\ref{sec:endpoint-choice}).

%% file: sections/03_covariant_propagation.tex
\section{Covariant propagation of a finite beam}\label{sec:theory}
The gravitational field changes the separation of nearby rays
launched in neighbouring directions. The local lens map follows by
propagating this separation to a specified source surface. We define
the finite separation using the short geodesic joining the rays, and
use \emph{force} below as shorthand for a contribution to its covariant
relative acceleration, \(D^2\xi^a/\dd\lambda^2\). Each photon
follows a geodesic; the relative acceleration describes how their
separation changes. The angular derivatives of the initial photon
direction supply the launch data. Propagating the separation and its rate
then gives the roulette maps degree by degree.

Let \(k^a=\dd x^a/\dd\lambda\) be the future-directed reference
tangent. A dot on a spacetime tensor denotes its covariant affine
derivative, \(\dot T=DT/\dd\lambda=k^b\nabla_bT\). Thus the
connector tangent \(\xi^a=\log_x(x')^a\), joining equal-affine
events, has separation rate \(v^a=\dot\xi^a\). In a
parallel-transported basis, dots give ordinary component derivatives.
Indices \(a,b,\ldots\) are spacetime indices, \(i,j,k\) are spatial,
and \(A,B,\ldots\) label the two screen directions. Spacetime and
screen indices are raised and lowered with \(g_{ab}\) and
\(\delta_{AB}\), respectively. We use signature \((-+++)\) and
\([\nabla_c,\nabla_d]w^a=R^a{}_{bcd}w^b\).

A vector written in an index slot means contraction into that slot;
for example,
\begin{equation}
\begin{aligned}
 R^a{}_{k\xi k}&=R^a{}_{bcd}k^b\xi^ck^d,&
 (\nabla_\xi R)^a{}_{k\xi k}
 &=\xi^e k^b\xi^ck^d\nabla_eR^a{}_{bcd}.
 \label{eq:contractions}
\end{aligned}
\end{equation}
In these force contractions, covariant derivatives act on curvature
with the contracted vectors held fixed. Ordered derivatives act from
right to left, so in \(\nabla_k\nabla_\xi R\) the
\(\xi\)-derivative acts first.

We parallel transport the observer's orthonormal transverse vectors
\(e_A^a\) along the reference ray, with
\(g_{ab}e_A^ae_B^b=\delta_{AB}\), \(k_ae_A^a=0\) and
\(\nabla_ke_A^a=0\). The screen components are
\(\xi_A=e_{Aa}\xi^a\) and \(v_A=e_{Aa}v^a\); the full separation
can also have longitudinal components (figure~\ref{fig:geometry}).
Following Papers~0, I and II, we write \(k^a=E(u^a-n^a)\), where
\(E=-u_ak^a>0\) and \(n^a\) points from observer towards source.
With \(\lambda_o=0\), the source lies at \(\lambda_s<0\), so
tracing rays from observer to source integrates towards decreasing
\(\lambda\). The positive affine distance in the examples is
\(L=-\lambda_s\). Table~\ref{tab:notation} collects these conventions
and the quantities introduced below.

\begin{table}\centering\small
\setlength{\tabcolsep}{3pt}
\begin{tabular}{p{0.21\columnwidth}p{0.73\columnwidth}}\toprule
Symbol&Meaning\\\midrule
\(\boldsymbol\theta,\boldsymbol\beta\)&Observed image and source positions\\[3pt]
\(\boldsymbol\theta_*\)&Observed direction chosen as the expansion centre\\[3pt]
\(\boldsymbol\zeta,\rho,\vartheta\)&Local angular offset, angular radius (radians) and azimuth\\[3pt]
\(k^a,\xi^a,v^a\)&Reference tangent, connector tangent and separation rate\\[3pt]
\(A,B\)&Components along the two screen vectors of an orthonormal tetrad; Cartesian angular or
lens-plane indices in thin-lens formulae\\[3pt]
\(m,n\)&Angular degree of the map; total degree in \((\xi^a,v^a)\) of a local force\\[3pt]
\(p,d,r\)&Numbers of curvature factors, curvature derivatives and relative-direction factors\\[3pt]
\(h,\epsilon\)&Scaling parameter; metric perturbation parameter\\[3pt]
\(\mathcal R_{AB}\)&Screen tidal matrix; \(R^a{}_{bcd}\) is spacetime curvature\\[3pt]
\(\mathcal F^a(\xi,v),\mathcal F_n^a(\xi,v)\)&Relative acceleration as a function of separation and rate; its homogeneous force-degree-\(n\) part\\[3pt]
\(\mathcal J,\mathcal K\)&Jacobi propagators with angular-launch and displacement initial data, respectively\\[3pt]
\(\mathcal G^a{}_b(\lambda,\lambda')\)&Green function for Jacobi propagation; its affine arguments distinguish it from spin-three flexion\\[3pt]
\(H_{ab},H_{\rm gw}\)&Magnetic Weyl tensor; plane-wave tidal profile\\[3pt]
\(w^a,w,w_{\rm iso}\)&Tangent along the connector; flat product coordinate; isothermal
pressure-to-density ratio\\[3pt]
\(\dot T\)&Covariant affine derivative \(DT/\dd\lambda=k^b\nabla_bT\); ordinary component
derivative in a parallel-transported basis\\[3pt]
\bottomrule
\end{tabular}
\caption{Notation and expansion labels. Indexed orientation tensors
\(\epsilon_{AB},\epsilon_{ijk}\) and metric perturbations \(h_{ij}\)
are distinguished from the scalar counting parameters. The Kerr polar
coordinate \(\theta\) is separate from the roulette azimuth \(\vartheta\).}
\label{tab:notation}\label{tab:four-labels}\end{table}

\subsection{From the Jacobi map and Sachs optics to a finite beam}
\label{sec:sachs}
For an infinitesimal bundle, the screen tidal matrix
\begin{equation}
\begin{aligned}{}
\mathcal R_{AB}&=-e_A^ae_B^cR_{abcd}k^bk^d
 \label{eq:tidal-definition}
\end{aligned}
\end{equation}
governs the linear angle-to-length map,
\begin{equation}
\begin{aligned}{}
\xi_A&=\mathcal J_{AB}\zeta^B+O(\rho^2),&
 \ddot{\mathcal J}_{AB}&=\mathcal R_{AC}\mathcal J^C{}_B,\\
 \mathcal J_{AB}(0)&=0,&
 \dot{\mathcal J}_{AB}(0)&=-E_o\delta_{AB} .
 \label{eq:sachs-jacobi}
\end{aligned}
\end{equation}
Here \(u_o^a\) is the observer four-velocity and
\(E_o=-u_{oa}k_o^a\) fixes the observed photon energy. At \(E_o=1\)
the initial derivative is \(-I\), as in Paper~II.
The endpoint value is the linear coefficient \(\mathcal M_{AB}\)
for this screen prescription.

Sachs optics describes the same beam using its fractional rate of change.
Where \(\mathcal J_{AB}\) is invertible, define
\begin{equation}
\begin{aligned}{}
\mathcal S^A{}_B&=\dot{\mathcal J}^A{}_C(\mathcal J^{-1})^C{}_B,\\
 \dot{\mathcal S}^A{}_B+\mathcal S^A{}_C\mathcal S^C{}_B
 &=\mathcal R^A{}_B .
 \label{eq:sachs-matrix}
\end{aligned}
\end{equation}
The trace, symmetric trace-free part and antisymmetric part of
\(\mathcal S_{AB}\) give expansion, shear and twist. These are rates;
\(\mathcal J_{AB}\) gives their accumulated linear map. The inverse fails
at a caustic, although the Jacobi map can still be propagated through it.
The product \(\mathcal S^A{}_C\mathcal S^C{}_B\) is nonlinear in
the optical rates, while the Jacobi approximation still has
\(\xi_A=\mathcal J_{AB}\zeta^B\), linear in the observed angular
offset. Finite-separation terms require the higher angular derivatives
supplied by the NLGDE.

A light cone from a point observer has zero optical twist. This does
not require its accumulated Jacobi map to be symmetric: successive
shears with different orientations can produce an antisymmetric map.
The barred linear amplitude describes this antisymmetric part
\citep{penmao,robertsonlewis2023}. We next construct the finite
separation, state its propagation law through cubic degree, and use it
to build the hierarchy at arbitrary degree.

\subsection{The connector: finite separation and relative direction}
\label{sec:bundle-geometry}
\begin{figure}[!htb]
  \centering
  \includegraphics[width=\textwidth,height=.72\textheight,keepaspectratio]{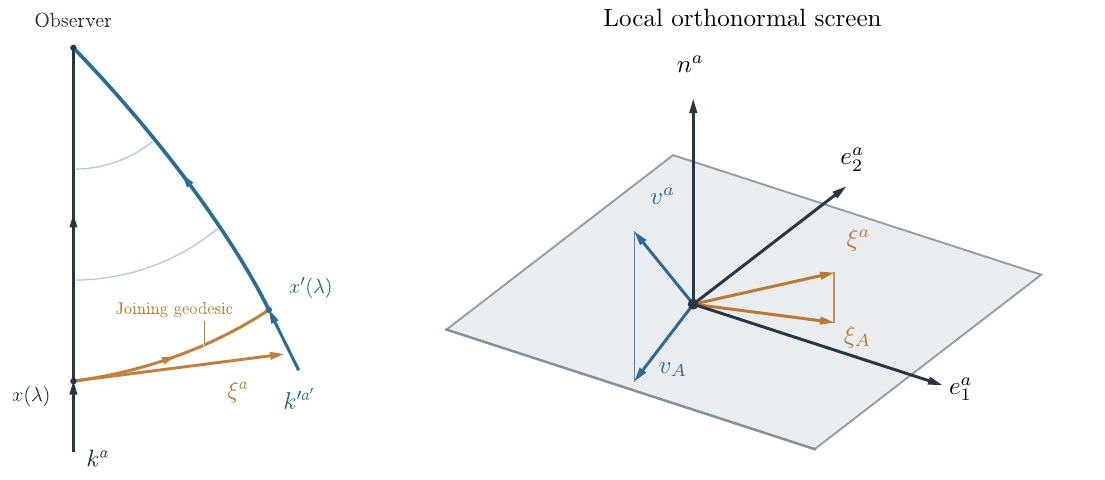}
  \caption{Finite separation and its projection onto a local screen.
Left: neighbouring rays are paired at equal affine parameter \(\lambda\).
The joining geodesic has initial tangent \(\xi^a\) at \(x(\lambda)\).
The arrows labelled \(k^a\) and \(k'^{a'}\) point towards the observer,
in the future direction; tracing the rays into the past decreases
\(\lambda\). The pale arcs pair equal affine parameters.
Right: the screen is perpendicular to \(n^a\) and spanned by
\(e_1^a,e_2^a\). The orange arrows show \(\xi^a\) and its screen
projection \(\xi_A=e_{Aa}\xi^a\); the blue arrows show
the separation rate \(v^a\) and its projection \(v_A=e_{Aa}v^a\).
The two panels illustrate the connector and the projection separately;
the right-hand vectors are schematic.}
  \label{fig:geometry}
\end{figure}

At equal affine parameter, join the reference event \(x\) to its
neighbour \(x'\) by the connector \(\gamma(\tau)\),
\(0\leq\tau\leq1\), with initial tangent \(\xi^a\):
\begin{equation}
\begin{aligned}{}
\frac{\dd^2\gamma^a}{\dd\tau^2}
 &+\Gamma^a{}_{bc}(\gamma)
       \frac{\dd\gamma^b}{\dd\tau}\frac{\dd\gamma^c}{\dd\tau}=0,\\
 \gamma^a(0)&=x^a,\qquad
 \left.\frac{\dd\gamma^a}{\dd\tau}\right|_0=\xi^a,\qquad
 \gamma^{a'}(1)=x'^{a'} .\label{eq:logmap}
\end{aligned}
\end{equation}
The exponential map constructs the second event from the first:
start at \(x\) with tangent \(\xi^a\), follow the connecting geodesic,
and take its endpoint at \(\tau=1\). We write \(x'=\exp_x\xi\).
In flat Cartesian coordinates this is simply \(x'^a=x^a+\xi^a\).
The tangent arrow in figure~\ref{fig:geometry} belongs to the tangent
space at \(x\); the curved connector it generates lies in spacetime.
Conversely, given the two events, the logarithm
\(\xi^a=\log_x(x')^a\) returns the initial tangent of their joining
geodesic. We work in a neighbourhood where this connector is unique.
Its screen components \(\xi_A\) are the output coordinates
of our equal-affine examples.

The separation rate \(v^a\) describes how that
initial tangent changes as both rays advance. Comparing photon directions
requires parallel transport of the neighbouring tangent back along the
connector before subtracting the reference tangent. In flat space the
difference equals \(v^a\). In curved spacetime, moving the connector's
initial event also changes the geodesic joining the two events, so its
endpoint variation cannot be obtained from the change of tangent alone.
Writing the transported direction as \(p^a\), the leading difference
is \(p^a-k^a-v^a=\tfrac12R^a{}_{\xi\xi k}+O(\rho^3)\)
when \(\xi,v=O(\rho)\). Thus this geometric correction already
enters at quadratic degree.

We can separate the two operations: move the initial event while carrying
the tangent in parallel, or change the tangent at a fixed event
(upper panels of figure~\ref{fig:connector-variations}). Both occur as the photons
advance. Successive connectors also sweep out a strip between the rays.
Curvature across this strip changes the transported photon direction,
even though the direction is parallel along the photon's own ray
(lower panel). Appendix~\ref{app:geometry}
derives these relations using the world function and Jacobi propagators
\citep{synge1931,synge1960,dewittbrehme,poisson2011,dixon1970,dixon1979},
with the transported null screen of \citet{sachs1961} and \citet{jordanehlerssachs}.

\subsection{The nonlinear geodesic deviation equation}
\label{sec:nlgde-result}
To describe the bending and changing width of an arc, we need the
relative acceleration at finite ray separation. \citet{vines} derived
an exact equation using covariant bitensors and expanded it through fourth
order; \citet{kim} organised the finite-separation dynamics on the tangent
bundle. We use these results to calculate the angular derivatives of the
lensing map.

We write the exact local relative acceleration as
\(\ddot\xi^a=\mathcal F^a(\xi,v)\), with separation \(\xi\)
as its first vector argument and covariant rate \(v\) as its second.
At a fixed reference event \(x\) and tangent \(k^a\), we define its
homogeneous degree-\(n\) part by scaling both arguments together:
\[
 \mathcal F_n^a(\xi,v)
 =\left.\frac1{n!}\frac{\dd^n}{\dd h^n}
                \mathcal F^a(h\xi,hv)\right|_{h=0},\qquad n\geq1.
\]
Thus \(\mathcal F_n^a(h\xi,hv)=h^n\mathcal F_n^a(\xi,v)\),
\(\mathcal F_0^a=0\), and the linear force is
\(\mathcal F_1^a(\xi,v)=-R^a{}_{k\xi k}\).
The tangent, curvature and its ordered derivatives are fixed coefficients;
appendix~\ref{app:force-recipe} gives the recurrence for these force
polynomials at any degree for which the curvature is sufficiently differentiable.
The indexed force \(\mathcal F_n^a\) is distinct from the complex
flexion \(\mathcal F\) of section~\ref{sec:degree-spin-review}.
When differentiating the force we can vary either vector argument while
holding the other fixed, then insert the separation and rate of the
propagated ray family. With the curvature convention above, positive
sectional curvature on a Riemannian factor makes
\(-R^a{}_{k\xi k}\) point towards the reference geodesic for
transverse \(\xi^a\), giving the expected focus.

Projecting the linear geodesic deviation equation onto the Sachs screen gives
\begin{equation}
\begin{aligned}{}
\ddot\xi_A&=\calR_{AB}\xi^B+O(\rho^2).
 \label{eq:linear-gde}
\end{aligned}
\end{equation}
The trace of \(\mathcal R_{AB}\) focuses the beam and its trace-free
part shears it. Moving this tidal term to the left defines the linear
Jacobi operator. The same operator propagates the unknown map at each
angular degree, with nonlinear forcing supplied by the lower-degree
solutions on the right.

The nonlinear force samples curvature variation across the separation,
variation along the ray and the changing relative directions. At cubic
degree, the geometry of the connector and its parallel transport also
produces local products of curvature. These contributions appear in the
complete equation through cubic degree in \(\xi^a\) and \(v^a\):
\begin{equation}
\begin{aligned}{}
\ddot\xi^a={}&\underbrace{-R^a{}_{k\xi k}}_{\substack{\text{linear tidal}\\\text{acceleration}}}
                +\calF_2^a+\calF_3^a+O(\rho^4),
 \quad \xi^a,v^a=O(\rho),\label{eq:nlgde-result}
\end{aligned}
\end{equation}
with the quadratic correction
\begin{equation}
\begin{aligned}{}
\calF_2^a(\xi,v)={}&-\tfrac12(\nabla_\xi R)^a{}_{k\xi k}
 +\tfrac12(\nabla_k R)^a{}_{\xi k\xi}
 -2R^a{}_{v\xi k},\label{eq:F2}
\end{aligned}
\end{equation}
and the cubic correction
\begin{equation}
\begin{aligned}{}
\calF_3^a(\xi,v)={}&-\tfrac16(\nabla_\xi\nabla_\xi R)^a{}_{k\xi k}
 +\tfrac16(\nabla_k\nabla_\xi R)^a{}_{\xi k\xi}\\
 &+\tfrac16R^a{}_{k\xi b}R^b{}_{\xi k\xi}
  +\tfrac12R^a{}_{b\xi k}R^b{}_{\xi k\xi}
  -\tfrac16R^a{}_{\xi b\xi}R^b{}_{k\xi k}\\
 &-(\nabla_\xi R)^a{}_{v\xi k}
  +\tfrac13(\nabla_k R)^a{}_{\xi v\xi}
  +\tfrac23R^a{}_{vv\xi}.\label{eq:F3}
\end{aligned}
\end{equation}
A quadratic force can still contribute to a cubic angular map when
one of its arguments contains a second-degree separation, as made
explicit below.

Restricting the contractions \((\nabla_\xi R)^a{}_{k\xi k}\) in \eqref{eq:F2}
and \((\nabla_\xi\nabla_\xi R)^a{}_{k\xi k}\) in
\eqref{eq:F3} to transverse separation and screen output gives the
leading transverse-gradient force. The remaining terms describe changes
along the ray and in its relative direction, together with local curvature
products at cubic degree. Longitudinal separation is propagated with the
screen components because it can contribute to their later evolution.

Appendix~\ref{app:geometry} derives these forces from the exact Vines--Kim
equation. The mixed cubic term in \eqref{eq:F3}
corrects the printed equation~(32) of \citet{kim}; the quartic coefficients
derived in appendix~\ref{app:quartic} correct four terms in the printed
equation~(83) of \citet{vines}. Three were already identified by
\citet{kim}; table~\ref{tab:vines-quartic} gives the attribution
and the additional fourth correction.

\subsubsection{The physical contributions across a finite beam}\label{sec:three-effects}
The force terms distinguish variation of curvature across and along the
beam, changes in relative ray direction, and the geometry of finite
separation. Figure~\ref{fig:geometry} shows the separation and propagation
directions entering these contractions. Angular degree counts powers
of the initial angular offset; section~\ref{sec:angular-degree-count}
gives the detailed count.
\begin{description}
\item[Variation across the separation.] The contraction
\((\nabla_\xi R)^a{}_{k\xi k}\) in \eqref{eq:F2} samples changes
of the tidal tensor between neighbouring rays. Its transverse part
contains the familiar curvature gradient responsible for flexion.
\item[Variation along the ray.] The contraction
\((\nabla_k R)^a{}_{\xi k\xi}\) in \eqref{eq:F2} differentiates
curvature along \(k^a\), following its variation as the beam propagates.
\item[Changing relative directions.] The contraction
\(R^a{}_{v\xi k}\) in \eqref{eq:F2} depends on \(v^a\) as well
as \(\xi^a\). Rays crossing a gravitational field at different
inclinations acquire different relative accelerations.
\item[Finite separation and parallel transport.] The three curvature
products in \eqref{eq:F3}, such as
\(R^a{}_{k\xi b}R^b{}_{\xi k\xi}\), arise because curvature
changes the connecting geodesic and the transport along it. They can
remain nonzero when curvature gradients vanish.
\end{description}
The cubic derivative and relative-direction terms resolve the same
physical effects to higher degree. Propagation also couples distortions
acquired earlier to the forces acting later: a quadratic change in the
beam separation enters the later quadratic force and contributes at
cubic angular degree. Section~\ref{sec:hierarchy} works through this
insertion using the approximations defined in
section~\ref{sec:general-force}. The complete quartic
force is in appendix~\ref{app:quartic}.

\subsubsection{The weak-field flexion limit}\label{sec:weak-flexion-limit}
The leading transverse term recovers the usual flexion map
\citep{hessian}. In this limit, we work to first order in scalar metric
perturbations. We also use the small-angle Born approximation and retain
the leading transverse derivatives. The conformal comoving radial distance
\(\chi\) then replaces the affine parameter in the reduced ray equation.
Let \(\Phi,\Psi\) be the scalar metric potentials in conformal
Newtonian coordinates. Then
\(\calR_{AB}\simeq-\nabla_A\nabla_B(\Phi+\Psi)\) and
\(\xi_{(1)A}=\chi\zeta_A\), with \(\nabla_A\) a comoving
transverse derivative. Matching the quadratic coefficient gives
\begin{equation}
\begin{aligned}{}
\frac{\dd^2\calM_{ABC}}{\dd\chi^2}
 &\simeq-\chi^2\nabla_A\nabla_B\nabla_C(\Phi+\Psi),\\
 \calM_{ABC}(\chi_S)
 &\simeq-\int_0^{\chi_S}(\chi_S-\chi)\chi^2
             \nabla_A\nabla_B\nabla_C(\Phi+\Psi)\,\dd\chi.
 \label{eq:flexion-born-bridge}
\end{aligned}
\end{equation}
The factor \(1/2\) in the force cancels the Taylor factor in
\(\xi_A=\calM_{AB}\zeta^B+\calM_{ABC}\zeta^B\zeta^C/2+\cdots\).
For the lensing potential
\[
 \psi(\boldsymbol\theta)=\int_0^{\chi_S}
 \frac{\chi_S-\chi}{\chi_S\chi}
 (\Phi+\Psi)\big|_{\rm Born}\,\dd\chi,
\]
the fields are evaluated at \(\eta=\eta_o-\chi\) and
\(\boldsymbol x=(\chi\theta_1,\chi\theta_2,\chi)\). Each angular
derivative contributes \(\chi\nabla_A\). Hence
\(\calM_{ABC}=-\chi_S\psi_{,ABC}\), with the sign and factorial
convention of \eqref{eq:map-review}. Its spin-one and spin-three parts
are the flexions defined in section~\ref{sec:review}. This limit uses
the comoving source-plane separation. Section~IV of Paper~0
\citep{hessian} gives the cosmological derivation;
appendix~\ref{app:born-roulettes} gives its leading-gradient roulette
limit. Physical optical fields additionally require the angular and
source-ruler conversions in appendix~\ref{app:physical-optical-fields}.

\subsubsection{Counting angular degree}\label{sec:angular-degree-count}
Scale the observed offset as \(\zeta^A\mapsto h\zeta^A\), keeping
the reference ray fixed. Both \(\xi^a\) and \(v^a\) start at
order \(h\). Curvature and its tensor derivatives, evaluated on that
ray, supply coefficients rather than angular powers:
\[
\begin{array}{c|c|c}
\text{factor}&\text{meaning}&\text{first degree}\\\hline
\xi^a&\text{separation}&1\\
v^a&\text{separation rate}&1\\
k^a&\text{reference direction}&0\\
\xi^a\nabla_a&\text{derivative along separation}&1\\
v^a\nabla_a&\text{derivative along separation rate}&1\\
k^a\nabla_a&\text{derivative along the ray}&0
\end{array}
\]
For example, \((\nabla_\xi R)^a{}_{k\xi k}\) in
\eqref{eq:F2} contains one \(\xi^a\) in \(\nabla_\xi\) and one
in a curvature slot; \((\nabla_kR)^a{}_{\xi k\xi}\) has two
separation slots, and \(R^a{}_{v\xi k}\) has one separation slot
and one rate slot. Each therefore begins at degree two,
irrespective of the number of curvature derivatives. A screen derivative is
\(e_A^a\nabla_a\); an along-ray derivative is \(k^a\nabla_a\).
The derivative \(\xi^a\nabla_a\) includes longitudinal separation
when present. Higher angular terms in \(\xi^a\) and \(v^a\) then
make each quadratic force contribute to later map degrees as well.
These derivatives of spacetime curvature differ from the angular
derivatives of the endpoint map in section~\ref{sec:review}.
This is an expansion in angular offset, with no requirement that the
curvature be weak.

\subsubsection{The general force and its couplings}\label{sec:general-force}
At force degree \(n\), the relative acceleration contains terms with
one Riemann tensor and, from \(n=3\), products of Riemann tensors.
Grouping by their number \(p\), we can write
\begin{equation}
 \mathcal F_n^a(\xi,v)
 =\left.\mathcal F_n^a\right|_R
  +\sum_{p=2}^{\lfloor(n+1)/2\rfloor}
       \left.\mathcal F_n^a\right|_{R^p},\qquad n\geq2.
 \label{eq:general-force-sectors}
\end{equation}
Covariant derivatives may act on each curvature factor, and every
contraction has \(n\) separation or rate factors in total. There are at
most two rate factors, because the acceleration is quadratic in the
endpoint tangents. The count in \eqref{eq:curvature-degree-count}
relates these numbers to the number of curvature derivatives.

The comparisons use the same linear tidal matrix and projected initial
data, with the nonlinear propagation specified as follows:
\begin{description}
\item[TG.] Transverse-gradient forcing acts on products of the linear
separation.
\item[TG+C.] Lower-degree distortions feed back into the same force,
beginning with the quadratic separation entering at cubic angular degree.
\item[Full.] We also retain the other forces and their subsequent
lower-degree couplings. Examples are the along-ray term
\((\nabla_kR)^a{}_{\xi k\xi}\) and relative-direction term
\(R^a{}_{v\xi k}\) in \eqref{eq:F2}, and the curvature product
\(R^a{}_{k\xi b}R^b{}_{\xi k\xi}\) in \eqref{eq:F3}.
\end{description}

Equation~\eqref{eq:one-curvature-all} gives the single-curvature terms
for \(n\geq3\), with the quadratic case in \eqref{eq:F2}. Its first contraction,
\((\nabla_\xi^{n-1}R)^a{}_{k\xi k}\), supplies the TG force when
the separation and output are transverse. The other contractions
contain an along-ray derivative \(\nabla_k\), a rate vector \(v^a\)
in a curvature slot, or an ordered derivative \(\nabla_v\).
They describe variation along the ray and changes in relative ray
direction. Evaluating them on the linear separation and rate gives
direct forcing at angular degree \(m=n\), without any feedback from
earlier nonlinear distortions. Even when curvature is covariantly
constant, the relative-direction terms \(R^a{}_{v\xi k}\) in
\eqref{eq:F2} and \(R^a{}_{vv\xi}\) in
\eqref{eq:F3} can remain, together with the curvature
products in \eqref{eq:F3}. These forces are included in
Full and tested in the constant-curvature example of
section~\ref{sec:fixed-degree-comparison}. The linear tidal acceleration
\(-R^a{}_{k\xi k}\) is retained in TG, TG+C and Full. It is distinct
from \(R^a{}_{\xi\xi b}\), the quadratic connector operator used in
appendix~\ref{app:geometry}.

Propagation couples the force degrees through their separation and
rate arguments. A degree-\(n\) force contributes to angular degree
\(m\) through products whose input degrees obey
\(j_1+\cdots+j_n=m\), with every \(j_i\geq1\). The direct term has
\(n=m\) and all \(j_i=1\), while \(n<m\) feeds lower-degree
nonlinear maps back into the force. C denotes this feedback within
the transverse-gradient force. It first appears at cubic angular
degree when the quadratic separation enters the quadratic force.
At degree four, that force receives either a cubic and a linear
separation or two quadratic separations; the cubic force also
receives a quadratic separation and two linear ones. The full
hierarchy includes the corresponding feedback through the
along-ray, relative-direction and curvature-product forces.

Starting from the leading transverse-gradient approximation of
Papers~I and II, C restores its nonlinear propagation. The other
single-curvature forces and the local curvature products then
complete the force at each degree. C can be large when a
distorted beam continues through an extended lens or encounters
another deflector. In the weak-lens geometries tested below, the along-ray and
relative-direction contributions are suppressed by the impact scale
relative to the nearer endpoint distance, and curvature products by
the potential depth
(section~\ref{sec:relevance}). Their relative importance also
depends on beam width, since the remaining forces change the
quadratic map while C first changes the cubic map. The compact lens
and wave examples test their contributions when weak-field
suppression fails. Section~\ref{sec:tg-benchmarks} gives the explicit
TG and TG+C equations, including the cubic C term.

\subsection{Matter focusing and the Weyl tidal matrix}\label{sec:ricci-weyl}
The linear optical tidal matrix separates into an isotropic Ricci part and a
trace-free Weyl part: the former changes the beam area, while the latter
produces anisotropic stretching and can be nonzero in vacuum. We denote the Weyl tensor by
\(C_{abcd}\), its contribution to
the screen tidal matrix by \(\mathcal W_{AB}\), and the complete
optical tidal matrix by \(\mathcal R_{AB}\). With the conventions above,
\begin{equation}
\begin{aligned}{}
R_{abcd}&=C_{abcd}+R^{\rm Ric}_{abcd},\\
 R^{\rm Ric}_{abcd}&=
 \tfrac12\bigl(g_{ac}R_{db}-g_{ad}R_{cb}
                -g_{bc}R_{da}+g_{bd}R_{ca}\bigr)\\
 &\quad-\tfrac R6\bigl(g_{ac}g_{db}-g_{ad}g_{cb}\bigr),\\
 \calR_{AB}&=-\tfrac12R_{ab}k^ak^b\delta_{AB}+\mathcal W_{AB},\\
 \mathcal W_{AB}&=-e_A^ae_B^c C_{abcd}k^bk^d,\qquad\delta^{AB}\mathcal W_{AB}=0 .
 \label{eq:ricci-weyl}
\end{aligned}
\end{equation}
In Einstein gravity, with \(c=1\), \(R_{ab}k^ak^b=8\pi G T_{ab}k^ak^b\).
It focuses the beam when \(T_{ab}k^ak^b>0\); a cosmological constant drops out
of this null contraction. The Weyl tidal matrix can shear a beam in vacuum.

Substituting this decomposition into \eqref{eq:F2}--\eqref{eq:F3}
separates the Ricci and Weyl contributions to each amplitude.
Appendix~\ref{app:ricci-weyl-explicit} gives the contraction rule for
Weyl-squared, mixed Weyl--Ricci and pure Ricci products.
The split into local products and derivatives refers to a fixed ordering
of covariant derivatives: commuting derivatives moves curvature products
between displayed terms. We keep that ordering fixed when comparing individual contributions.

\subsection{From ray separation to the lens map}\label{sec:hierarchy}
To obtain the angular derivatives of the map, expand a family of rays
about the reference direction. Fix an angular offset
\(\zeta^A\) in the observer's coordinates. The dimensionless parameter \(h\) labels the
family of angular offsets \(h\zeta^A\): \(h=0\) is the reference
ray and \(h=1\) the chosen neighbouring ray. Varying \(h\) changes angular distance along a fixed azimuth on
the observer's sky, with the spacetime and affine parameter held fixed.
Here we expand the full spacetime separation \(\xi^a\), whose screen
components are \(\xi_A=e_{Aa}\xi^a\). The longitudinal components
must also be propagated. The Taylor expansion in this angular scale is
\begin{equation}
\begin{aligned}{}
\xi^a(\lambda,h\boldsymbol\zeta)
 &=\sum_{m\geq1}\frac{h^m}{m!}\xi_{(m)}^a(\lambda;\boldsymbol\zeta),\\
 \xi_{(m)}^a
 &=\left.\frac{\partial^m\xi^a(\lambda,h\boldsymbol\zeta)}{\partial h^m}
       \right|_{h=0},\qquad v_{(m)}^a=\frac{D\xi_{(m)}^a}{\dd\lambda}=\dot\xi_{(m)}^a.
 \label{eq:coupling-example}
\end{aligned}
\end{equation}
These directional coefficients contract the map tensors with the chosen
angular offset:
\[
 \begin{aligned}
 \xi_{(m)}^a&=\calM^a{}_{B_1\cdots B_m}\zeta^{B_1}\cdots\zeta^{B_m},\\
 \xi_{(m)A}&=e_{Aa}\xi_{(m)}^a
   =\calM_{AB_1\cdots B_m}\zeta^{B_1}\cdots\zeta^{B_m}.
 \end{aligned}
\]
For \(\zeta^A=\rho n^A(\vartheta)\), these coefficients include the
angular radius: \(\xi_{(m)A}=\rho^mP_A^{(m)}(\vartheta)\).
The patterns \(P_A^{(m)}\) and their spin amplitudes use the unit
direction \(n^A\), with this factor of \(\rho^m\) removed.
The rate \(v^a\) has the same expansion with coefficients
\(v_{(m)}^a\). The reference tangent and curvature are held fixed.

Substitute these series into \eqref{eq:F2}--\eqref{eq:F3} and
equate equal powers of \(h\). The first two equations are
\begin{equation}
\begin{aligned}{}
\ddot\xi_{(1)}^a+R^a{}_{k b k}\xi_{(1)}^b&=0,\\
 \ddot\xi_{(2)}^a+R^a{}_{k b k}\xi_{(2)}^b
 &=2\mathcal F_2^a(\xi_{(1)},\dot\xi_{(1)})\\
 &=-\xi_{(1)}^b\xi_{(1)}^c(\nabla_bR)^a{}_{k c k}
   +\xi_{(1)}^b\xi_{(1)}^c(\nabla_kR)^a{}_{b k c}
   -4\dot\xi_{(1)}^b\xi_{(1)}^cR^a{}_{b c k}.
 \label{eq:hierarchy-howto}
\end{aligned}
\end{equation}
For the cubic map, each quadratic product can contain one first-degree
and one second-degree factor. Explicitly,
\[
 \xi^b\dot\xi^c=h^2\xi_{(1)}^b\dot\xi_{(1)}^c
 +\frac{h^3}{2}\bigl(\xi_{(1)}^b\dot\xi_{(2)}^c+
                        \xi_{(2)}^b\dot\xi_{(1)}^c\bigr)+O(h^4).
\]
Applying this rule to all three quadratic-force terms gives
\begin{equation}
\begin{aligned}{}
\ddot\xi_{(3)}^a+R^a{}_{k b k}\xi_{(3)}^b={}&
 -\tfrac32\bigl(\xi_{(1)}^b\xi_{(2)}^c+\xi_{(2)}^b\xi_{(1)}^c\bigr)
             (\nabla_bR)^a{}_{k c k}\\
 &+\tfrac32\bigl(\xi_{(1)}^b\xi_{(2)}^c+\xi_{(2)}^b\xi_{(1)}^c\bigr)
             (\nabla_kR)^a{}_{b k c}\\
 &-6\bigl(\dot\xi_{(1)}^b\xi_{(2)}^c+\dot\xi_{(2)}^b\xi_{(1)}^c\bigr)
             R^a{}_{b c k}
   +6\mathcal F_3^a(\xi_{(1)},\dot\xi_{(1)}).
 \label{eq:cubic-hierarchy-main}
\end{aligned}
\end{equation}
The factors follow from the Taylor factorials. In particular, the last
term is the cubic force evaluated on the linear solution, while the
preceding terms insert the quadratic solution into the quadratic force.
At every degree the unknown separation is propagated by the linear
Jacobi operator, with a source calculated from the lower-degree solutions
along the ray and initial data fixed by the observer's angular coordinates.

The same substitution gives any finite angular degree for which the
curvature is sufficiently differentiable. It distributes the previously
calculated maps among the separation and rate slots of each force term.
Appendix~\ref{app:maps} gives the general rule, its factorials and
its Green-function solution. Propagation followed by the chosen
source-surface intersection fixes the endpoint map. Its screen
projection with \(e_{Aa}\) and Fourier decomposition
\eqref{eq:fourier} then give the roulette amplitudes.

\subsubsection{The transverse-gradient benchmarks}\label{sec:tg-benchmarks}
Applying the TG prescription of section~\ref{sec:general-force}
to the angular hierarchy, the highest transverse derivative at each
degree acts on products of the linear map. For \(m\geq2\),
\begin{equation}
\begin{aligned}{}
\ddot\xi_{(m)}^A-\calR^A{}_B\xi_{(m)}^B
 &=\xi_{(1)}^{B_1}\cdots\xi_{(1)}^{B_m}
   \nabla_{B_1}\cdots\nabla_{B_{m-1}}\calR^A{}_{B_m}.
 \label{eq:tg-main}
\end{aligned}
\end{equation}
TG+C also inserts lower-degree maps into the transverse-gradient force.
The quadratic equation is unchanged; the cubic equation becomes
\begin{equation}
\begin{aligned}{}
\ddot\xi_{(3)}^A-\calR^A{}_B\xi_{(3)}^B
 &={}\xi_{(1)}^B\xi_{(1)}^C\xi_{(1)}^D
       \nabla_D\nabla_C\calR^A{}_B\\
 &\quad+3\xi_{(1)}^{(B}\xi_{(2)}^{C)}\nabla_C\calR^A{}_B.
 \label{eq:tgc-main}
\end{aligned}
\end{equation}
The last term is the \(n=2\), \(j_1+j_2=1+2\) coupling C described
in section~\ref{sec:general-force}. Equation~\eqref{eq:tg-main} is the
leading-gradient hierarchy of Paper~II's equation~(7), while
\eqref{eq:tgc-main} is the consistent angular expansion of its force
equation~(4). The quadratic Ba\.za\'nski equation used in Paper~0 also
included the along-ray gradients and relative-direction terms retained
by Full.

\subsection{Choosing observed angles and a source endpoint}
\label{sec:endpoint-choice}
\begin{figure}[!htbp]
\centering
\begin{minipage}[c]{0.31\textwidth}\centering
\includegraphics[width=\linewidth,trim=7bp 9bp 36bp 13bp,clip]{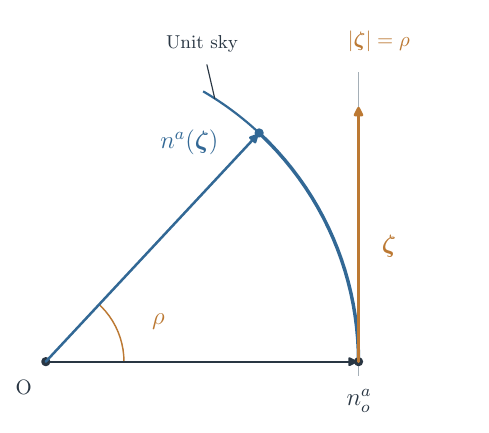}
\end{minipage}\hfill
\begin{minipage}[c]{0.345\textwidth}\centering
\includegraphics[width=\linewidth,trim=7bp 3bp 275bp 3bp,clip]{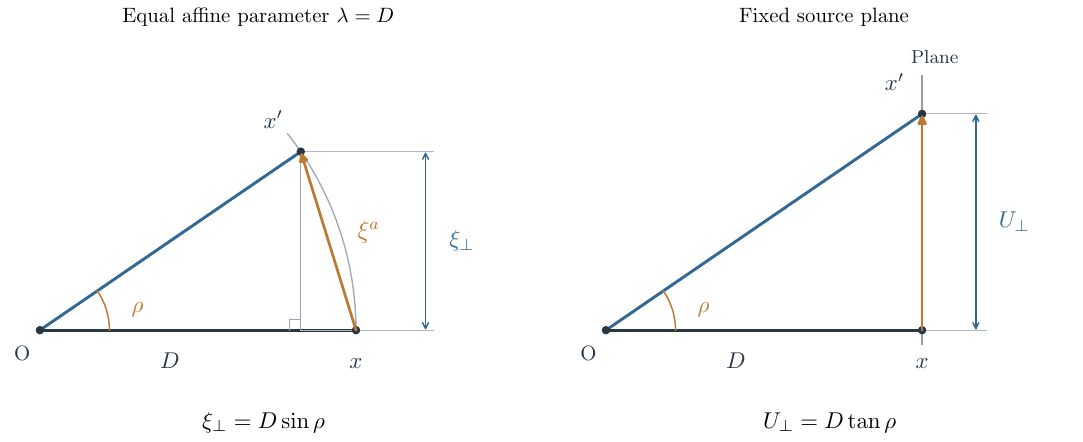}
\end{minipage}\hfill
\begin{minipage}[c]{0.325\textwidth}\centering
\includegraphics[width=\linewidth,trim=277bp 3bp 12bp 3bp,clip]{figures/fig06b_source_endpoints.pdf}
\end{minipage}
\caption{Sky directions and endpoint prescriptions. Left: the great-circle
arc from \(n_o^a\) to \(n^a(\boldsymbol\zeta)\) has angular length
\(\rho=|\boldsymbol\zeta|\), equal to the norm of the ochre tangent
vector. With \(e^a=e_A^a\zeta^A/\rho\), the selected direction is
\(n^a(\boldsymbol\zeta)=\cos\rho\,n_o^a+\sin\rho\,e^a\).
The point \(O\) is the origin in direction space; subtracting this direction from \(u_o^a\)
and multiplying by \(E_o\) gives \eqref{eq:sky}.
Centre: pairing rays at affine distance \(D\) in flat spacetime gives
transverse separation \(\xi_\perp=D\sin\rho\) and longitudinal
separation \(D(\cos\rho-1)\). Right: a fixed source plane gives
\(U_\perp=D\tan\rho\). The transverse maps agree at linear degree
but have cubic terms \(-D\rho^3/6\) and \(+D\rho^3/3\), respectively.
The marked lengths are transverse projections; angles are enlarged for clarity.}
\label{fig:sky-launch}\label{fig:endpoints}
\end{figure}

The angular coordinates at the observer and the source-surface
intersection both contribute to the nonlinear map derivatives. We use geodesic normal coordinates on the
observer's unit sky: \(\rho\) is the angular distance from the reference
direction in radians, and \(\vartheta\) is its azimuth. Thus
\(\zeta^A=\rho n^A(\vartheta)\) is dimensionless. The exact initial null tangent is
\begin{equation}
\begin{aligned}{}
k_o^a(\boldsymbol\zeta)&=E_o\left[u_o^a-\cos\rho\,n_o^a
                  -\frac{\sin\rho}{\rho}e_A^a\zeta^A\right] .
 \label{eq:sky}
\end{aligned}
\end{equation}
Here \((u_o^a,n_o^a,e_1^a,e_2^a)\) is an orthonormal observer tetrad:
\(u_o^a\) is the four-velocity, and the three other vectors form its
spatial frame. The unit vector \(n_o^a\) points along the reference
spatial direction from observer to source. The sine ratio is regular at
\(\rho=0\), and converts the normal sky-coordinate offset into a
unit-sphere direction (figure~\ref{fig:sky-launch}). Together the sine
and cosine keep every launched ray null and at the common observed
energy \(E_o=-u_{oa}k_o^a\). This fixes the relative affine normalisation
of the rays; equal affine parameter is not in general equal emission
time or equal source redshift. The same angular dependence fixes the
higher-degree initial data, which cannot be set to zero while retaining
these sky coordinates.

All position maps vanish initially because the rays leave one observer.
Their initial affine derivatives are the angular derivatives of the
initial tangent. Expanding \eqref{eq:sky} gives
\begin{equation}
\begin{aligned}{}
\dot\calM^a{}_B|_o&=-E_o e_B^a,&
 \dot\calM^a{}_{BC}|_o&=E_o n_o^a\delta_{BC},\\
 \dot\calM^a{}_{BCD}|_o&=E_o e^a{}_{(B}\delta_{CD)} .
 \label{eq:initial}
\end{aligned}
\end{equation}
Differentiating the exact direction fixes the initial data at every
higher degree (appendix~\ref{app:sky-data}). These initial
terms describe the geometry of the observer's sky before lens curvature
is encountered.

A flat-spacetime example explains why we retain the full vector. With
\(E_o=1\), straight rays at equal affine parameter satisfy
\begin{equation}
\begin{aligned}{}
\xi^a&=-\lambda\left[(\cos\rho-1)n_o^a+
                    \frac{\sin\rho}{\rho}e_A^a\zeta^A\right],\\
 k_a\xi^a&=-\frac12\lambda\rho^2+O(\rho^4),\\
 \xi_A&=-\lambda\zeta_A+\frac16\lambda\rho^2\zeta_A+O(\rho^5).
 \label{eq:flat}
\end{aligned}
\end{equation}
The endpoints lie on a sphere around the observer, rather than on a flat
screen. Their joining vector therefore has a longitudinal component
beginning at quadratic degree. In curved spacetime that component can affect
later transverse propagation. We evolve the spacetime vector and project
it afterwards. The cubic term in \eqref{eq:flat} comes entirely from the sky coordinates and
the equal-affine endpoint. In the present convention it
gives \(\alpha_0^1=-\lambda\) and \(\alpha_0^3=\lambda\).

Now stop the same straight rays at a plane a distance \(D\) along the
reference direction. The neighbouring ray reaches that plane at
\(\lambda=-D/\cos\rho\). Its transverse endpoint is therefore
\begin{equation}
\begin{aligned}{}
U_A^{\rm plane}&=D\frac{\tan\rho}{\rho}\zeta_A
 =D\zeta_A+\frac D3\rho^2\zeta_A+O(\rho^5).
 \label{eq:flat-plane}
\end{aligned}
\end{equation}
In the plane convention the cubic roulette is \(\alpha_0^3[U]=2D\),
whereas the equal-affine result is \(\alpha_0^3[\xi]=-D\) at \(\lambda=-D\).
The two prescriptions agree at linear degree and differ at cubic degree
even in flat spacetime. Subtracting a flat contribution therefore requires
the same sky coordinates and endpoint as the curved-spacetime map
(figure~\ref{fig:endpoints}).

For a specified emitting surface \(F(x)=0\), the arrival parameter is
itself a function of observed direction:
\begin{equation}
\begin{aligned}{}
F\!\left(x'(\lambda_s(\boldsymbol\zeta),\boldsymbol\zeta)\right)&=0 .
 \label{eq:endpoint-condition}
\end{aligned}
\end{equation}
We compose the ray solution with this intersection before extracting the
source-coordinate amplitudes. This local intersection requires
\(F_{,a}k^a\ne0\): a ray tangent to the chosen source surface can
invalidate the arrival-parameter expansion even when the ray flow is smooth
(appendix~\ref{app:endpoint-composition}).

The examples use two endpoint prescriptions. Pairing rays at equal
affine parameter, taking their logarithmic separation and projecting it
onto the Sachs screen defines an \emph{equal-affine logarithmic screen}.
The constant-curvature test uses this prescription. Kerr uses a fixed
source plane, and the Schwarzschild example compares both choices.
Within each comparison, the input coordinates and endpoint prescription are
held fixed.

The flat examples first differ at cubic degree, giving a relative
transverse correction of \(O(\rho^2)\). That is a property of
these particular endpoints: a tilted, curved or redshift-selected
source surface can change lower-degree coefficients through the
derivatives of its arrival parameter
(appendix~\ref{app:endpoint-composition}).
Changing the sky coordinates also changes the nonlinear coefficients, since
a nonlinear coordinate transformation mixes angular degrees. To
interpret a coefficient as an image distortion, we must therefore
convert coordinate increments into measured angles and physical source
separations. Appendix~\ref{app:physical-optical-fields} gives this
conversion using the angular metric and source scale.

\subsection{Roulette amplitudes as integrals along the ray}
\label{sec:amplitude-integrals}
Curvature generates distortions along the ray, and the Jacobi propagator
carries each contribution to the final screen. We can project this
propagation directly into roulette spins. Here we use the equal-affine
logarithmic map with the already specified source-screen axes.

All angular input indices refer to the same observer sky. The source
\(S^a{}_{B_1\cdots B_m}\) is the angular derivative of the nonlinear
forces in \eqref{eq:amplitude-force-tensor}. It uses the lower-degree
separation and rate maps at the integration point. Equation~\eqref{eq:all-orders}
shows the Taylor factorials, and \eqref{eq:connector-low-degree-sources}
gives the quadratic and cubic insertions explicitly.

We use the same Jacobi notation for the full spacetime propagators,
\(\mathcal J^a{}_{b_o}\) and \(\mathcal K^a{}_{b_o}\), where
\(b_o\) is an index at the observer. Their observer data are
\(\mathcal J(0)=0\), \(\dot{\mathcal J}(0)=-E_o I\),
\(\mathcal K(0)=I\) and \(\dot{\mathcal K}(0)=0\), with
\(I\) the identity. Their transverse projection gives the screen map
of \eqref{eq:sachs-jacobi}:
\[
 \mathcal J_{AB}(\lambda)
 =e_{Aa}(\lambda)\mathcal J^a{}_{b_o}(\lambda)e_B^{b_o}(0)
 =\mathcal M_{AB}(\lambda).
\]
The full propagators also carry the longitudinal components of the
nonlinear sources. Their Green function is
\begin{equation}
 \mathcal G(\lambda_s,\lambda)
 =\frac1{E_o}\left[
       \mathcal K(\lambda_s)\mathcal J(\lambda)^\dagger
       -\mathcal J(\lambda_s)\mathcal K(\lambda)^\dagger\right].
\label{eq:amplitude-jacobi-green}
\end{equation}
Here \(\dagger\) is the metric adjoint between the full tangent
spaces, reducing to a transpose on the Sachs screen.
Appendix~\ref{app:ray-propagation} derives this expression and its
relation to the reciprocal-Jacobi form of Paper~II.
The launch derivatives follow from the exact sky direction
\eqref{eq:sky}. At linear degree,
\(\dot{\mathcal M}^a{}_B(0)=-E_o e_B^a\), so
\(\mathcal M^a{}_B(\lambda)=\mathcal J^a{}_{b_o}(\lambda)e_B^{b_o}(0)\).
Higher launch derivatives retain the geometry of the unit sky.
With the future-directed tangent used here, the source lies at
\(\lambda_s<0\), and the integral retains its observer-to-source orientation.

The equal-affine screen map is therefore
\begin{equation}
\begin{aligned}
 \mathcal M_{AB_1\cdots B_m}(\lambda_s)
 =e_{Aa}(\lambda_s)\Biggl[
 &-\frac1{E_o}\mathcal J^a{}_{b_o}(\lambda_s)
       \dot{\mathcal M}^{b_o}{}_{B_1\cdots B_m}(0)\\
 &+\int_0^{\lambda_s}\mathcal G^a{}_b(\lambda_s,\lambda)
       S^b{}_{B_1\cdots B_m}(\lambda)\,\dd\lambda\Biggr].
 \\[-1pt]
 S^a{}_{B_1\cdots B_m}
 ={}&\left.\partial_{B_1}\cdots\partial_{B_m}
       \sum_{n=2}^{m}\mathcal F_n^a(\xi,\dot\xi)
       \right|_{\boldsymbol\zeta=0}.
 \label{eq:force-integral-map}
\end{aligned}
\end{equation}
Here \(\partial_B=\partial/\partial\zeta^B\). The first term propagates
the initial angular derivatives, and the integral accumulates the
nonlinear forcing along the reference ray.
At \(m=1\), the source vanishes and this reduces to the Jacobi map.
At every higher degree, the source is calculated from the maps of
lower degree. The fixed spin projector then extracts an amplitude,
\begin{equation}
\begin{aligned}
 \alpha_s^m+i\beta_s^m
 &=\Pi_s^{AB_1\cdots B_m}\mathcal M_{AB_1\cdots B_m},\\
 \Pi_s^{AB_1\cdots B_m}
 &=\frac1{\pi(1+\delta_{s0})}\int_0^{2\pi}
       e^{is\vartheta}n^A n^{B_1}\cdots n^{B_m}\,\dd\vartheta .
\end{aligned}
\label{eq:amplitude-spin-projector}
\end{equation}
Substituting \eqref{eq:force-integral-map} gives the force integral
for each spin,
\begin{equation}
\begin{aligned}
 \alpha_s^m+i\beta_s^m
 =\Pi_s^{AB_1\cdots B_m}e_{Aa}(\lambda_s)\Biggl[
 &-\frac1{E_o}\mathcal J^a{}_{b_o}(\lambda_s)
       \dot{\mathcal M}^{b_o}{}_{B_1\cdots B_m}(0)\\
 &+\int_0^{\lambda_s}\mathcal G^a{}_b(\lambda_s,\lambda)
       S^b{}_{B_1\cdots B_m}(\lambda)\,\dd\lambda\Biggr].
\end{aligned}
\label{eq:force-integral-amplitudes}
\end{equation}
The real and imaginary parts take the radial cosine and sine
projections. At zero spin the cosine weight is \(1/(2\pi)\) and
the sine vanishes. No additional factorial enters the amplitudes.
Allowed spins have \(m+s\) odd and \(s\leq m+1\). The barred
amplitudes are the corresponding Fourier projections of
\(H'_m/(m+1)-K_m\), using the radial and tangential components
of the same map, and end at \(s=m-1\).
For the exact wave in section~\ref{sec:wave}, this equal-affine
construction reproduces all unbarred and barred amplitudes through
cubic degree.

This generalises the screen propagation integral in Paper~II,
equation~(15), where transverse-gradient forces act on the linear
map \citep{clarksonII}. Paper~0 already includes the quadratic
along-ray and relative-direction forces, with
\(\mathcal M_{ABC}=2\mathcal H_{ABC}\) for matched sky, screen
and endpoint conventions \citep{hessian}. The scalar,
first-metric-order, background-ray limit with leading transverse
gradients gives the potential roulettes of Papers~I and II
\citep{clarksonI,clarksonII}, with the coordinate-map assumptions
and dimensionless normalisation in
appendix~\ref{app:born-roulettes}.

\begingroup
\begin{figure}[!tp]\centering
\begin{minipage}[t]{\textwidth}\centering
\includegraphics[width=\textwidth]{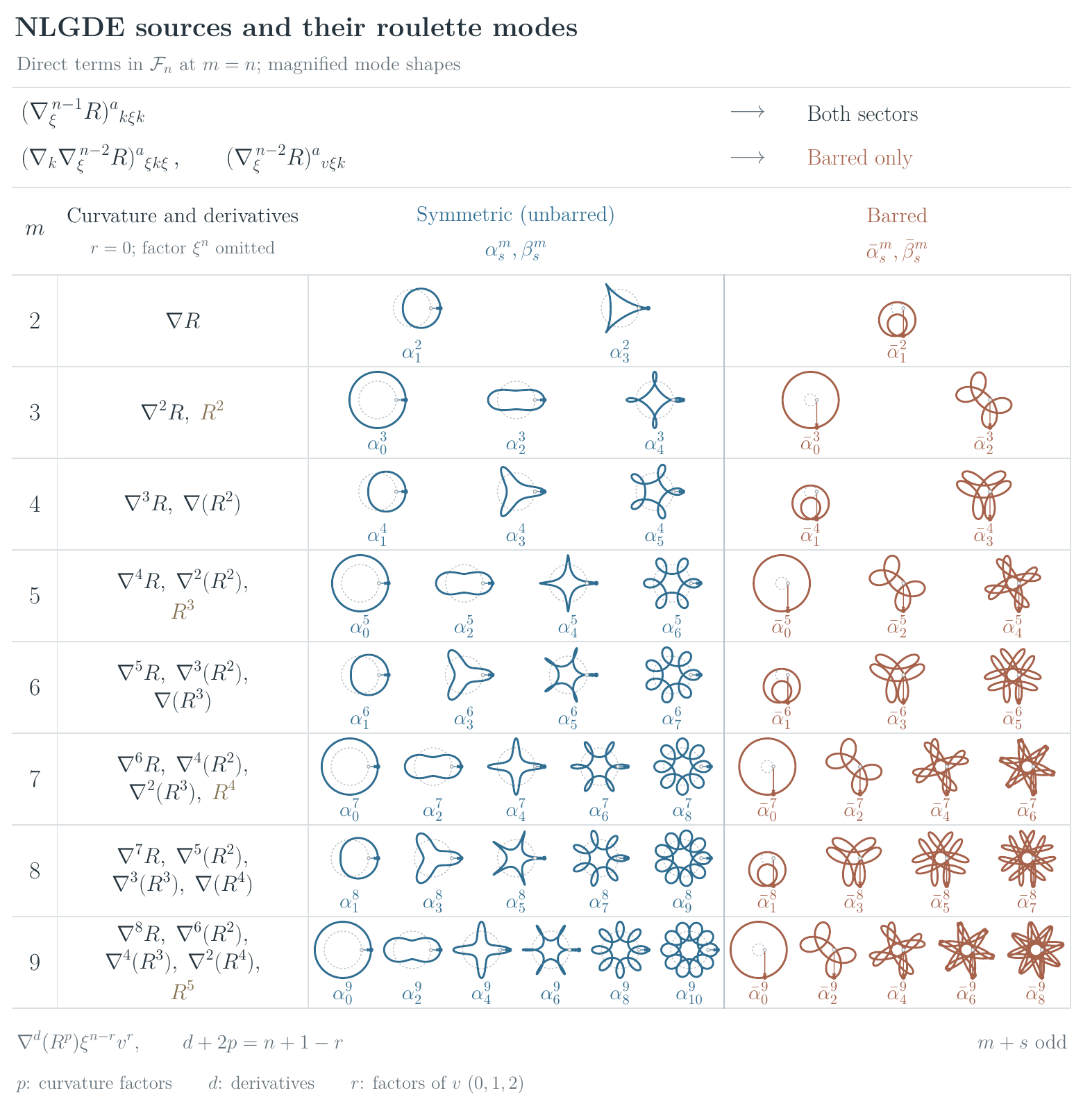}
\caption{Curvature sources and elementary roulette patterns. Rows increase
angular degree; the left column lists the available curvature derivatives
and products. Blue curves show symmetric basis patterns and muted red
curves their barred partners. Matching colours follow the same input
direction. Display gains are 0.5 for the symmetric family and 4.5 for
the barred family, with one isotropic rescaling within each.
\protect\\[0.35em]
The upper strip shows direct forcing of a round linear beam at equal
force and map degree. The displayed along-ray and relative-direction
terms drive tangential displacement; transverse gradients can also
change the radius. Anisotropic propagation and insertions of lower-degree
maps can mix these assignments in the final image.}
\label{fig:roulette-source-flowers}\end{minipage}\end{figure}\endgroup
\subsection{Curvature terms at each degree}\label{sec:curvature-orders}
Higher force degrees contain both further derivatives of curvature and
products of curvature. A term such as \(\nabla_\xi\nabla_\xi R\) contains one
curvature factor, however many covariant derivatives act on it; a
product \(RR\) or \(R\nabla R\) contains two. Appendix~\ref{app:curvature-orders-details}
derives the single-curvature force at arbitrary degree, retaining every
ordered placement of a relative-direction derivative. The same connector
recursion supplies the products needed, for example, to complete the
quartic force. We can constrain which structures appear by counting
their dimensions.

If \(r\) counts relative-direction factors, \(p\) curvature factors and
\(d\) derivatives on curvature, the allowed terms satisfy
\begin{equation}
\begin{aligned}{}
d+2p&=n+1-r,\qquad r=0,1,2 .
 \label{eq:curvature-degree-count}
\end{aligned}
\end{equation}
With affine parameter and \(\xi\) measured in length,
\(k,v\) are dimensionless. A degree-\(n\) acceleration with \(r\)
rate factors scales as \(L^{n-r-d-2p}=L^{-1}\), giving the count.
It restricts possible terms; the recurrence fixes their coefficients.
Completeness through \(m_{\max}\) requires forces through
\(n=m_{\max}\), the earlier-map couplings, and the correct launch and
endpoint data.

Figure~\ref{fig:roulette-source-flowers} isolates what each direct force
source can excite before anisotropic propagation or lower-degree
insertions mix its modes. This interpretation applies at \(m=n\),
with a round linear input beam, \(\xi_{(1)}^a=\rho\mathcal J e_A^a n^A\) and
\(v_{(1)}^a=\rho\dot{\mathcal J}e_A^a n^A\). Here
\(\mathcal J_{AB}(\lambda)=\mathcal J(\lambda)\delta_{AB}\) is the
round-beam Jacobi map.

For the notation in the figure, the left column lists \(r=0\) terms with \(\xi^n\) and numerical
prefactors suppressed; the upper strip also includes \(r=1\) terms.
Olive entries mark products containing no derivatives of curvature.
The notation \(\nabla^d(R^p)\) includes derivatives
distributed among the \(p\) curvature factors. For \(r=1,2\),
the corresponding derivative count is \(d=n+1-2p-r\geq0\).
The count first permits three undifferentiated curvature factors at
degree five. Appendices~\ref{app:quartic} and
\ref{app:curvature-grading} give the complete quartic force and
this first \(R^3\) sector explicitly.

For the round linear input shown in
figure~\ref{fig:roulette-source-flowers}, the along-ray and
relative-direction sources have zero radial contraction by curvature
antisymmetry. They therefore directly excite barred modes. This statement
applies at equal force and angular degree; anisotropic propagation and
lower-degree insertions mix the sectors. The polynomial proof is in
appendix~\ref{arch-matrix-proof}.

\endgroup

\FloatBarrier

%% file: sections/04_propagation_examples.tex
\section{Testing the propagation hierarchy}
\label{sec:spacetime}
The hierarchy in \eqref{eq:hierarchy-howto}--\eqref{eq:cubic-hierarchy-main}
distinguishes feedback of earlier distortions into the transverse
tidal-gradient force from the additional along-ray, relative-direction
and curvature-product forces. Both act continuously as the beam
propagates through a gravitational field. The TG and TG+C restrictions
in \eqref{eq:tg-main}--\eqref{eq:tgc-main} separate their effects.
We use two lens planes to test the transverse coupling C, then test
the remaining forces in spherical and flattened spacetimes and in
a finite gravitational-wave burst. Within each comparison the
reference ray, launch coordinates and endpoint prescription are held
fixed, so the image changes follow from the retained propagation terms.

\subsection{Coupling and the remaining spacetime terms}
Suppose the transverse tidal gradient has already bent a small beam away
from its linear ellipse. Farther along the ray, the same quadratic force
acts on this distorted separation. Inserting its quadratic part into
either separation slot of the quadratic force produces a cubic term.
This is the coupling in \eqref{eq:cubic-hierarchy-main}. It accumulates
while the distorted beam continues through a varying tidal field, and
becomes more relevant as the beam is widened enough for the quadratic
distortion to develop.

Full propagation also changes the quadratic distortion itself.
Equation~\eqref{eq:F2} contains the along-ray curvature derivative and
the effect of neighbouring rays crossing the field at different angles.
A static lens still has these terms: its curvature varies along a
photon's trajectory, and neighbouring photons acquire different
directions. At cubic degree, curvature products in \eqref{eq:F3}
account for the geometry of finite separation and parallel transport.
The lower-degree distortions generated by these forces enter the
couplings in turn.

All these contributions change the amplitudes of the existing roulette
patterns. For the complete and TG+C maps, whose linear coefficients
agree, define \(\Delta P^{(m)}=P^{(m)}_{\rm full}-P^{(m)}_{\rm TG+C}\).
Their difference through cubic degree is
\begin{equation}
\begin{aligned}{}
\Delta\xi_A^{[3]}&=\frac{\rho^2}{2}\Delta P_A^{(2)}
                   +\frac{\rho^3}{6}\Delta P_A^{(3)} .
 \label{eq:amplitude-to-image-change}
\end{aligned}
\end{equation}
The angular weights explain why the remaining quadratic terms can
matter more than C across a narrow beam. C first enters at cubic degree, weighted by \(\rho^3/6\), while
the remaining terms already alter the quadratic coefficient, weighted
by \(\rho^2/2\). For a narrow beam the latter can dominate; as it
widens, cubic contributions and eventually higher degrees become
necessary.

\subsection{Two lens planes: rotation and nonlinear coupling}\label{sec:two-plane-coupling}
Two separated deflectors show how successive distortions generate
rotation and nonlinear coupling. In reduced angular coordinates, their
potentials are \(\phi_1,\phi_2\), the dimensionless source-map derivative is
\(\mathcal A_{AB}=\partial U_A/\partial\theta^B\), and the length map is
\(\mathcal M_{AB}=\chi_S\mathcal A_{AB}\). The map and its derivative obey
\begin{equation}
\begin{aligned}{}
\theta_{2A}&=\theta_A-b_{12}\,\nabla_A\phi_1(\theta),\\
 U_A(\theta)&=\theta_A-\nabla_A\phi_1(\theta)
                    -\nabla_A\phi_2(\theta_2),\\
 \mathcal A&=I-H_1-H_2+b_{12}H_2H_1,\\
 \frac{\mathcal A-\mathcal A^{\mathsf T}}{2}&=\frac{b_{12}}{2}[H_2,H_1] .
 \label{eq:two-plane}
\end{aligned}
\end{equation}
Here \(b_{12}\) is the distance factor and
\(H_i=\nabla\nabla\phi_i\) is evaluated on its own lens plane.

\subsubsection{Rotation from a weak lens pair}
\label{sec:weak-two-lens}

\begin{figure}[!htb]
  \centering
\includegraphics[width=\textwidth,height=.74\textheight,keepaspectratio]{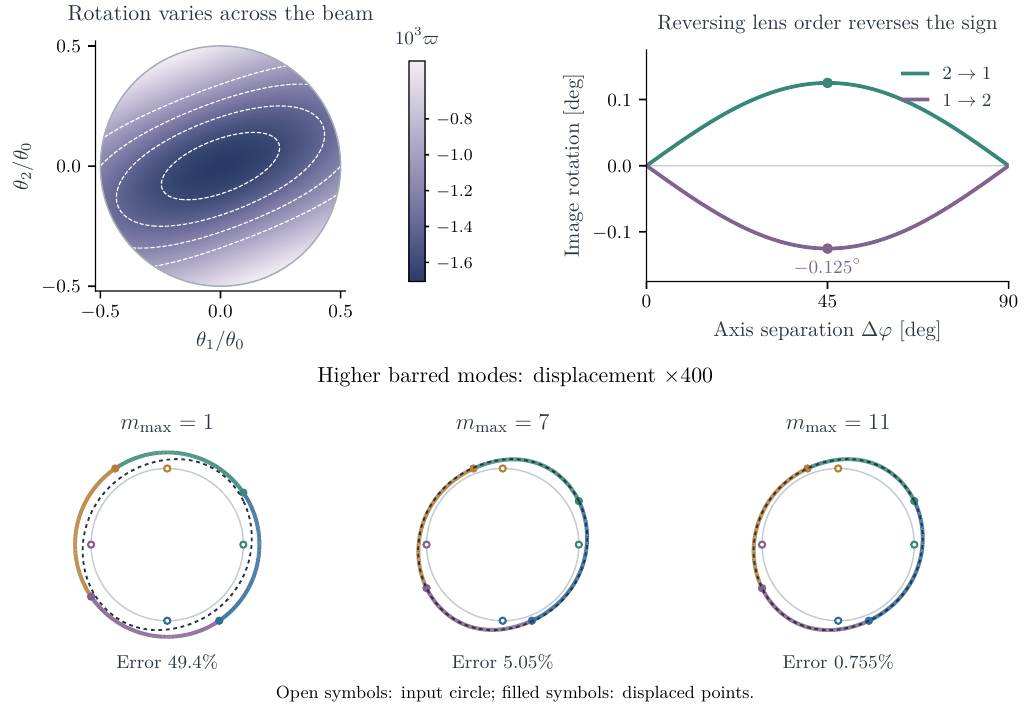}
  \caption{Two individually weak lenses generate rotation and its spatial
variation. Upper left: the exact antisymmetric Jacobian component
\(\varpi\), negative throughout the displayed disc. Upper right:
reversing the lens order at fixed reduced potentials reverses the central inverse-map
rotation, whose magnitude reaches \(0.125^\circ\) at a
\(45^\circ\) axis separation. The model is globally single-image.
Below: successive barred reconstructions approach the dashed displacement
obtained by integrating the exact curl. This is the independent barred
part, after subtracting the fully symmetric sector. Colours follow initial
direction and the grey circle marks the input of radius \(0.5\theta_0\),
where \(\theta_0\) is the lenses' common angular core scale.
Only the lower-row displacements are
enlarged by 400; errors are relative to the exact displacement's peak
on that circle.}
  \label{fig:weak-two-lens}
\end{figure}
Two scalar lenses each have a symmetric local distortion matrix.
Successive stretches about different axes need not commute, so the
combined map depends on their order. In reduced angular
coordinates the output and input are both angles. Accordingly,
\(\bar\alpha_0^1=\varpi\) below; multiplying the output by \(\chi_S\)
would give the length convention \(\bar\alpha_0^1=\chi_S\varpi\)
of section~\ref{sec:review}. For the map \eqref{eq:two-plane},
with \(\epsilon_{12}=+1\), the local barred linear amplitude is
\begin{equation}
\begin{aligned}{}
\bar\alpha_0^1&=\varpi\equiv\frac{\mathcal A_{12}-\mathcal A_{21}}{2} .
 \label{eq:rotation}
\end{aligned}
\end{equation}
This is the lens--lens rotation mechanism
\citep{penmao,robertsonlewis2023}. It already occurs in ordered linear
Jacobi propagation; it is distinct from optical twist and from the local
\(R^2\xi^3\) force in \eqref{eq:F3}.

We choose a weak pair to show that this rotation does not require a
critical curve or multiple images. The positive-density lenses share a
core scale \(\theta_0\), central convergence \(0.113\) and shear
magnitude \(0.0533\), with axes separated by \(45^\circ\) and
\(b_{12}=0.6\). Their potentials are specified in
appendix~\ref{app:weak-two-lens-inputs}.

Each positive semidefinite Hessian obeys \(\|H_i\|\leq1/6\).
Hence \(\|I-\mathcal A\|\leq1/3+b_{12}/36=0.35\) globally, and the
smallest singular value is at least \(0.65\). For any source \(U_s\),
\(\theta=U_s+\nabla\phi_1(\theta)+
\nabla\phi_2(\theta-b_{12}\nabla\phi_1(\theta))\) is a contraction
on \(\mathbb R^2\), proving that there is exactly one image.

At the centre, reversing these lenses while keeping their reduced potentials fixed
transposes \(\mathcal A\), preserving its symmetric part while reversing
\(\varpi\). The rotation angle is
\begin{equation}
\begin{aligned}{}
\phi_{\rm img}
 &=\operatorname{atan2}(\mathcal A_{12}-\mathcal A_{21},\operatorname{tr}\mathcal A)\\
 &=\begin{cases}-0.1252^\circ,&1\ \hbox{then}\ 2,\\
                 +0.1252^\circ,&2\ \hbox{then}\ 1.
   \end{cases}
 \label{eq:weak-image-angle}
\end{aligned}
\end{equation}
This is the rotation angle in the polar decomposition of the inverse map,
\(\mathcal A^{-1}=R_{\rm img}S_{\rm img}\), with \(S_{\rm img}\) symmetric
and positive definite. An arbitrary source's major-axis orientation also
depends on that stretch.

To extract the higher barred modes, we use the spatial variation of
\(\varpi\). Taking the curl of the radial and tangential components
of the degree-\(m\) map gives
\(\rho^{m-1}[(m+1)K_m-H'_m]/m!\). Comparing this expression with
the Fourier decomposition relates each barred degree to derivatives of
\(\varpi\).
Using \eqref{eq:fourier} and
\(\varpi=(\partial_2U_1-\partial_1U_2)/2\) gives
\begin{equation}
\begin{aligned}{}
\varpi(\theta_*+\rho n)
 &=\sum_{m\geq1}\frac{m+1}{2m!}\rho^{m-1}
                    \sum_s\bar\sigma_s^m(\vartheta) .
 \label{eq:curl-barred}
\end{aligned}
\end{equation}
Thus the constant rotation gives the barred \(m=1\) mode; its spatial
gradient gives \(m=2\), its Hessian gives \(m=3\), and so on.
The sum over \(s\) uses the allowed barred spins, and \(\vartheta\)
is the azimuth of \(n\).

The exact independent barred displacement about the origin follows by
integrating the curl along each radial line:
\begin{equation}
\begin{aligned}{}
\overline{\delta U}(\rho,\vartheta)
 &=-\frac{2t(\vartheta)}{\rho}
             \int_0^\rho r\,\varpi(rn(\vartheta))\,\dd r .
 \label{eq:weak-curl-integral}
\end{aligned}
\end{equation}
Here \(t(\vartheta)\) is the tangential unit vector of
section~\ref{sec:review}; the fully symmetric sector has been removed.
The centred model is inversion symmetric, so its even map degrees vanish.
Figure~\ref{fig:weak-two-lens} compares the odd barred hierarchy with this exact integral. The
barred-displacement error falls from \(49\%\) at degree
one to \(5.1\%\) at seven and \(0.76\%\) at eleven, relative to its
exact peak on the displayed circle. The full degree-eleven map has sampled
position error \(5.8\times10^{-5}\theta_0\) over
\(\rho\leq0.5\theta_0\).

Reversing the lens order changes the tangential distortion within the
same roulette basis. The rotation starts at second order in lens
strength, while its spatial derivatives populate several angular
degrees. These are different expansion orders.

\subsubsection{From weak rotation to multiple images}
\begin{figure}[!htb]
  \centering
\includegraphics[width=\textwidth,height=.72\textheight,keepaspectratio]{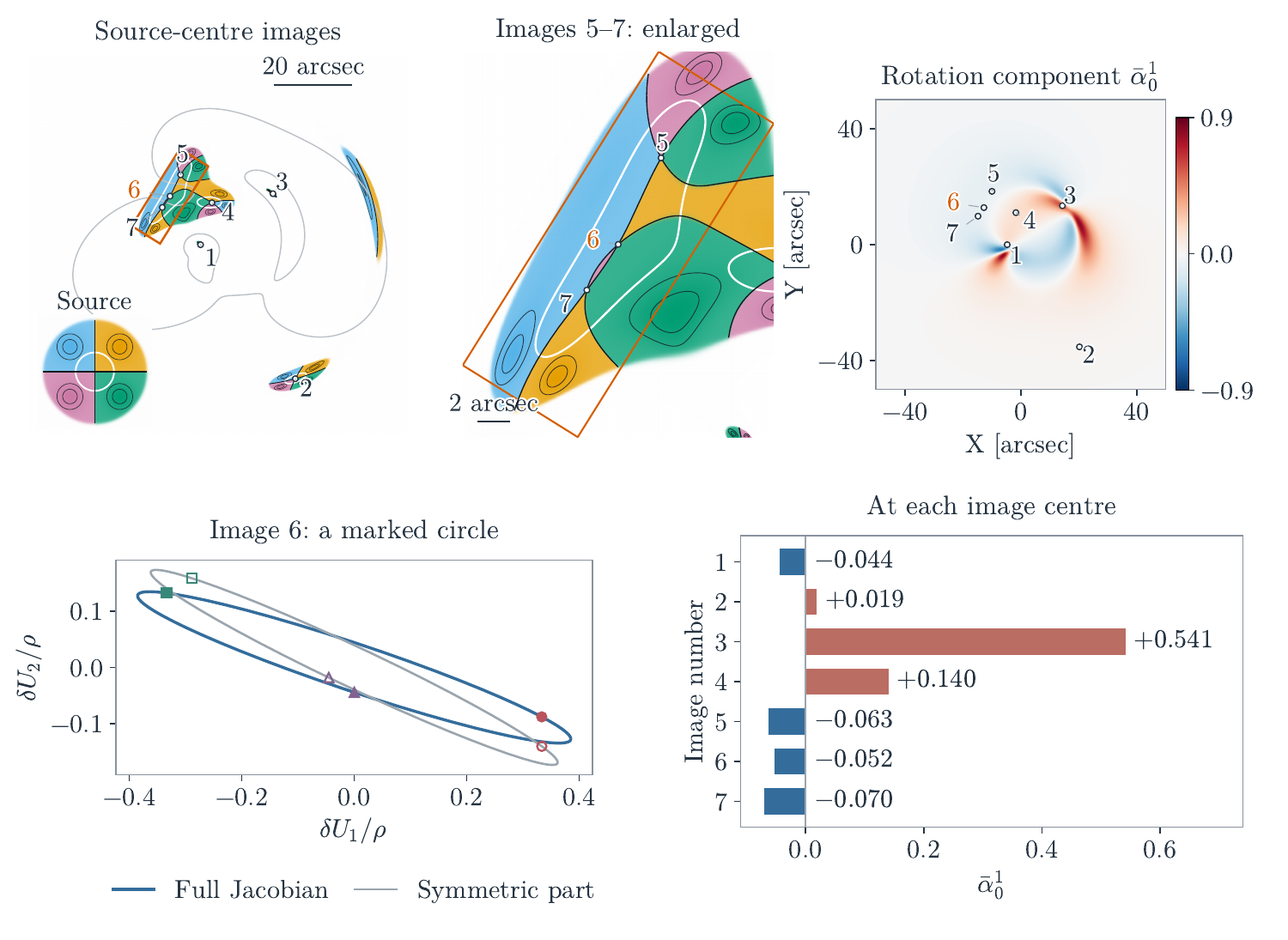}
  \caption{Rotation from two misaligned lens planes. The upper-left panel shows
the seven images of the source centre and the critical curves in grey.
The orange window selects images 5--7, enlarged beside it and used
again in figure~\ref{fig:two-plane-coupling}. The colour map shows the
antisymmetric Jacobian component
\(\bar\alpha_0^1=(\mathcal A_{12}-\mathcal A_{21})/2\); red is positive and blue
negative, and the bars give its value at each numbered image. Opposite
colours indicate opposite senses of the local tangential displacement.
At lower left, the blue curve maps a small circle of directions around
image 6 using the full Jacobian. The grey curve removes its
antisymmetric part. Filled and open markers follow the same incoming
directions in the two maps, showing the displacement introduced by the
barred linear roulette. Displacements are divided by the input angular
radius \(\rho\). This coefficient is a map component, not a rotation
angle. Both lenses and the source are unchanged from
figure~\ref{fig:model-gallery}; their parameters are in
appendix~\ref{app:two-plane-strong-control}.}
  \label{fig:two-plane}
\end{figure}

The same mechanism operates when the lens pair produces multiple
images. Figure~\ref{fig:two-plane} uses two stronger, misaligned cored
elliptical potentials and our common structured source; the parameters
are in appendix~\ref{app:two-plane-strong-control}. Its seven source-centre
images pass through different parts of the two lenses and therefore
acquire different antisymmetric Jacobian components. To see what this
component does, we choose image 6 and follow a small circle of directions
around its ray. The lower-left panel compares the source-plane
displacements obtained with the full Jacobian and with its antisymmetric
part removed. Matching markers follow the same incoming directions, so
their separation shows the tangential displacement supplied by the
barred linear roulette.

This displacement acts together with the stretching of the beam.
Consequently, \(\bar\alpha_0^1=(\mathcal A_{12}-\mathcal A_{21})/2\) measures an
antisymmetric contribution to the map, rather than an angle through
which the whole ellipse has turned. The weak-lens example above
separates a rotation angle using the polar decomposition. Here the
coefficient map and the values at the seven image centres show where
the antisymmetric contribution is generated and how it varies between
images; higher barred roulettes describe its variation across each
finite image.

\subsubsection{Nonlinear coupling across an extended arc}
\begin{figure}[!htb]
  \centering
\includegraphics[width=\textwidth,height=.72\textheight,keepaspectratio]{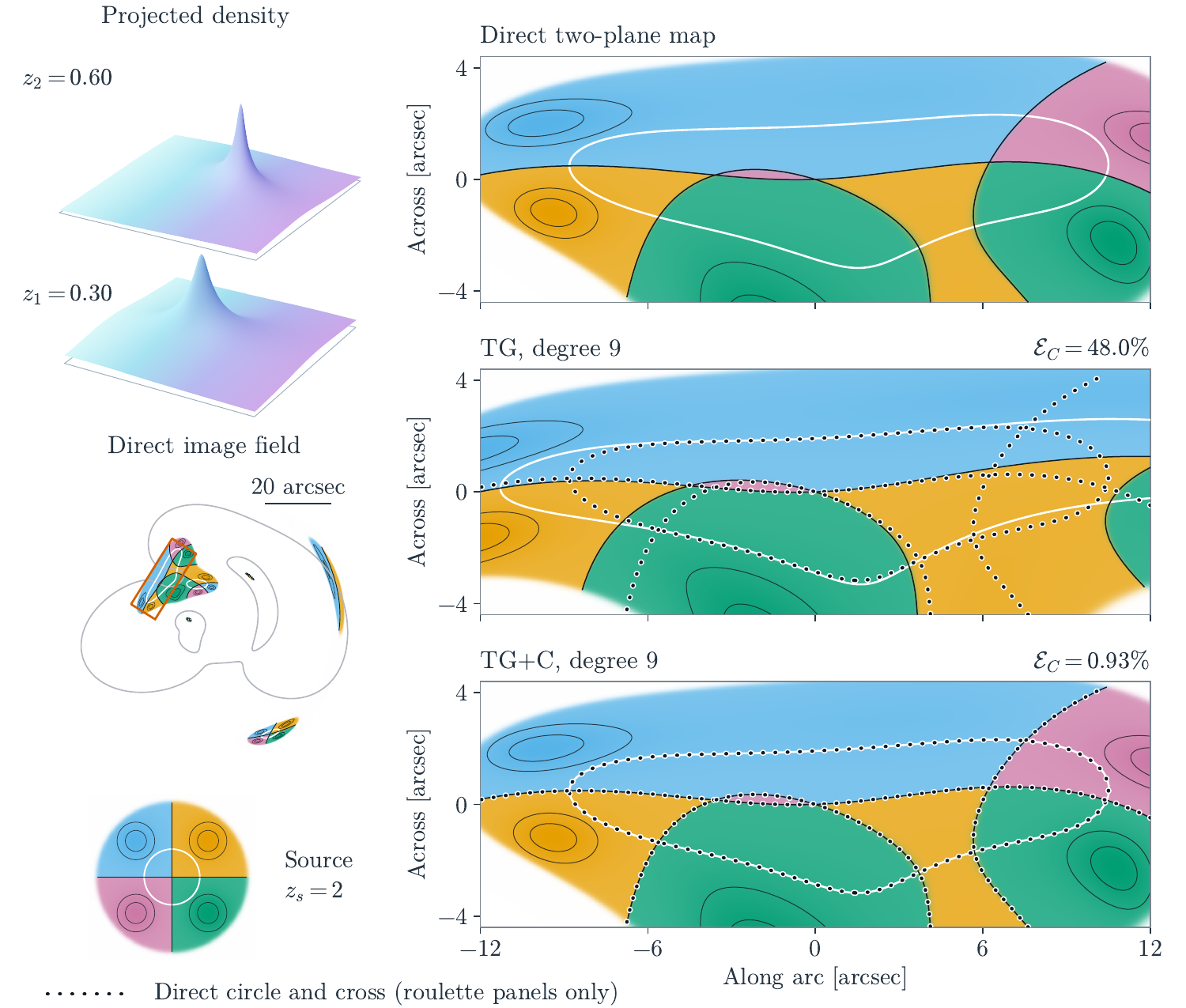}
  \caption{Nonlinear coupling across the
three-image arc of the two-plane lens introduced in
figure~\ref{fig:two-plane}. Left: projected-density surfaces for the two
deflectors, the full direct-image field and the common four-colour source
of radius \(0.12\theta_0=2.4\,\mathrm{arcsec}\). Surface heights use a
common \(\log(1+\kappa_i)\) scale; the separation of the sheets is
schematic. Grey curves are critical curves. The orange box selects
images 5--7 in a \(24\times8.8\,\mathrm{arcsec}\) window centred on
image 6. Right, from top to bottom: the direct map, TG and TG+C at
angular degree nine. Coordinates are relative to image 6, along the
most stretched linear image direction and perpendicular to it.
Solid guides follow the source cross, white ring and quadrant contours
through each map. Dark dotted curves with white outlines mark the
direct positions of the cross and ring; these reference curves appear
only in the roulette panels. TG retains the full linear rotation, but
displaces these features because it omits nonlinear coupling. The
labels give the colour error \(\mathcal E_C\), excluding guides;
the source, lens parameters and image window are identical in all
three panels.}
  \label{fig:two-plane-coupling}
\end{figure}

We now use the seven-image lens just introduced to test C. A finite beam's distortion at the first plane
changes its deflection at the second. We keep both lenses
and the source fixed, and compare TG with TG+C in the enlarged arc
window of figure~\ref{fig:two-plane-coupling}. We selected the reference direction, numbered 6
in the figure, for a
large TG beam error while keeping the degree-five coupled beam error
below \(2\%\). The source was applied afterwards. One expansion about that image covers the
three source-centre images in this window. Both
approximations retain the full Jacobian in \eqref{eq:two-plane},
including the antisymmetric component examined in
figure~\ref{fig:two-plane}. Their difference measures the subsequent
nonlinear coupling, C, defined in section~\ref{sec:general-force}.

Write \(\alpha_i=\nabla\phi_i\) and
\((\mathsf L_{12})^A{}_B=\delta^A{}_B-b_{12}H_1{}^A{}_B\), the linear map from
the observer's sky to the second plane. The two hierarchies agree through
quadratic degree. Their first difference is the insertion of the first
lens's quadratic distortion into the nonlinear deflection at the second:
\begin{equation}
\begin{aligned}{}
\Delta_C U^A(\boldsymbol\zeta)
 &\equiv U^A_{\text{TG+C}}(\theta_*+\boldsymbol\zeta)
             -U^A_{\mathrm{TG}}(\theta_*+\boldsymbol\zeta)\\
 &=\frac{b_{12}}{2}
   (\partial_B\partial_C\alpha_2^A)(\mathsf L_{12})^B{}_D\zeta^D
   (\partial_E\partial_F\alpha_1^C)\zeta^E\zeta^F
   +O(\rho^4).
 \label{eq:two-plane-cubic}
\end{aligned}
\end{equation}
All derivatives are evaluated where the central ray crosses their own
lens plane. The first lens supplies the quadratic displacement, and the
spatially varying tidal field of the second acts on that displacement
and the linear beam together. This is the coupling C in
\eqref{eq:tgc-main}; it vanishes when the planes coincide
and \(b_{12}=0\).

The source cross and white ring make the change visible. In the roulette
panels, dotted curves mark their direct-image positions, while solid
guides follow the reconstructed map. At degree nine the colour error
falls from \(48.0\%\) to \(0.93\%\) when C is included. The closed
TG expression retains an error of \(47.9\%\): increasing degree cannot
supply the omitted feedback. Standard multi-plane lensing already
contains this coupling \citep{schneider1992,penmao}; the comparison shows how it is
lost and recovered by the two roulette hierarchies. The following spacetime examples test the
terms beyond TG+C.
Appendix~\ref{app:two-plane-strong-control} specifies the selected
alignment and window, and gives the degree convergence.

\subsection{Weak-field size of the additional terms}\label{sec:relevance}
The weak-field members of our spherical families give the scale of
the corrections from TG+C to Full. The quantity \(d_m\) in
\eqref{eq:coefficient-difference} measures the maximum change around a
degree-\(m\) coefficient pattern relative to the maximum full pattern:
\begin{equation}
\begin{aligned}{}
d_m&=\frac{\max_\vartheta|P_{\rm full}^{(m)}-P_{\rm TG+C}^{(m)}|}
               {\max_\vartheta|P_{\rm full}^{(m)}|} .
 \label{eq:coefficient-difference}
\end{aligned}
\end{equation}
The isothermal sweep uses \(w_{\rm iso}=10^{-4},10^{-3},3\times10^{-3},10^{-2},3\times10^{-2}\),
with an additional weak member at \(w_{\rm iso}=5.39\times10^{-7}\).
The Schwarzschild sweep uses the same five values of \(\mathsf M/D\).
Both families place the observer at \((-D,b_o,0)\), launch the reference
ray along the local positive \(x\) direction with \(E_o=1\), and end
at \(\lambda_s=-2D\). We keep \(b_o/(D\theta_E)=1.10\) for the
isothermal family and \(1.2\) for Schwarzschild, holding their leading
thin-lens focusing fixed while varying the strength of the spacetime
curvature. The nominal distances remain \(D_L=D_{LS}=D\) and
\(D_S=2D\), as specified in appendices~\ref{app:sis-propagation}
and~\ref{app:schwarz-propagation}.
For the isothermal sphere with \(\sigma=220\,\mathrm{km\,s^{-1}}\),
\((d_2,d_3)=(1.9,3.6)\times10^{-6}\). The weakest Schwarzschild
member has \(\theta_E=0.0141\) rad and
\((d_2,d_3)=(0.0084,0.0153)\). Although these Einstein angles are
very different, the weak members of both families give approximately
\begin{equation}
\begin{aligned}{}
d_2&\simeq(0.55\text{--}0.60)\theta_E,&
 d_3&\simeq(1.05\text{--}1.1)\theta_E,
 \qquad \theta_E\ \hbox{in radians}.
 \label{eq:relevance-scaling}
\end{aligned}
\end{equation}
These numerical scalings describe the weak members at the fixed
reference-ray and endpoint geometries above. Across the full sweeps,
\((d_2,d_3)\) increase from \((0.00035,0.00066)\) to
\((0.080,0.156)\) for the isothermal family and from
\((0.0084,0.0153)\) to \((0.101,0.234)\) for Schwarzschild.
Extrapolating the weak-member scalings within their
weak-field regime gives changes of a few parts in \(10^6\) for
arcsecond Einstein angles and of order \(10^{-4}\) for angles of
tens of arcseconds. These are changes in coefficient patterns, before
the angular weights in \eqref{eq:amplitude-to-image-change} are
applied. The image error also depends on beam size and source structure.
In these families, strong magnification does not imply large corrections
beyond TG+C.

A local estimate explains this suppression. In a weak, localised lens
with impact scale \(b\), a transverse curvature derivative has size
\(R/b\). In the quadratic along-ray term, integration by parts moves
one derivative from curvature to the propagation weight. If boundary
terms vanish and the beam and weight vary over an endpoint distance
\(D_{\rm near}\gg b\), the integrated contribution is smaller by
roughly \(b/D_{\rm near}\). The relative-direction term is controlled
by \(b|v_{(1)}|/|\xi_{(1)}|\), which has the same estimate for a slowly
varying weak-field beam. Local curvature products relative to second
curvature derivatives have size \(Rb^2\sim\mathsf M/b\). Near the
Einstein radius of a far-field compact lens with comparable distances,
both small parameters are of order \(\theta_E\).

Near a focus or an endpoint close to the lens, the beam can vary too
rapidly for these estimates to apply. The spherical comparison in
section~\ref{sec:spherical-comparison} uses larger parameters, for which
these forces are visible. Appendix~\ref{app:schwarz-propagation} repeats
the compact Schwarzschild comparison on a fixed source plane. The wave example below provides a different
case, with time-dependent curvature and flat observer and source
regions, in which the omitted terms produce a visible image change.

\subsection{Isothermal and Schwarzschild lenses}
\label{sec:sis-propagation}\label{sec:spherical-comparison}
The isothermal sphere and Schwarzschild lens give a simple comparison
between a ray passing through matter and one passing through the
surrounding gravitational field of a compact mass. In the weak-deflection
limit the isothermal deflection is nearly independent of impact
parameter, whereas the point-mass deflection falls as its inverse.
Both can produce a tangential critical circle. Near that circle the
isothermal lens leaves the radial image width approximately unchanged,
while the point mass compresses it. Both can form long tangential
images with quite different distortions across the beam.

\begin{figure}[!htb]\centering
\includegraphics[width=\textwidth]{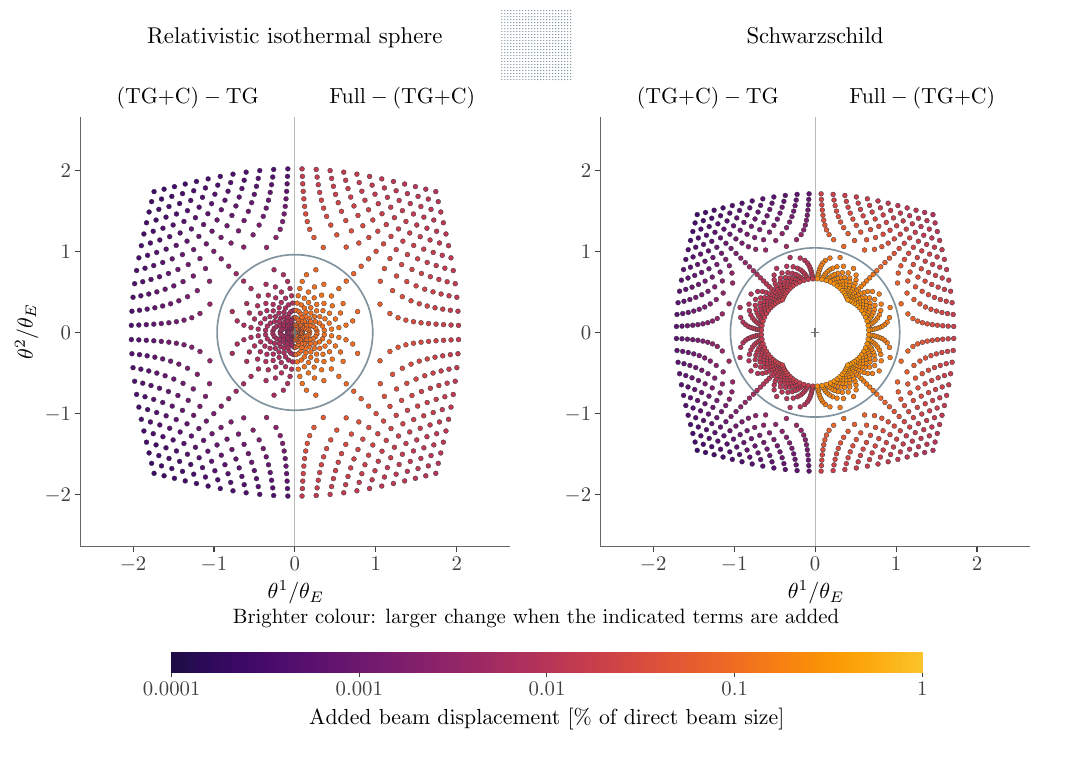}
\caption{Changes to local light beams behind an isothermal sphere and a
Schwarzschild lens. Each dot is a directly traced image of a point in the
common \(24\times24\) source grid shown above; there are 924 isothermal
and 1152 Schwarzschild image positions. Around each image ray we launch
a small circle of neighbouring directions. On the left side of each
field, colour gives the RMS change in their predicted separation when
C is added to TG. On the right side, it gives the further change when
the complete cubic propagation is used. The single logarithmic colour
scale expresses both changes as a percentage of the directly traced
beam size. The warm sequential scale measures added displacement: brighter dots
mean a larger change, with the two differences written above each half.
The remaining propagation terms are larger than C for these beams,
especially on the inner image branches. Grey circles mark critical
curves. The angular beam radius is
\(0.04\min(\theta_E,|\boldsymbol\theta|)\).
The common beam-size normalisation is defined in appendix~\ref{app:accuracy-measures}.}
\label{fig:spherical-dots}\end{figure}

Figure~\ref{fig:spherical-dots} places the same regular grid of source
points behind the two lenses. We use the relativistic perfect-fluid
isothermal sphere, \(p=w_{\rm iso}\varrho\)
\citep{remmen2021sis}, and the Schwarzschild spacetime, matching their
nominal Einstein angles,
\begin{equation}
\begin{aligned}{}
\theta_E&=2\pi w_{\rm iso}=\sqrt{2\mathsf M/D},
 \qquad w_{\rm iso}=0.01 .
 \label{eq:spherical-field-match}
\end{aligned}
\end{equation}
The observer and source plane are at \(x=-D\) and \(x=D\).
The full-ray critical radii are \(0.962\theta_E\) and
\(1.046\theta_E\), respectively: the matching fixes the leading
Einstein angle, while the two metrics determine their actual critical
circles. Appendix~\ref{app:spherical-dot-inputs} gives the inputs.
Direct null rays determine the image positions. We then launch a small
circle of neighbouring directions around each image ray and compare
three predictions for its separation at the central ray's arrival
parameter: TG, TG+C and the complete cubic map. The dot stays at the direct image position. Its
colour shows how much the local beam prediction changes when the
additional propagation terms are included, using one colour scale for
both lenses and both comparisons.

The image geometry distinguishes the mass profiles. The compact
mass gives an inner partner to every source in the displayed grid.
The isothermal inner images extend towards the centre and disappear
for the more distant source points. Both kinds of added displacement
grow towards the inner image region, away from the weakly distorted
outer images. Their maxima need not follow the critical circle: that
circle marks a degeneracy of the integrated linear map, whereas the
nonlinear terms depend on the curvature and beam shape along the whole
path.

The larger changes on the right include a correction to the quadratic
distortion, which C cannot supply. A quadratic displacement bends
opposite sides of a beam in the same direction, whereas a cubic
displacement reverses direction between opposite sides. Adding C first
changes the cubic pattern; completing the propagation changes the
quadratic bending as well. Both additions survive in Schwarzschild,
where Ricci focusing vanishes, because the Weyl tidal field and the
changing ray geometry still act across the beam.

Strong stretching alone therefore does not determine which propagation
terms are needed. Reducing \(w_{\rm iso}\) while keeping the image
close to the isothermal critical circle preserves strong tangential
stretching but makes the additional terms small. For the galaxy-scale
member in section~\ref{sec:relevance}, following a
long arc mainly requires more angular degrees or more expansion centres.
The relative importance of the propagation terms instead changes with
the beam and its path: a wider beam increases cubic coupling relative
to quadratic displacement, while a path deeper through a compact lens
encounters stronger variation of the tidal field.

\subsection{A flattened lens: different paths through the tidal field}
\label{sec:elliptical-propagation}
Spherical symmetry repeats the same experiment around each annulus.
Flattening the lens lets beams at the same angular radius encounter
different density and tidal gradients. We use the static relativistic
Miyamoto--Nagai model~I \citep{vogt2005galaxies}, viewed perpendicular
to its symmetry axis. Its finite thickness and anisotropic pressure
specify a spacetime through which we propagate each beam; the metric
and parameters are in appendix~\ref{app:elliptical-propagation}.

\begin{figure}[!htbp]\centering
\begin{minipage}[c]{.64\textwidth}\centering
\includegraphics[width=\linewidth]{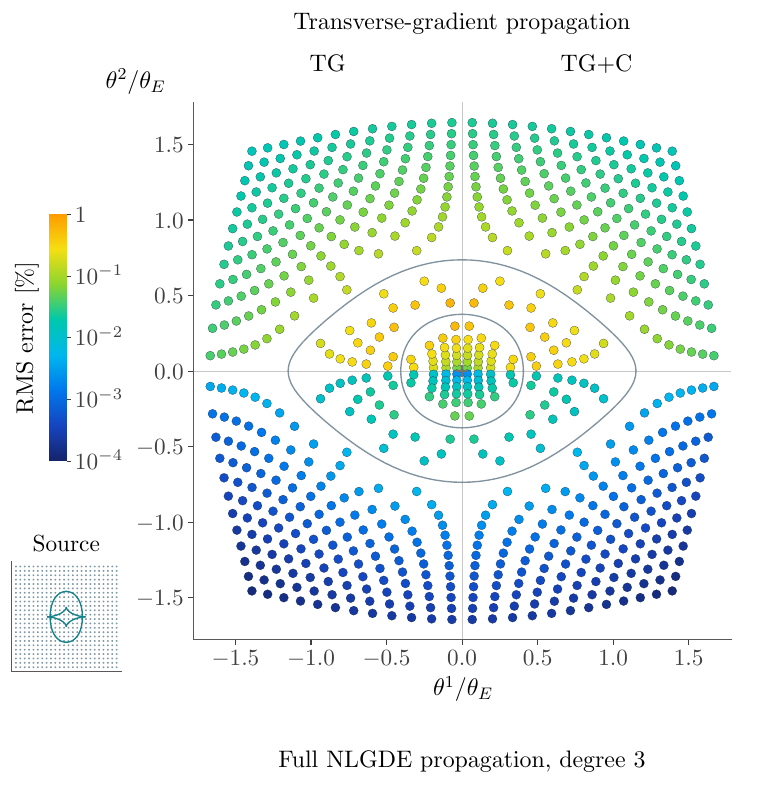}
\end{minipage}\hfill
\begin{minipage}[c]{.33\textwidth}
\caption{Accuracy of local beam propagation through a flattened matter lens.
The upper-left quadrant uses TG, the upper-right TG+C, and both lower
quadrants Full cubic propagation, as indicated by the row labels.
Reflection symmetry makes reflected dots a comparison of equivalent
beams. All four quadrants share one continuous logarithmic colour scale,
from \(10^{-4}\%\) to \(1\%\). Each dot gives the RMS error in the
predicted separation of neighbouring rays, as a percentage of the
directly traced beam size. The darker colours across the lower half
show the reduction in error when the remaining propagation terms are
included. Dot positions are the 728 direct images of the source grid
in the inset. Grey curves mark critical curves; the teal curves in
the source inset are caustics. Each local sky circle has radius
\(0.04\theta_E\), and its rays are compared at equal affine parameter.
This scale measures error against direct rays, while the warm scale in
figure~\ref{fig:spherical-dots} measures changes between approximations.
The error norms are defined in appendix~\ref{app:accuracy-measures}.}
\label{fig:elliptical-propagation}\end{minipage}
\par\smallskip
\begin{minipage}[c]{.59\textwidth}\centering
\includegraphics[width=\linewidth]{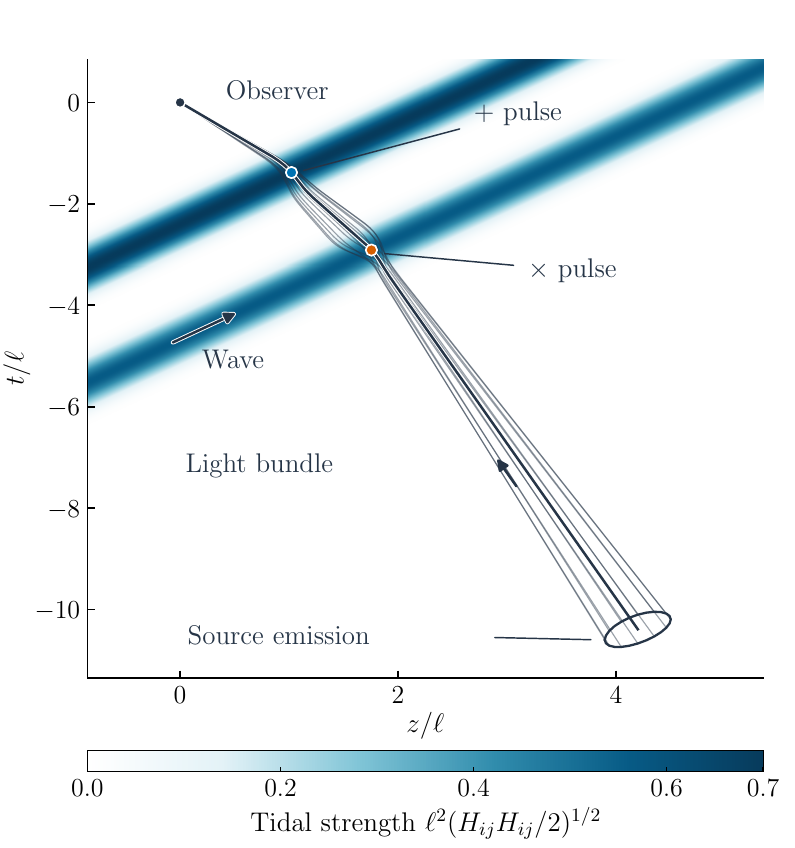}
\end{minipage}\hfill
\begin{minipage}[c]{.37\textwidth}
\caption{Light crossing the gravitational-wave burst used in
figure~\ref{fig:wave}. Future time runs upwards, and
\(u=(t-z)/\sqrt2\) labels the wave phase. Colour shows the tidal
strength \(\ell^2(H_{ij}H_{ij}/2)^{1/2}\), where
\(H_{ij}=(H_{\rm gw})_{ij}\) is the Brinkmann profile and
\(R_{uiuj}=-H_{ij}\). White denotes the flat regions outside
the two pulses. The wave itself is the lens: light emitted below
the burst crosses the cross pulse and then the plus pulse before
reaching the observer. The black curve follows the reference ray,
and the grey curves join the observer to the boundary of the
source circle. Blue and orange points mark the reference ray's
crossings of the plus and cross pulse peaks. The lower oval consists
of emission events on the stationary source plane; the rays leave
that plane at different times. Both source and observer are outside
the burst. The plane is \(\boldsymbol n\cdot\boldsymbol r=8\ell\),
the source radius is \(0.5\ell\), and the viewing inclination is
\(0.65\) rad. Peak tidal amplitudes are \(0.7\ell^{-2}\) and
\(0.56\ell^{-2}\). The curves are calculated from the exact
transfer map of appendix~\ref{app:wave-model}.}
\label{fig:wave-field}\end{minipage}
\end{figure}

The flattened lens produces outer images and additional inner branches
around a noncircular critical structure. As in the spherical comparison,
each dot in figure~\ref{fig:elliptical-propagation} is the centre of a
small neighbouring beam. We now compare each approximation directly
with the traced beam, so colour measures reconstruction error. Reflection
symmetry lets us compare corresponding beams in the four quadrants of
figure~\ref{fig:elliptical-propagation}: TG at upper left, TG+C at
upper right and Full in both lower quadrants. The row labels identify
the retained propagation, and one continuous colour scale applies
throughout. The
complete map reduces the errors throughout this field, including the
inner branches where the transverse approximations leave the largest
discrepancies.

The small change from the upper-left to the upper-right region follows
from what C adds. It lets the quadratic distortion feed into subsequent
cubic evolution, but that quadratic distortion was itself calculated
using only transverse gradients. The complete NLGDE first corrects this
bending and then propagates its effect at cubic degree, together with
the explicit cubic forces. The improvement across the lower half is
the combined result of these changes.
The equal-affine comparisons test propagation before conversion to
a physical source surface. A physical source surface requires the composition in
section~\ref{sec:endpoint-choice}; the Schwarzschild fixed-plane test
and the wave example below show that the improvement survives this
operation.

\subsection{An exact gravitational-wave burst}\label{sec:wave}
A gravitational wave changes the tidal field while light crosses it.
An exact plane wave lets us follow that change without expanding in
the wave amplitude. Strong lensing by such waves and their nonlinear
optics have been studied in detail \citep{harte2013,harte2015}; here
we use the exact ray map to test the finite-beam propagation hierarchy.
The observer and emitting surface lie in flat regions on opposite
sides of a finite burst, so their angles and rulers have an immediate
physical interpretation. Figure~\ref{fig:wave-field} places the
source events, wave pulses and observation event in one spacetime view.

\begin{figure}[!htb]\centering
\includegraphics[width=.97\textwidth]{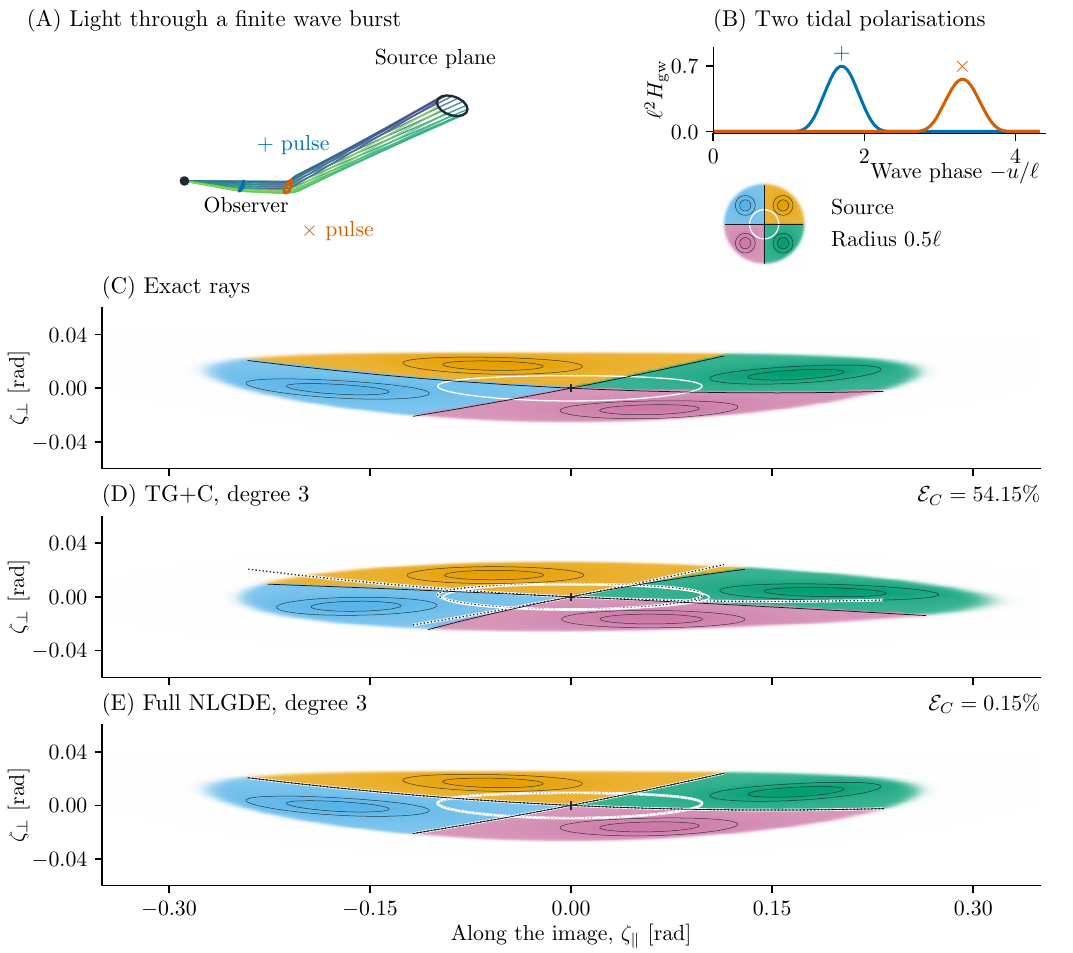}
\caption{An extended image seen through an exact gravitational-wave burst.
(A) Spatial projection of rays joining the observer to the source
boundary; the blue and orange rings locate equal-phase sections of the
plus and cross pulses encountered when tracing backwards. The plotted
paths are calculated rays. (B) The two tidal profiles, in units set by
\(\ell\); these are curvature profiles, not strain. The observer is
after the burst and the stationary source plane is before it, with
\(D=8\ell\) and inclination \(0.65\) rad. The inset shows the common
source of radius \(0.5\ell\). (C)--(E) The exact image, TG+C and full
NLGDE propagation through angular degree three, all with the same
normal-angle launch coordinates and physical source-plane intersection.
The displayed sky coordinates are rotated along and across the
stretched image, with equal angular scale on both axes. Solid guides
follow each reconstruction. Dark dotted curves edged in white mark
the exact source cross and ring in the two approximate panels.
TG+C and full share the longitudinal coefficients and endpoint
composition, isolating the change in transverse propagation. Colour
errors use the larger \(0.7\times0.7\) rad sky window specified in
appendix~\ref{app:wave-model}, excluding the guides. The colours label
source positions; frequency shifts and flux weighting are omitted.
The improvement comes from correcting coefficients already present
at quadratic and cubic degree.}
\label{fig:wave}\end{figure}

The burst contains successive plus and cross tidal pulses. We view a
stationary source plane obliquely through them, using the same
four-colour source as in the lens examples. Figure~\ref{fig:wave}
shows how its circle becomes a thin, bent image. TG+C reproduces much
of the linear stretching, but its cross and ring miss the exact
positions. With full propagation through the same cubic degree, these
features nearly coincide. The colour error falls from \(54.1\%\)
to \(0.15\%\). The wave amplitude is deliberately large enough to
resolve this effect; this is an exact spacetime test, rather than a
forecast for a detected astrophysical wave.

\FloatBarrier

There is a useful exact control behind this comparison. In Brinkmann
coordinates the transverse geodesic equation is linear,
\(\boldsymbol X''=H_{\rm gw}(u)\boldsymbol X\). At a common wave
phase \(u\), the transverse map is therefore exactly linear in the
initial slope \(\boldsymbol p=\dd\boldsymbol X/\dd u\). The
observed map instead starts from directions on the unit sky and ends
where each ray meets the oblique physical source plane. The phase of
that intersection varies across the beam. For this map the covariant
finite-separation and relative-direction terms are needed, together
with the launch and endpoint conversions. Both comparisons include
the same conversions, so those choices cannot account for the
difference between their images.

\Needspace{4\baselineskip}
We also calculate the physical map directly from the plane-wave
transfer matrix and expand it by independent polynomial arithmetic.
Its coefficients agree with the cubic NLGDE, and the full map's
source-position remainder falls as \(\rho^4\), compared with
\(\rho^2\) for TG and TG+C. The independent transfer map is given
in appendix~\ref{app:wave-model}. The wave samples curvature variation
as well as finite-separation geometry. Section~\ref{sec:fixed-degree-comparison}
compares these results with a control in which curvature gradients vanish.

\subsection{Comparison at fixed angular degree}
\label{sec:fixed-degree-comparison}
Table~\ref{tab:propagation-results} collects the image comparisons for
which the additional terms have a substantial effect. The two-plane
case isolates C; the Schwarzschild and wave cases test the remaining
spacetime terms on fixed source planes. Their errors refer to different
specified images, so only entries in the same row should be compared.

\begin{table}[!htbp]\centering\small
\setlength{\tabcolsep}{4pt}
\begin{tabular}{llccccc}\toprule
Example & Parameter & Degree & Error & TG & TG+C & Full\\\midrule
Two planes & \(b_{12}=0.596\) & 9 & \(\mathcal E_C\) & \(48.0\%\) & \(0.93\%\) & \(0.93\%\)\\
Schwarzschild & \(\mathsf M/D=0.07\) & 3 & \(\mathcal E_I\) & \(33\%\) & \(32\%\) & \(2.2\%\)\\
Wave burst & \(\ell^2\max|H_{{\rm gw},11}|=0.7\) & 3 & \(\mathcal E_C\) & \(53.4\%\) &
\(54.1\%\) & \(0.15\%\)\\\bottomrule
\end{tabular}
\caption{Image errors at fixed angular degree. The two-plane comparison
is in figure~\ref{fig:two-plane-coupling}, the Schwarzschild fixed-plane
comparison uses the source in appendix~\ref{app:schwarz-propagation}, and the wave is in
figure~\ref{fig:wave}. The colour error \(\mathcal E_C\) and brightness
error \(\mathcal E_I\) use the full windows and norms specified in
appendix~\ref{app:accuracy-measures}. For the reduced two-plane model,
TG+C gives the complete coefficients. The Schwarzschild and wave
benchmarks share the full temporal and longitudinal coefficients and
apply a common source-surface conversion, as specified in
appendices~\ref{app:schwarz-propagation} and \ref{app:wave-model}.}
\label{tab:propagation-results}\end{table}

The sphere-product control tests a different limit: all curvature
gradients vanish, while nonlinear propagation remains. Its cubic
\(\bar\beta_2^3\) coefficient changes from \(-0.808L\), using launch
and explicit cubic-force terms, to \(3.115L\) when lower-degree
coupling is included. Keeping the full linear and quadratic maps
but only the cubic launch term gives a \(96\%\) brightness error;
the full cubic map reduces it to \(3.5\%\). The geometry and source
are specified in appendix~\ref{app:sphere-inputs}. These two
comparisons distinguish the coefficient sign reversal from its effect
on an extended image.

\FloatBarrier

%% file: sections/05_extended_images.tex
\section{Reconstructing extended images from several expansion centres}\label{sec:lenses}
A correct local map can still fail to describe an extended image if
the image reaches too far from the reference ray. We now examine this
limitation using complete coefficients for each specified lens map.
A fold pair shows that crossing a critical curve need not obstruct the
expansion, and a compact component shows how nearby lens structure
can require new centres. We then follow more complicated maps
across a cluster field and around a rotating black hole. These examples test how angular degree and the
choice of reference rays work together to recover the image.

For the analytic thin lenses, completeness refers to the chosen reduced
lens equation. Their series and remainder bounds let us control the
error across whole patches. The Kerr coefficients instead come from the
full metric geodesic flow and include the observer and source-plane maps.
We quote RMS brightness errors \(\mathcal E_I\) relative to the direct
intensity, and colour errors \(\mathcal E_C\) relative to its contrast
against the white background. Both use the whole stated image window;
the norms are given in appendix~\ref{app:accuracy-measures}.

\subsection{A fold pair from one expansion}
A cored elliptical potential with external shear lets us test
whether one expansion can recover both images of a fold pair:
\begin{equation}
\begin{aligned}{}
\chi_L^2\psi_1(R)&=R_{\rm E}f(R)+\tfrac12R^A\Gamma_{AB}R^B,\\
 f^2(R)&=c^2+R^AQ_{AB}R^B,\\
 Q&=\mathsf R(\varphi)\begin{pmatrix}q&0\\0&q^{-1}\end{pmatrix}\mathsf R(\varphi)^T,\\
 \Gamma&=\gamma\begin{pmatrix}\cos2\varphi_\gamma&\sin2\varphi_\gamma\\
                         \sin2\varphi_\gamma&-\cos2\varphi_\gamma\end{pmatrix} .
 \label{eq:elliptical}
\end{aligned}
\end{equation}
The core radius is \(c\). The parameter \(q\) sets the ellipticity
of the potential, and \(\mathsf R(\varphi)\) rotates its axes by
\(\varphi\). The external shear has amplitude \(\gamma\) and
orientation \(\varphi_\gamma\). We choose
\(q=0.8\), \(c=0.035R_{\rm E}\), \(\varphi=18^\circ\),
\(\gamma=0.07\), and \(\varphi_\gamma=-28^\circ\).

Here ellipticity is specified for the potential. Its surface density
follows by differentiation; the resulting lensing convergence is positive:
\begin{equation}
\begin{aligned}{}
\kappa_1(R)&=\frac{R_{\rm E}}{2f^3}
                    \left[c^2\operatorname{tr}Q+R^2\right] .
 \label{eq:kappa-ellipse}
\end{aligned}
\end{equation}
We used \(\det Q=1\); the external shear is trace free.

\begin{figure}[!htb]
  \centering
  \includegraphics[width=\textwidth,height=.72\textheight,keepaspectratio]{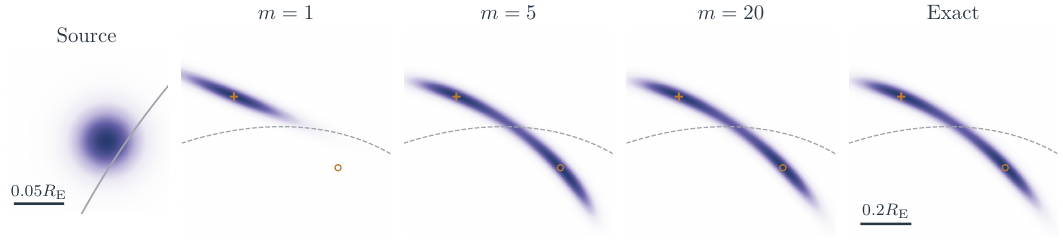}
  \caption{One forward expansion recovers both members of a fold pair.
The source and caustic are followed by the same image window at degrees
one, five and twenty, then the exact lens map. The cross marks the fixed
expansion centre and the open circle the other source-centre image.
Dashed grey curves are critical curves, not convergence boundaries.
All image panels share spatial and colour scales; the source has
its own scale bar. Increasing degree recovers the pair and its connecting
arc without moving the expansion centre.}
  \label{fig:fold}
\end{figure}

The critical curve makes the inverse map singular, but the forward map
of this smooth potential remains regular across it. One expansion can
therefore describe both members of a nearby fold pair. Its coefficients
vary with image position, while the barred sector remains zero because
the map derives from a scalar potential. In
figure~\ref{fig:fold} we set the source centre to
\((0.28,0.08)R_{\rm E}\) and its Gaussian width to \(0.025R_{\rm E}\).
The expansion is centred on the positive-parity fold image. The linear map describes
only its local elongation, while degrees five and twenty progressively recover
the pair and the connecting extended arc. The corresponding brightness errors \(\mathcal E_I\) over the window
are \(73\%\), \(12\%\) and \(0.026\%\).

\subsection{A compact component: the limit of one centre}
A small softened component can restrict the convergence domain while
leaving the real lens map smooth. We add the positive-density potential
\begin{equation}
\begin{aligned}{}
\chi_L^2\psi_p(R)&=\frac{b_p^2}{2}
        \ln\!\left(\frac{a_p^2+|R-R_p|^2}{R_{\rm E}^2}\right),\\
 \kappa_p(R)&=\frac{b_p^2a_p^2}{(a_p^2+|R-R_p|^2)^2} .
 \label{eq:compact}
\end{aligned}
\end{equation}
The component is centred at \(R_p\), has softening radius \(a_p\),
and has mass scale set by \(b_p^2\). We set \(b_p=0.09R_{\rm E}\), \(a_p=0.008R_{\rm E}\), and
\(R_p=(0.20,1.10)R_{\rm E}\). A source at
\((0.22,0.08)R_{\rm E}\), of width \(0.025R_{\rm E}\), has nine
resolved image centres in the full lens. Several are close to the compact
component and experience sharply different distortions.

For a patch centred on \(R\), two dimensionless ratios compare the
offset with the local scales limiting the Taylor series:
\begin{equation}
\begin{aligned}{}
\eta_1^2(x;R)&=\frac{x^AQ_{AB}x^B}{c^2+R^AQ_{AB}R^B},&
 \eta_p^2(x;R)&=\frac{|x|^2}{a_p^2+|R-R_p|^2},\\
 \eta(x;R)&=\max(\eta_1,\eta_p) .
 \label{eq:rho}
\end{aligned}
\end{equation}
The intersection \(\eta<1\) is a convergence domain of the compound
map, as follows from the analytic generating functions in
appendix~\ref{app:bounds}. Softening makes the map smooth at every real
point in the window, but does not give a Taylor series about one point
an unlimited domain. Here the compact component sets the smaller disc,
of radius only \(0.165R_{\rm E}\) about
\(R\simeq(0.360,1.06)R_{\rm E}\). Its boundary is a limit of that
analytic series, with no mass singularity on the real circle.
Figure~\ref{fig:atlas} keeps this centre fixed through
\(m_{\max}=100\); substantial light remains outside the domain.

Only \(54.3\%\) of the exact sampled flux in this window lies inside
\(\eta<1\). The degree-100 whole-window brightness error remains
\(66\%\), despite convergence to near machine precision well inside the
disc. At 1.2 times the compact component's convergence radius, the
compact-map error grows from \(0.193R_{\rm E}\) at degree ten to
\(2.58\times10^6R_{\rm E}\) at degree one hundred. Higher-precision
evaluation confirms this growth: the point lies outside the series'
convergence domain.

\subsection{Adding centres to meet an error target}\label{sec:changing-centres}

\begin{figure}[!htb]
  \centering
\includegraphics[width=\textwidth,height=.72\textheight,keepaspectratio]{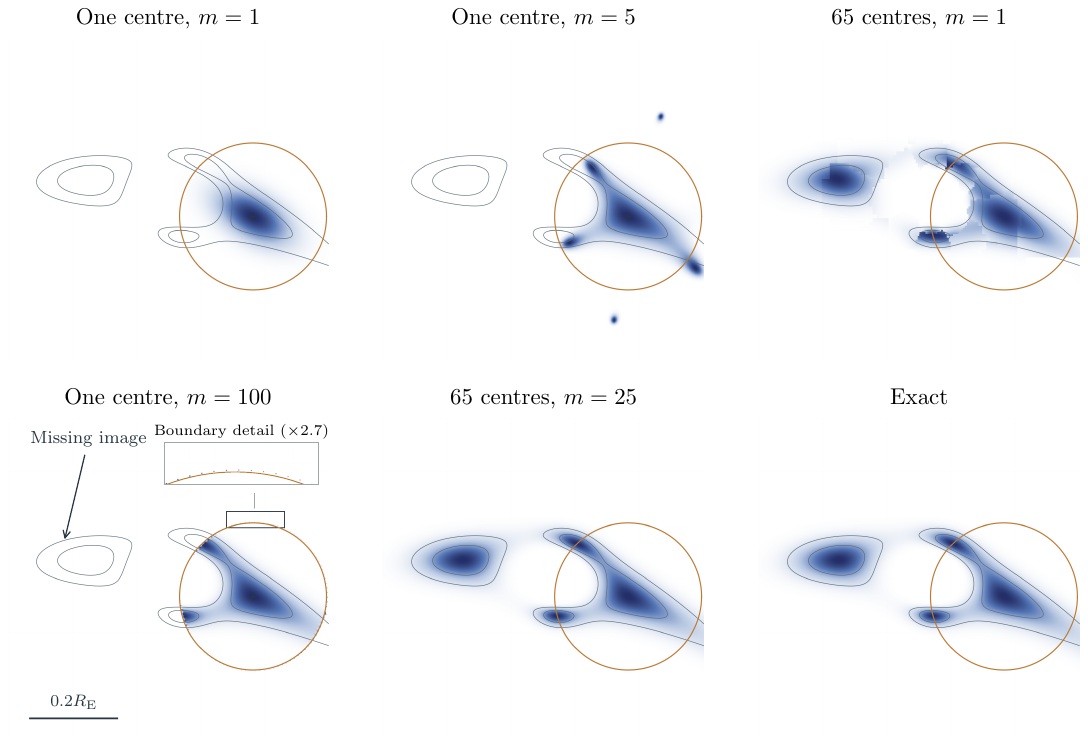}
  \caption{Additional expansion centres recover light outside a single Taylor
domain. The upper row compares one centre at degrees one and five with
65 centres at degree one; the lower row compares one degree-100 map,
65 degree-25 maps and the exact lens equation. Every panel uses the same
window, source and colour scale. The orange boundary encloses the
original domain; the arrow marks a missed image. Outside this boundary
lies \(45.7\%\) of the sampled exact flux. The degree-25 cover has a
uniform map-error bound below \(10^{-6}R_{\rm E}\) and brightness
error \(3.0\times10^{-7}\%\) on the independent diagnostic grid.
Grey contours show common exact brightness levels. The inset resolves
false images near the degree-100 boundary. Jumps at degree one are patch edges.
Appendix~\ref{app:bounds} gives the cover and its error bound.}
  \label{fig:atlas}\label{fig:compact}
\end{figure}
We now keep the same lens, source and window and calculate new local
maps about additional rays. Their collection is an atlas. Each patch
retains its own central source position and predicts the same absolute
source coordinates. Paper~II already proposed a separate series centred
on each image and identified the ring of spurious images near the Taylor
radius \citep{clarksonII}. Here we assign the patches by an explicit
error criterion. If \(U_j^{[M_j]}\) is the polynomial about centre \(j\), truncated at
degree \(M_j\), and \(B_j\) bounds its position error, we require
\begin{equation}
\begin{aligned}{}
|U_j^{[M_j]}(R)-U(R)|&\leq B_j(R,M_j)\leq\delta_U .
 \label{eq:patch-criterion}
\end{aligned}
\end{equation}
The tolerance \(\delta_U\) has units of source position. The
derivatives at each new centre are recalculated from the lens.
The new derivatives supply local information that extrapolating the
original polynomial cannot recover.

We first refine square tiles across the window, then select expansion
centres whose patches cover them. The analytic tail bounds in
appendix~\ref{app:bounds} control the error throughout each closed tile.
Greedy selection adds centres until every tile meets \(\delta_U\),
without using source brightness. Appendix~\ref{app:compact-greedy}
specifies the tile refinement and selection rule.

Figure~\ref{fig:atlas} shows a cover with 65 centres, each of degree 25,
for \(\delta_U=10^{-6}R_{\rm E}\). Its largest analytic tile bound
is \(9.15\times10^{-7}R_{\rm E}\). On the independent image grid the
maximum map error is \(1.07\times10^{-8}R_{\rm E}\), and the brightness
error is \(3.0\times10^{-7}\%\). Multiple centres recover the part of the
image inaccessible to the original centre.

There is also a choice between increasing degree and adding centres.
For the compact-component example at a common position-error target,
degrees 10, 25, 50 and 100 require 343, 65, 38 and 25 centres.
A dense two-component polynomial stores \(M(M+3)+2\) numbers, including
its endpoint, giving the totals in figure~\ref{fig:atlas-cost}. Raising the
degree reduces the number of centres but eventually increases the total
coefficient storage. The useful balance depends on the
map, required accuracy and representation of its coefficients.

The position bound also controls the discontinuity introduced by
assigning each tile to one local map. Both sides of a patch boundary
approximate the same exact map within \(\delta_U\), so their jump is
at most \(2\delta_U\). For a Gaussian of unit peak and width
\(\sigma\), \(|\nabla I_s|\leq e^{-1/2}/\sigma\) converts these
position errors into absolute brightness bounds. A smoothly varying
magnification or a reconstructed critical curve additionally requires
derivative bounds or controlled smooth overlap at patch boundaries.

\subsection{A cluster lens with crossing caustics}
The preceding examples follow a few image branches. A cluster with
several mass centres presents a different coverage problem: a single
extended source can cross many caustics, producing images over widely
separated angular scales. We use a thin lens with several mass scales
to test whether one prescribed accuracy target can control the map
across this whole field.

\begin{figure}[!htb]
  \centering
  \includegraphics[width=\textwidth,height=.72\textheight,keepaspectratio]{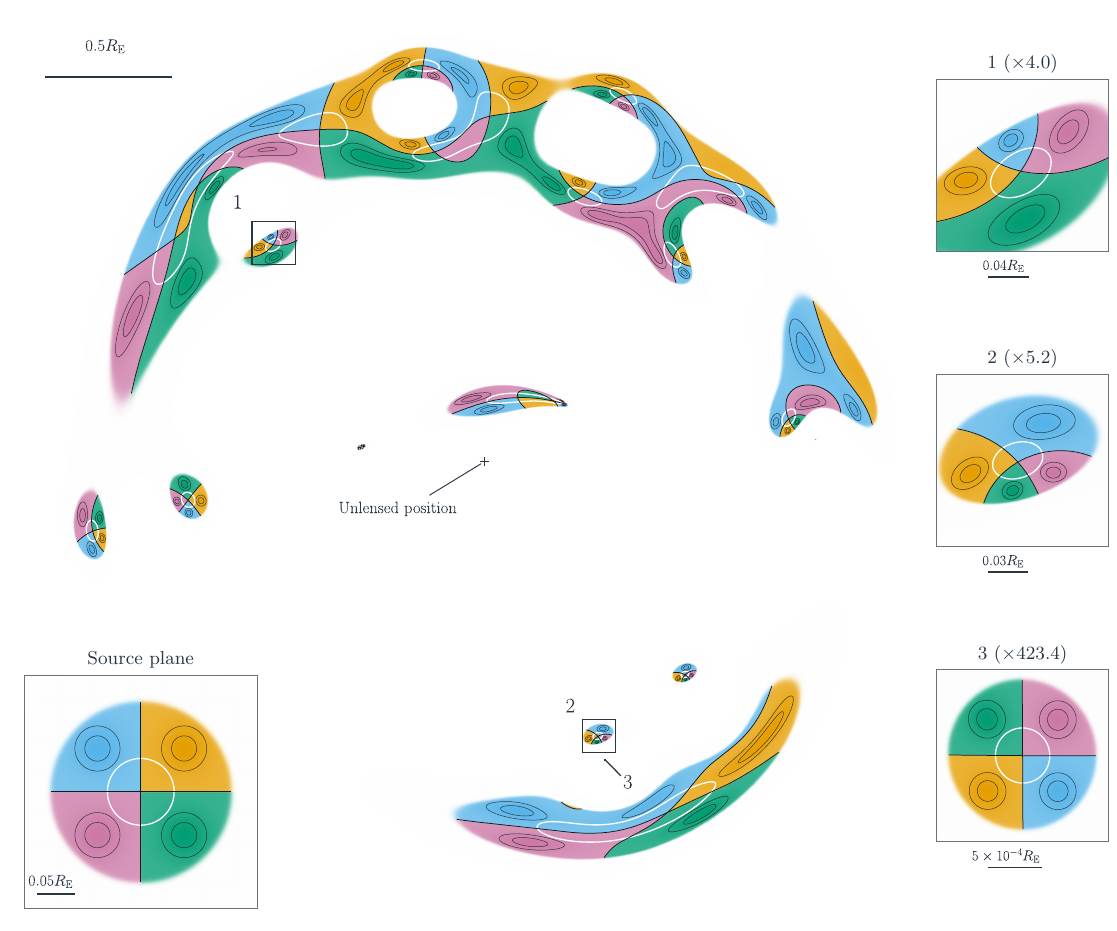}
  \caption{A network of crossing caustics produces many images of one structured
source. The lens contains three extended and eight compact components;
27 source-centre images are resolved. The source inset is at bottom left,
and the cross marks its projected unlensed centre,
\((0.075,0.015)R_{\rm E}\). Boxes 1--3 locate the enlargements at
right, with displayed linear magnifications 4.0, 5.2 and 423.4 and
individual scale bars. Each box has its crop's exact extent; a leader
locates box 3, which is smaller than a printed pixel. The main field is
\([-1.8,1.8]^2R_{\rm E}\), with \(R_1\) horizontal and \(R_2\)
vertical. All panels use direct rays and the same four-quadrant source as
figure~\ref{fig:model-gallery}. The cross, white circle and quadrant
contours follow the lens map.
Figure~\ref{fig:model-gallery} shows the density, and
figure~\ref{fig:cluster-atlas} tests the reconstruction.}
  \label{fig:cluster}
\end{figure}

\begin{figure}[!htb]
  \centering
\includegraphics[width=\textwidth,height=.755\textheight,keepaspectratio]{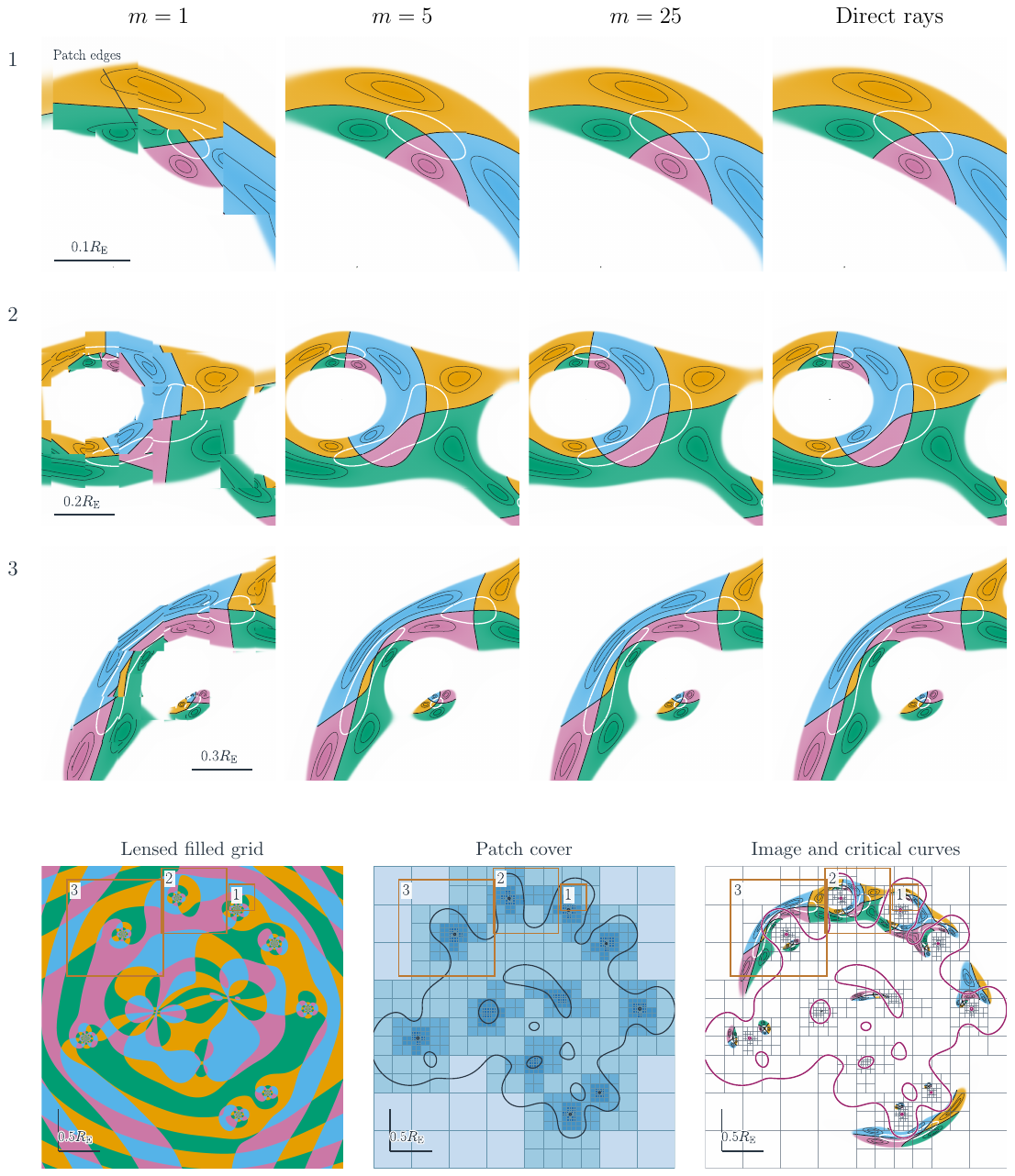}
  \caption{Increasing degree resolves the cluster images on one fixed cover of
1,015 patches. Columns compare degrees one, five and 25 with direct rays.
Rows 1--3 show a compact northern image, the curved northern group and
the north-western arc, in windows of width \(0.31\), \(0.78\) and
\(1.15R_{\rm E}\). Each row shares its window and colour scale, with
a scale bar in the first column. Black curves separate the four
source-colour regions. The cover was selected using the degree-25 bound
and then held fixed. Contours are drawn within each patch; the
degree-one jumps occur at patch boundaries.
\protect\\[0.35em]
The lower panels show, from left to right, direct rays for a source
plane filled by repeating coloured squares of side \(0.20R_{\rm E}\);
the patch cover, with darker cells finer and critical curves in dark ink;
and the compact-source image with grey tile boundaries and magenta
critical curves. They share the window \([-1.8,1.8]^2R_{\rm E}\),
with numbered boxes locating rows 1--3. The critical curves satisfy
\(\det(\partial U/\partial R)=0\) and need not follow the illuminated
arcs. Refinement is also independent of source brightness.
\protect\\[0.35em]
For the compact source, whole-window colour errors at degrees 1, 5, 9
and 25 are \(23\%\), \(1.0\%\), \(0.059\%\) and
\(1\times10^{-6}\%\). The cover bounds the position error by
\(10^{-6}R_{\rm E}\) throughout every tile, including unilluminated
regions.}
  \label{fig:cluster-atlas}
\end{figure}

This model combines three extended and eight compact positive-density
components with external shear. Its potential is a sum of the two forms
already used in \eqref{eq:elliptical} and \eqref{eq:compact}, with parameters
in appendix~\ref{app:cluster-inputs}. The common colour source has radius
\(0.12R_{\rm E}\) and centre \((0.075,0.015)R_{\rm E}\).
Its 27 resolved source-centre images locate only part of the structure:
different source features cross different caustic branches
(figure~\ref{fig:cluster}).

Because the potential is a sum of the two analytic forms already used,
we can add their roulette coefficients and error bounds. This gives a
way to choose expansion centres before specifying which parts of the
field are illuminated. Starting with \([-1.8,1.8]^2R_{\rm E}\),
we expand each tile about its centre and split it into four whenever its
uniform degree-25 bound exceeds \(10^{-6}R_{\rm E}\). The resulting
1\,015 centres control the position error throughout every closed tile
in exact arithmetic. Here refinement creates both the tiles and their
centres; section~\ref{sec:changing-centres} instead selected a greedy
set of patches over a fixed tiling.

Figure~\ref{fig:cluster-atlas} holds this cover fixed to test angular
degree across the field. At degree 25 the largest analytic tile bound is
\(9.32\times10^{-7}R_{\rm E}\), while the largest sampled error is
\(1.24\times10^{-8}R_{\rm E}\). The first controls every point,
including unilluminated regions; the second measures the smaller error
actually found on the evaluation grid.

\begin{figure}[!htb]
  \centering
\includegraphics[width=\textwidth,height=.72\textheight,keepaspectratio]{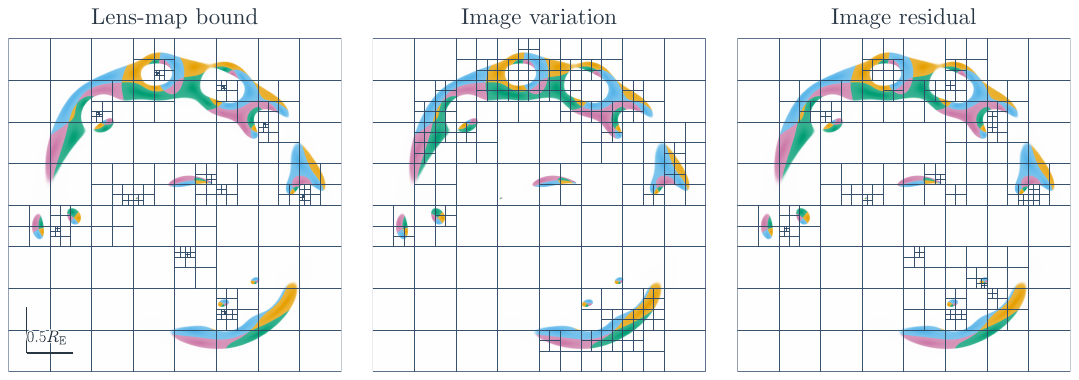}
  \caption{Where different refinement criteria add centres.
Each cluster cover starts from an \(8\times8\) grid and has 256
patches after 64 single-tile splits. The degree-25 lens-map bound refines
around lens structure even without source light; image variation and
the degree-25 image residual allocate patches using this source's
synthetic pixels. All panels span \([-1.8,1.8]^2R_{\rm E}\).
The fixed number of splits lets us compare where each criterion places
the additional centres.}
  \label{fig:cluster-covers}
\end{figure}

Figure~\ref{fig:cluster-covers} shows why the choice of refinement
criterion matters. A bound on the lens map selects structure even where
the particular source emits no light. Image variation favours tiles with
large deviations from their mean RGB value, while the image-residual
criterion favours differences between the polynomial prediction and the
synthetic observed pixels. The latter two can concentrate effort on
visible arcs, but depend on the source and pixel sampling. All three
still calculate their coefficients from the prescribed lens.


\subsection{Kerr: spin and extended arcs}\label{sec:kerr}
Kerr lets us test extended-image reconstruction in a rotating
spacetime using the full metric. Its long curved arcs have strongly varying distortions,
so the reconstruction must balance angular degree against the number of
reference rays. We calculate complete endpoint coefficients by
differentiating the full metric geodesic flow, including the observer
and source-plane maps. This gives the same endpoint derivatives as the
covariant hierarchy for the same launch and endpoint. The comparison
therefore tests angular truncation and coverage.

Figure~\ref{fig:kerr-separated} first changes the spin while holding
the source fixed, then compares reconstructions within the
\(a/\mathsf M=0.9\) spacetime. The first comparison includes all
changes in the geometry and endpoint map. In particular, both the
electric and magnetic Weyl fields in figure~\ref{fig:model-gallery}
enter the screen tidal matrix through
\eqref{eq:connector-weyl-electric-magnetic}, with their split defined by
the chosen locally nonrotating (ZAMO) observer. A change in the image
therefore cannot be assigned to magnetic curvature alone. The
two-plane test in section~\ref{sec:weak-two-lens} provides the more
restricted comparison in which successive shears generate an
antisymmetric map component.

\subsubsection{Geometry and source}
We use the Cartesian ingoing Kerr--Schild metric \citep{visser}, with
mass \(\mathsf M\), spin length \(a\), and third spatial coordinate
\(w\). The spin-dependent caustic geometry \citep{bozza} motivates
keeping the observer and source fixed when comparing different spins.

A static observer is at \((-D,0.148D,0.02D)\), with
\(\mathsf M/D=0.01\). The reconstruction tests use \(a/\mathsf M=0.9\). We orthonormalise the
positive Cartesian \(x,y,w\) directions in that order,
after projection into the static observer's rest space, to fix its
spatial frame. The first vector defines the reference
direction; the other two define the sky axes. Equal-energy future-directed
tangents use the exact angular coordinates \eqref{eq:sky}. We trace each ray backwards to the stationary
source plane \(x=D\) and write its endpoint as
\(U_A=(y_s,w_s)\). This fixes both the observed-angle and endpoint
conventions before any derivatives are taken.

Two prescribed sources select separated images and long arcs, with
centres \(U_A/D=(-0.105,0.055)\) and \((-0.150,0.035)\),
respectively. Both use the four-quadrant profile of
figure~\ref{fig:model-gallery}, with radius \(0.045D\) and a common
colour scale on a white background. The centred white circle and
quadrant contours mark shape landmarks. These sources test the
geometrical map; the calculation does not model emission spectra or
accretion-flow dynamics.

\Needspace{6\baselineskip}
\subsubsection{Changing spin}
\begin{figure}[!t]\centering
\includegraphics[width=0.92\textwidth,height=0.73\textheight,keepaspectratio]{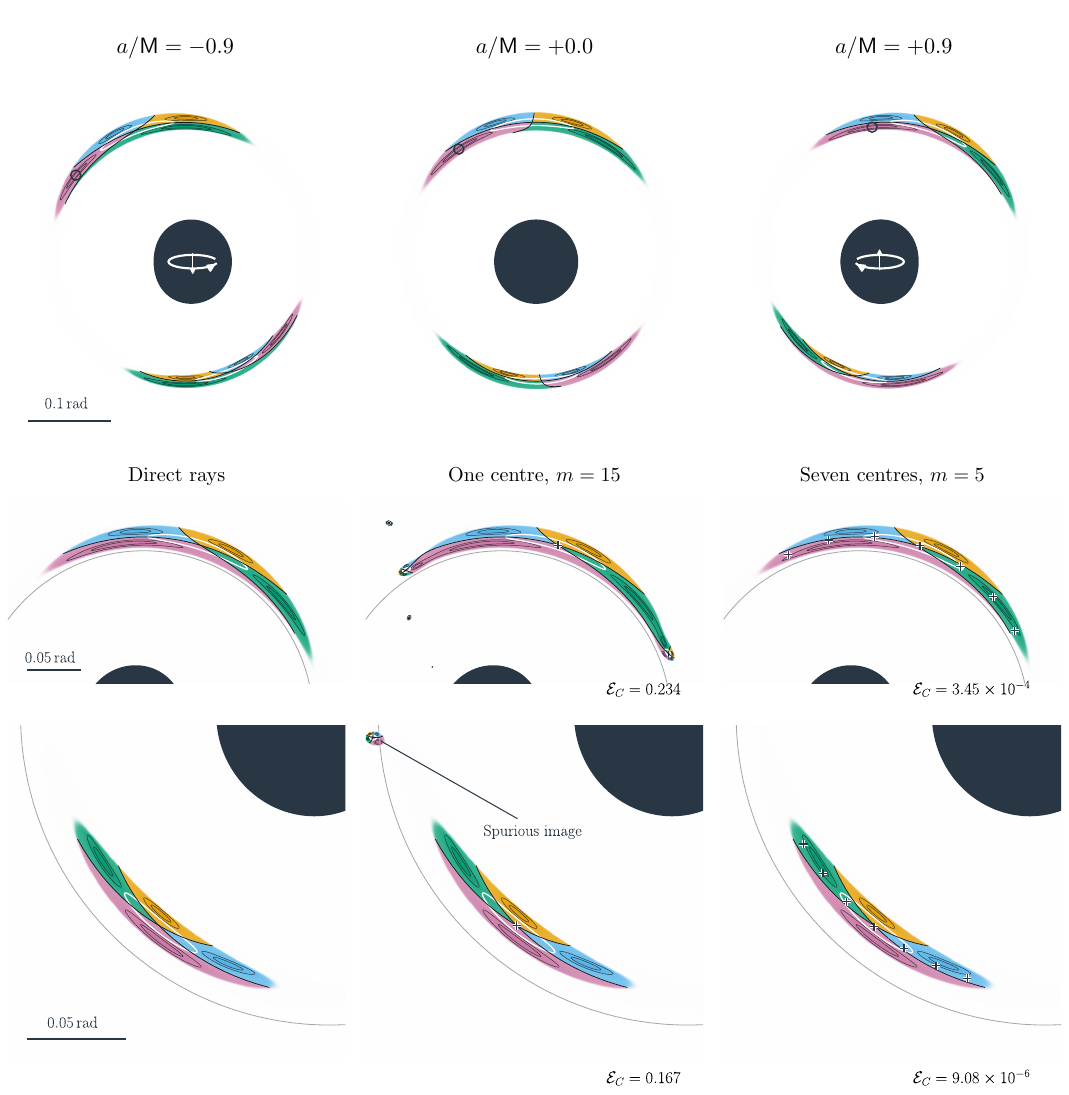}
\caption{Kerr spin changes the images, and additional expansion centres improve
their reconstruction. Top row: direct images of the source at
\((-0.150,0.035)D\) for three spins, with a common angular origin,
orientation and scale. The open circle follows one source landmark;
white loops indicate spin direction. These panels compare different
spacetimes.
\protect\\[0.35em]
The lower rows compare approximations in the fixed
\(a/\mathsf M=0.9\) spacetime. Columns show direct rays, one
degree-fifteen map and seven degree-five maps. The middle row is the
positive-parity long arc from the top-row source; the bottom row is a
negative-parity image of the source at \((-0.105,0.055)D\). Both
profiles have radius \(0.045D\), and each row retains its full
comparison window and angular scale. The reconstructions use 270 and
280 roulette amplitudes, respectively, plus constant endpoints.
Centres are chosen from the resolved reference arcs; their derivatives
are calculated from the metric. The lower-middle false image is included
in the colour error.
\protect\\[0.35em]
Crosses mark expansion centres, grey curves are critical curves and black
curves separate source-colour regions. Dark regions are shadows
calculated from spherical photon orbits (appendix~\ref{app:kerr-inputs});
untraced pixels in the common computational exclusion are white.}
\label{fig:kerr-separated}\label{fig:kerr-long}\label{fig:kerr-context}
\end{figure}

The top row of figure~\ref{fig:kerr-separated} keeps the longer-arc
source fixed while changing \(a/\mathsf M\) from \(-0.9\) through
zero to \(+0.9\). The observer stays at the same coordinate position,
and its frame is orthonormalised in each metric by the same prescription.
The four source landmarks move by about \(29\)--\(75\) mrad relative
to the zero-spin image; the open circle follows one of them. Source
features move along the arcs and change shape as the caustics move.
This comparison includes all changes in the endpoint map under the fixed
frame prescription. It does not represent a single rigid image rotation.

\subsubsection{Local roulette maps}
For each arc we first choose the ray that reaches the source centre.
Call its angular position \(\boldsymbol\zeta_c\), and write
\(\zeta^A-\zeta_c^A=\rho_c n^A(\vartheta_c)\). The local approximation is
\begin{equation}
\begin{aligned}{}
U_A^{[m_{\max}]}(\boldsymbol\zeta;\boldsymbol\zeta_c)
 &=U_A(\boldsymbol\zeta_c)+\sum_{m=1}^{m_{\max}}
       \frac{\rho_c^m}{m!}P_A^{(m)}(\vartheta_c;\boldsymbol\zeta_c),\\
 P_A^{(m)}&=\calM_{A B_1\cdots B_m}(\boldsymbol\zeta_c)
                       n^{B_1}\cdots n^{B_m},\\
 \calM_{A B_1\cdots B_m}(\boldsymbol\zeta_c)
 &=\left.\nabla_{B_1}\cdots\nabla_{B_m}U_A\right|_{\boldsymbol\zeta_c}.
 \label{eq:kerr-local-map}
\end{aligned}
\end{equation}
Here \(\nabla_B\) is the flat derivative in the one fixed observer
coordinate plane introduced in section~\ref{sec:review}. Its repeated
action gives the ordinary Taylor derivatives of the component functions
\(U_A\), as calculated by the Taylor expansion of the flow. It is distinct from
covariant differentiation on the unit sky. The tensor \(\calM\) differentiates the full endpoint map, including
its linear background contribution. Its amplitudes therefore follow the
full-map convention of section~\ref{sec:review}.

In a coordinate system, first write
\[
 \frac{\dd x^a}{\dd\lambda}=k^a,\qquad
 \frac{\dd k^a}{\dd\lambda}=-\Gamma^a{}_{bc}(x)k^bk^c.
\]
Next expand each of these eight coordinate components in the two launch
angles about the chosen direction \(\boldsymbol\zeta_c\), held fixed. Let \(q^I=(x^a,k^a)\),
\(I=1,\ldots,8\), collect them,
and let \(f^I\) denote the right-hand side just displayed. Then
\begin{equation}
\begin{aligned}{}
q^I(\lambda,\boldsymbol\zeta)
 &=\sum_{i+j\leq m_{\max}}q^I_{ij}(\lambda)
                                  (\zeta_1-\zeta_{c1})^i(\zeta_2-\zeta_{c2})^j,\\
 \frac{\dd q^I_{ij}}{\dd\lambda}
 &=[(\zeta_1-\zeta_{c1})^i(\zeta_2-\zeta_{c2})^j]f^I(q).
 \label{eq:coordinate-jets}
\end{aligned}
\end{equation}
The square bracket means: substitute the polynomial into the geodesic
equations and read off the indicated coefficient. Products and functions are evaluated by
multivariate truncated
power-series arithmetic (Taylor-mode automatic differentiation),
discarding degrees above \(m_{\max}\).
The label \(I\) numbers coordinate components; it is not a spacetime
tensor index. This procedure differentiates the flow rather than fitting
a polynomial to sampled rays. All retained Kerr rays have monotonic
Cartesian \(x\), so we integrate the coefficient equations using
\(x\) as the independent variable and stop at \(x=D\). This includes
the angle-dependent arrival affine parameter.

At each degree we decompose \(P_A^{(m)}\) into the original
\(\alpha_s^m,\beta_s^m,\bar\alpha_s^m,\bar\beta_s^m\) basis,
using equations~\eqref{eq:spins} and \eqref{eq:fourier}. The barred
amplitudes give the independent tangential part of the endpoint map in
the chosen sky coordinates and source coordinates. Their physical curvature
interpretation is discussed in appendix~\ref{app:weyl-projection};
appendix~\ref{app:optical-twist} explains the relation to endpoint
antisymmetry and optical twist.

To compare increasing degree with adding centres, we give the two
reconstructions similar numbers of amplitudes. A complete vector map
has \(2(m+1)\) real roulette amplitudes at degree \(m\), so one
expansion through degree fifteen uses 270, while seven through degree
five use 280. Each centre also carries a separate two-component
constant endpoint.

We place the seven centres along each resolved reference arc. Seven
azimuthal bins about the inward sky direction each supply a centre at
their median angle and radius. Appendix~\ref{app:kerr-inputs} specifies
that origin, the windows and the common validity mask. All retained
coefficients are complete, so differences between the reconstructions
measure angular truncation and coverage. The fixed-degree propagation
tests are in section~\ref{sec:spacetime}.

\subsubsection{Coverage of longer arcs}\label{sec:kerr-coverage}
The source at \(U_A/D=(-0.150,0.035)\) on the plane \(x=D\)
produces long arcs spanning a wide range of observed directions.
For its positive- and negative-parity images, one degree-fifteen map
leaves colour errors of about \(23\%\) and \(29\%\), respectively.
Seven degree-five maps reduce these to approximately \(0.034\%\)
and \(0.077\%\). The middle row of
figure~\ref{fig:kerr-separated} shows the positive-parity comparison.
Additional centres keep each evaluation point closer to its expansion
ray, allowing the maps to follow the changing distortion along the arc.

\subsubsection{Local accuracy and extrapolation}
The separated-source test uses \(U_A/D=(-0.105,0.055)\).
We again compare direct rays, one degree-fifteen map and seven
degree-five maps. For the positive-parity image, the whole-window
colour error falls from about \(55\%\) at degree three to
\(0.014\%\) at degree fifteen, whose internal landmarks closely match
the direct image. The lower row of figure~\ref{fig:kerr-separated}
shows the negative-parity image, where extrapolation has a different
effect.

In the negative-parity case, degree fifteen recovers the true arc but
also predicts a false image near the upper-left corner of the window.
The polynomial extrapolation produces this false image, which is
included in the whole-window colour error of about \(17\%\).
The error exceeds the degree-nine value of \(2.0\%\): increasing
Taylor degree improves a local approximation, but can worsen its
extrapolation across a finite window.

Additional centres reduce the distance over which any one polynomial is
extrapolated. Seven fifth-degree expansions reduce the two colour errors to roughly
\(0.001\%\). At each pixel we use the nearest centre and its complete
local polynomial, including the constant endpoint. All centres are translated origins within
the same observer coordinates.

\FloatBarrier

%% file: sections/06_discussion.tex
\section{Discussion}\label{sec:discussion}
The roulette hierarchy connects the local shape of an image to the
Riemann curvature along its rays. The Jacobi operator propagates each
angular degree, while the nonlinear force acts on the distortions
accumulated at lower degrees. Appendix~\ref{app:geometry} derives the corrected mixed cubic and
quartic forces, the ordered single-curvature terms and the fifth-degree
\(R^3\) terms, and relates them to the Vines and Kim equations.
Initial data from the observer's sky and composition with the source
surface convert the propagated separation into the chosen image map.
These steps distinguish the curvature that drives the beam from the
angular coordinates and endpoint at which its image is measured.

An already distorted beam responds differently to a later deflection.
TG+C retains this nonlinear composition, as illustrated by the recovery
of the two-plane image. The same composition is present in standard
multi-plane lensing and in the ordered deflections of the reduced scalar
cosmological ray equation (appendix~\ref{app:ordered-deflections}).
For a single reduced thin plane, its exact lens equation gives TG,
TG+C and the complete coefficients together. Beyond that limit, the
remaining force terms describe the effects of along-ray curvature
variation, relative ray direction and local curvature products. Their
contribution is only a few parts in \(10^6\) in our weak isothermal
example, even though the image is strongly stretched. The weak-field
family scalings in \eqref{eq:relevance-scaling} give roughly
\(10^{-4}\) at cluster Einstein angles. The compact Schwarzschild,
sphere-product and wave calculations explore regimes where these terms
become appreciable, changing both the local coefficients and the
reconstructed images.

The angular extent covered by those coefficients depends on the Taylor
remainder and nearby lens structure. A smooth forward expansion can
recover a fold pair across a critical curve, whereas a compact lens
component can place part of the image outside one centre's convergence
domain. Additional centres supply the local information needed there.
For the analytic thin lenses we bound the position error uniformly over
the resulting covers; for Kerr we compare several lower-degree maps
with one higher-degree map at similar amplitude counts. Differentiating
the exact geodesic flow supplies these higher angular degrees directly,
while the covariant force expansion identifies their curvature sources.
The wave calculation checks the agreement of the two routes through
cubic degree.

Fitting roulette maps across resolved images would express their local
distortions in a common angular expansion and relate them to the
gravitational tidal field along the rays. Such an inference includes the source structure,
observer and source-surface conversions, and lensing degeneracies.
The examples show how to propagate and combine the local maps
when the lens is specified, giving controlled tests for fits to observed
images. Wavelet representations can describe either the source or the
lens map in those fits; appendix~\ref{app:wavelets} gives their relation
to roulette amplitudes.

%% file: appendices/a_connector_geometry.tex
\section{The relative acceleration from connector geometry}
\label{app:geometry}\label{app:construction}
A photon direction is parallel transported along its own ray. To compare
it with a second photon direction, we must also transport it across the
beam. As the photons advance, this connecting path changes. The relative
acceleration follows by accounting for both the change of the connector
and the change of transport along it.

The same connector can be specified by its two endpoints \((x,x')\),
or by its initial event and tangent \((x,\xi)\), with
\(x'=\exp_x\xi\). Vines differentiates the two-endpoint world function;
Kim uses the initial event and tangent \citep{vines,kim}. Passing between
them changes the variables used to describe the same geometry.

\begin{figure}[!htb]\centering
\includegraphics[width=\textwidth]{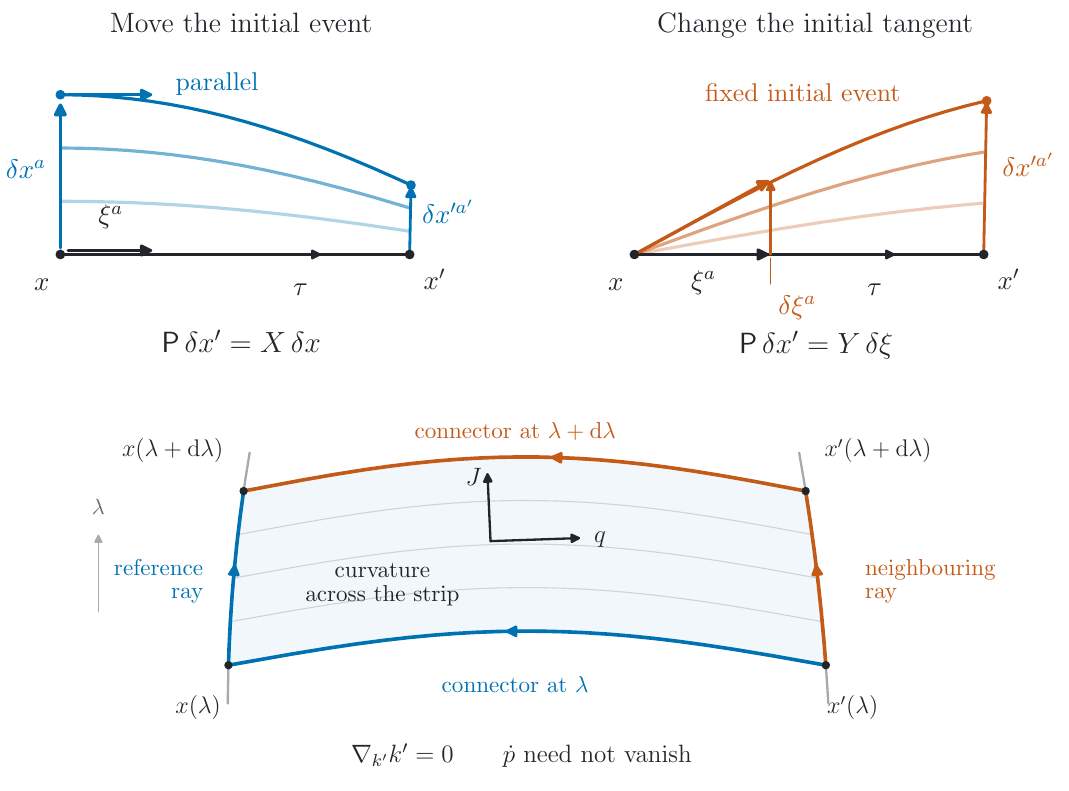}
\caption{How the connector changes as neighbouring rays advance.
Upper left: move its initial event by \(\delta x^a\), carrying the
initial tangent \(\xi^a\) in parallel. Upper right: keep that event
fixed and change the tangent by \(\delta\xi^a\). The endpoint
displacements are \(K^{a'}{}_b\delta x^b\) and
\(H^{a'}{}_b\delta\xi^b\), respectively; transport back to the
initial event gives the corresponding maps \(X\) and \(Y\).
These panels use great-circle connectors on a unit sphere, with
transverse variations enlarged. Below: successive connectors bound a
strip between the rays. The blue and orange paths compare transport
across the connector before or after advancing the photon. Their
difference is curvature integrated across the strip. Future affine
parameter increases upwards. The connector tangent \(q\) points
towards the neighbouring ray, opposite to the cross-connector transport
arrows; \(J\) points along the variation between connectors.
Appendix~\ref{app:technical} derives these endpoint and transport maps.}
\label{fig:connector-variations}\label{fig:connector-strip}
\end{figure}

As \(\lambda\) changes, the initial event moves with tangent \(k\)
and the connector's initial tangent changes at rate \(v\). Their
separate contributions to the other endpoint are the two Jacobi
propagators shown in figure~\ref{fig:connector-variations}. Once their
output is transported back to \(x\), the neighbouring direction is
\(p=Xk+Yv\). The separation rate \(v\) is therefore determined by
how the entire connecting geodesic changes. The change of \(p\) also
depends on the strip between successive connectors
(figure~\ref{fig:connector-strip}). Differentiating the endpoint relation
and imposing geodesic motion at the neighbouring endpoint gives the
nonlinear deviation equation. We derive these endpoint variations and
their transport before expanding the force.

The flat-space examples in section~\ref{sec:endpoint-choice} isolate
the effects of the angular coordinates and endpoint choice. The sphere-product
test in section~\ref{sec:fixed-degree-comparison} shows how lower-degree coupling
changes the cubic coefficient even when curvature gradients vanish.

\subsection{Conventions and derivative directions}
We use signature \((-+++)\), an affine parameter \(\lambda\) increasing
towards the future, and
\begin{equation}
 [\nabla_c,\nabla_d]w^a=R^a{}_{bcd}w^b,
 \qquad R_{ab}=R^c{}_{acb}.
 \label{s:conventions}
\end{equation}
Coordinates, \(\lambda\), and the separation \(\xi\) may be assigned
units of length; the reference tangent \(k^a\) and separation rate
\(v^a\) are then dimensionless. Curvature has units
length\(^{-2}\), and an acceleration has units length\(^{-1}\).
Angles are dimensionless but are evaluated numerically in radians.

The slot convention is
\begin{equation}
 R(p,q,r)^a=R^a{}_{bcd}p^bq^cr^d,
 \qquad R(k,\xi,k)^a=R^a{}_{k\xi k}.
\end{equation}
For positive sectional curvature \(K\) on a Riemannian factor,
\(R(p,q,r)=K[q(p\cdot r)-r(p\cdot q)]\). Thus
\(-R(k,\xi,k)\) points towards the reference geodesic for transverse
\(\xi\): the linear Jacobi equation focuses a positive-curvature beam.
This check fixes the sign before any nonlinear calculation.

A horizontal covariant derivative moves the base point while parallel
transporting its tangent arguments. In curvature contractions those
arguments are therefore held covariantly fixed. Write
\(\nabla_p\nabla_qR=p^aq^b\nabla_a\nabla_bR\), with the
\(q\)-derivative acting first. Every tensor index, including a derivative
index, participates in a further covariant derivative. In particular
\(\nabla_p\nabla_qR\) is not just two coordinate derivatives of its
components. A vertical derivative with respect to \(\xi\) is different:
it changes the tangent at a fixed base point, acting on every explicit
occurrence of \(\xi\), including those inside derivative directions.

Table~\ref{tab:four-labels} distinguishes force degree, angular degree, curvature grading and
metric perturbation order.
In the angular hierarchy, \(h\) is a formal, dimensionless scaling parameter,
set to one after collecting powers; \(\epsilon\) counts metric
perturbations. The tensor \(h_{ij}\) in the cosmological appendix is
the usual metric tensor perturbation.
We reserve \(z\) for redshift and use \(w\) for the third spatial
coordinate in explicitly Cartesian model coordinates.

The derivative \(\nabla_a\) is the spacetime Levi--Civita derivative.
On an explicitly flat lens or angular coordinate plane, \(\nabla_A\)
means its ordinary coordinate derivative. We use \(\bar\nabla_a\) for the flat
conformal-background derivative in the cosmological calculation and
\(\widetilde\nabla_a\) for the perturbed conformal connection.
The symbol \(\mathrm d_\vartheta\) differentiates a one-variable angular
pattern. Partial derivatives with respect to auxiliary parameters or the
independent components of \(\xi\) retain their ordinary mathematical
meaning; they are not spacetime covariant derivatives.

The endpoint bitensors below follow Vines's notation. An unprimed
index belongs to the reference event \(x\); a primed index belongs to
\(x'\). In \(H^{a'}{}_b\) and \(K^{a'}{}_b\), the lower index
labels initial data at \(x\), and the upper index is the resulting
vector at \(x'\). Multiplication by \(\mathsf P^a{}_{a'}\) transports that
output back to \(x\), giving Kim's \(Y^a{}_b\) and \(X^a{}_b\).

\subsection{Connector geometry and endpoint differentials}\label{app:technical}
We obtain the endpoint differentials and the change in their parallel
transport, referring every index to the reference ray. The two independent
variations change the connector's base point and its initial tangent.
Table~\ref{tab:connector-dictionary} collects the notation and its
relation to the propagators used by Vines.

Synge's world function and the connector tangent are related by
\begin{equation}
\begin{aligned}{}
\sigma(x,x')&=\tfrac12g_{ab}(x)\xi^a\xi^b,&
 \xi^a&=-g^{ab}(x)\nabla_b\sigma(x,x').\label{eq:world-function}
\end{aligned}
\end{equation}
The gradient acts at the reference event with the other endpoint fixed.
To compare the photon tangents at one event, transport the neighbouring
tangent back along the connector:
\begin{equation}
\begin{aligned}{}
p^a&=\mathsf P^a{}_{a'}(x,x')k'^{a'}.\label{eq:parallel-comparison}
\end{aligned}
\end{equation}
The separation rate \(v^a\) describes
how the connector's initial tangent changes as both rays advance.
In flat space \(p^a-k^a=v^a\); curvature modifies the relation.
Varying the connector's initial point and tangent gives
\begin{equation}
\begin{aligned}{}
p^a&=X^a{}_b k^b+Y^a{}_b v^b,\\
 X^a{}_b&=\mathsf P^a{}_{a'}K^{a'}{}_b,&
 Y^a{}_b&=\mathsf P^a{}_{a'}H^{a'}{}_b.
 \label{eq:connector-dictionary-main}
\end{aligned}
\end{equation}
Vines's endpoint Jacobi propagators \(K,H\) become \(X,Y\)
after transport back to \(x\); these are the propagators used in
Kim's tangent-bundle expansion \citep{vines,kim}.
These short-connector matrices reduce to the identity in flat space.
In general their difference from the identity measures why the transported
direction difference is not the separation rate:
\[
 p^a-k^a-v^a=(X^a{}_b-\delta^a{}_b)k^b
                    +(Y^a{}_b-\delta^a{}_b)v^b.
\]
The optical Jacobi map instead propagates angular separations along the
light ray. The derivation below obtains \(X\) by varying the connector's
base point and \(Y\) by varying its initial tangent, before collecting
the exact dictionary in section~\ref{app:vines-kim-dictionary}.

\subsubsection{The connector and the world function}
Let \(x(\lambda)\) and \(x'(\lambda)\) be two affinely parametrised
geodesics. Equal values of their affine parameters define the comparison;
for a null bundle a common observer energy fixes their relative scaling.
At each \(\lambda\), join the events by
\begin{equation}
 \gamma(\tau,\lambda)=\exp_{x(\lambda)}[\tau\xi(\lambda)],
 \qquad 0\leq\tau\leq1.
\end{equation}
The joining geodesic compares events on the two photon trajectories
and can be spacelike, timelike or null. We work on a smooth branch of
the logarithm, normally inside a convex normal neighbourhood.
Its tangent \(q=\partial_\tau\gamma\) is parallel along the connector.
The geodesic action gives
\begin{equation}
 \sigma(x,x')=\frac12\int_0^1g(q,q)\,\dd\tau
             =\frac12g_x(\xi,\xi).
\end{equation}
Varying this action and integrating the interior Euler--Lagrange term
by parts leaves only endpoint terms:
\begin{equation}
 \delta\sigma=q_{a'}(1)\delta x'^{a'}-\xi_a\delta x^a.
\end{equation}
Therefore \(\xi_a=-\nabla_a\sigma\), with the primed event fixed in
that derivative. This is the logarithm vector, not a coordinate difference.
The argument works also when the connector is null: \(\sigma=0\) does
not imply \(\nabla\sigma=0\).

Parallel transport the neighbouring tangent back along this connector:
\begin{equation}
 p^a=\mathsf P^a{}_{a'}k'^{a'},\qquad
 \nabla_qw=0,\qquad v^a=\nabla_k\xi^a.
\end{equation}
Both vectors in \(p-k\) now belong to the same tangent space.
In curved spacetime \(p-k\) generally differs from \(v\) because
motion of the initial point and a change of initial tangent produce
different variations of the connecting geodesic.

\subsubsection{Deriving the two Jacobi propagators along the connector}\label{sec:generator}
Figure~\ref{fig:connector-variations} separates the two independent changes
of the connector. Moving its initial event, with \(\xi\) carried in
parallel, gives a variation with nonzero initial displacement and zero
initial change of tangent. Changing \(\xi\) at a fixed event gives
zero initial displacement and a nonzero change of tangent. Every small
change of the connector is a sum of these two variations.

Let \(J=\partial_\lambda\gamma\) be the connecting variation field.
Since \([q,J]=0\) and the connection is torsion free,
\(\nabla_qJ=\nabla_Jq\). Commuting derivatives and using
\(\nabla_qq=0\) gives
\begin{equation}
 \nabla_q^2J^a=R^a{}_{bcd}q^bq^cJ^d.
\end{equation}
The Jacobi equation is linear in the infinitesimal variation \(J\).
The connector being varied can have finite length: no expansion in
\(\xi\) has been made here. Its curvature is evaluated along the
whole connector, so solving this linear variation problem retains the
nonlinear dependence on the separation.

Resolve all vectors in a parallel frame along the connector. Its tangent
has constant components \(\xi\). With
\(\mathcal Q(\tau)^a{}_b=R^a{}_{\xi\xi b}(\gamma(\tau))\),
let \(A\) propagate a change of the initial point at fixed parallel
initial tangent, and \(B\) a change of that tangent at a fixed initial
point. Their two independent Jacobi initial conditions obey
\begin{equation}
\begin{aligned}{}
\frac{\dd^2A}{\dd\tau^2}&=\mathcal Q A,& A(0)&=I,&
 A'(0)&=0,\\
 \frac{\dd^2B}{\dd\tau^2}&=\mathcal Q B,& B(0)&=0,&
 B'(0)&=I.\label{eq:connector-connector-Jacobi}
\end{aligned}
\end{equation}

The variation generated by advancing the two rays has
\(J(0)=k\) and \(\nabla_qJ(0)=v\). Hence
\begin{equation}
 J(\tau)=A(\tau)k+B(\tau)v,\qquad
 p=Xk+Yv,\qquad X=A(1),\quad Y=B(1).
 \label{s:tangent}
\end{equation}
The \(A,B\) here propagate variations along the short connector.
They are not the observer-to-source optical Jacobi matrix. In flat
space \(A=I,B=\tau I\), so \(p=k+v\).

A positively curved spatial surface gives a simple example of the
difference. Let the connector have length \(\ell\) on a surface of
constant Gaussian curvature \(K>0\). For a variation perpendicular to
the connector within this surface, the Jacobi equation is
\(\dd^2J_\perp/\dd\tau^2=-K\ell^2J_\perp\). Its two initial
conditions give the transverse eigenvalues
\begin{equation}
 X_\perp=\cos(\sqrt K\,\ell),\qquad
 Y_\perp=\frac{\sin(\sqrt K\,\ell)}{\sqrt K\,\ell}.
 \label{eq:connector-constant-curvature}
\end{equation}
An initially displaced connector begins with its separation already
present; one with a perturbed tangent first develops a separation.
Curvature acts on these different histories, producing the cosine and
sine solutions. Both approach unity as \(K\ell^2\) tends to zero,
but already differ at order \(K\ell^2\), although every curvature
gradient vanishes. Figure~\ref{fig:connector-variations} shows the
corresponding connector families.

\subsubsection{The curvature integral from varying the comparison path}
Starting with \(k'\) at \(x'(\lambda)\), we can first advance along
the neighbouring ray and then transport across the new connector.
Alternatively, we can transport across the old connector and then along
the reference ray. Both paths end at \(x(\lambda+\dd\lambda)\), and
together they bound the strip in figure~\ref{fig:connector-strip}.
Their difference is the curvature accumulated over the strip, which
changes \(p\) even though \(k'\) is parallel along its own ray.

To obtain this contribution, carry the neighbouring
tangent parallel along each connector, giving a field \(w(\tau,\lambda)\)
with \(\nabla_qw=0\) and \(w(1)=k'\). At \(\tau=1\), advancing \(\lambda\) follows the neighbouring photon,
so \(\nabla_Jw|_{\tau=1}=\nabla_{k'}k'=0\). Commutation gives
\begin{equation}
 \nabla_q\nabla_Jw^a=R^a{}_{bcd}w^bq^cJ^d.
\end{equation}
Integrating from zero to one gives the endpoint derivative minus the
base derivative. The endpoint term vanishes, leaving minus the change
of the transported tangent at the base. Pulling every index back gives
\begin{equation}
 \dot p^a=-\int_0^1R^a{}_{b\xi d}J^d\,\dd\tau\,p^b
          =-(\widetilde X^a{}_{bc}k^c+
                       \widetilde Y^a{}_{bc}v^c)p^b,
\end{equation}
where
\begin{equation}
 \widetilde X^a{}_{bc}=\int_0^1R^a{}_{b\xi d}A^d{}_c\,\dd\tau,
 \qquad
 \widetilde Y^a{}_{bc}=\int_0^1R^a{}_{b\xi d}B^d{}_c\,\dd\tau.
 \label{s:transport-integrals}
\end{equation}
Each element of the strip contributes curvature contracted with its
connector tangent and its displacement as \(\lambda\) changes.
The minus sign follows because the vanishing derivative at the
neighbouring endpoint equals the derivative at the reference event plus
this integral. Substituting \(J=Ak+Bv\) splits that integral according
to the two ways of varying the connector. This is why the same \(A,B\)
occur in both the endpoint relation and the transport terms.
The tildes denote transport-variation tensors; later, tildes on metrics
denote conformal quantities.

\subsubsection{Horizontal and vertical differentiation, including the mixed identity}
The endpoint changes when the connector's base point moves or its
initial tangent changes. A \emph{horizontal} variation moves the base
point with \(\xi\) held parallel and is generated by \(X\); a
\emph{vertical} variation changes \(\xi\) at that point and is
generated by \(Y\). These terms refer to the tangent bundle, rather
than directions on the optical screen. Along the ray pair both changes
occur, with rates \(k\) and \(v\), so a connector coefficient has
derivative contributions \(\nabla_k\) and \(\partial_v\).
Differentiating \eqref{s:tangent} along the reference ray gives
\begin{equation}
 \dot p=(\nabla_kX+\partial_vX)k+
        (\nabla_kY+\partial_vY)v+Y\ddot\xi,
 \qquad \partial_v=v^c\frac{\partial}{\partial\xi^c}.
\end{equation}

To combine the mixed terms, we need the relation between the two orders
of endpoint variation. Their coefficients are
\((\partial_cX^a{}_d+\widetilde Y^a{}_{ec}X^e{}_d)v^ck^d\)
and
\((\nabla_cY^a{}_d+\widetilde X^a{}_{ec}Y^e{}_d)k^cv^d\).
Let \(s\) and \(t\) label the horizontal and vertical variations.
Torsion freedom gives
\(\nabla_s\partial_t x'=\nabla_t\partial_s x'\).
Transporting this equality back to the base point introduces the
\(\widetilde X\) and \(\widetilde Y\) terms and gives
\begin{equation}
\begin{aligned}{}
\frac{\partial X^a{}_d}{\partial\xi^c}
       +\widetilde Y^a{}_{ec}X^e{}_d
 &=\nabla_dY^a{}_c+\widetilde X^a{}_{ed}Y^e{}_c .
 \label{eq:connector-torsion-identity}
\end{aligned}
\end{equation}

Relabelling the dummy indices in the first mixed term then gives precisely
the second. The two add, explaining the factor of two in
\begin{equation}
\begin{aligned}{}
-Y^a{}_b\ddot\xi^b={}&
 (\nabla_cX^a{}_d+\widetilde X^a{}_{ec}X^e{}_d)k^ck^d\\
 &+2(\nabla_cY^a{}_d+\widetilde X^a{}_{ec}Y^e{}_d)k^cv^d\\
 &+\left(\frac{\partial Y^a{}_d}{\partial\xi^c}
                    +\widetilde Y^a{}_{ec}Y^e{}_d\right)v^cv^d .
 \label{eq:connector-propagator-gde}
\end{aligned}
\end{equation}

The three tangent products correspond to moving the initial event
twice, moving it while changing the initial tangent, and changing that
tangent twice. Their form makes the force at most quadratic in \(v\)
at every separation degree. Solving for \(\ddot\xi\) requires
\(Y^{-1}\), because \(Y\) converts a change of the initial
connector tangent into a displacement of the other endpoint. The
equation is exact wherever this differential is invertible and applies
to null rays without any division by \(k^2\).

A useful leading-order check is
\begin{align}
 \partial_cX^a{}_d+\widetilde Y^a{}_{ec}X^e{}_d
 &=\tfrac12[R^a{}_{c\xi d}+R^a{}_{\xi cd}+R^a{}_{d\xi c}]
    +O(\xi^2)\\
 &=R^a{}_{c\xi d}+O(\xi^2),
\end{align}
using the algebraic Bianchi identity. The right side of
\eqref{eq:connector-torsion-identity} has the same leading term. An untransposed
\(c,d\) pair on one side instead produces \(R^a{}_{d\xi c}\),
which is generally different. The mixed slots in the published equation~(33b) of \citet{kim}
therefore require the transposition shown in \eqref{eq:connector-torsion-identity}.

\subsubsection{The direct Vines--Kim dictionary}\label{app:vines-kim-dictionary}
The geometric identities above can also be obtained by differentiating
the two-endpoint world function. Here \(\sigma^a{}_b=\nabla_b\nabla^a\sigma\) and
\(\sigma^a{}_{b'}=\nabla_{b'}\nabla^a\sigma\); each primed derivative
acts on the neighbouring endpoint while the other endpoint is held fixed.
In that language \citep{vines},
\begin{equation}
\begin{aligned}{}
v^a&=-\sigma^a{}_b k^b-\sigma^a{}_{b'}k'^{b'},\\
 \sigma^a{}_{c'}H^{c'}{}_b&=-\delta^a{}_b,&
 K^{a'}{}_b&=H^{a'}{}_c\sigma^c{}_b,\\
 k'^{a'}&=K^{a'}{}_b k^b+H^{a'}{}_b v^b.
 \label{eq:connector-vines-tangent}
\end{aligned}
\end{equation}
\begin{equation}
\begin{aligned}{}
X^a{}_b&=\mathsf P^a{}_{a'}K^{a'}{}_b,\\
 Y^a{}_b&=\mathsf P^a{}_{a'}H^{a'}{}_b,\\
 p^a&=X^a{}_b k^b+Y^a{}_b v^b .
 \label{eq:connector-vines-kim-dictionary}
\end{aligned}
\end{equation}

Vines's exact equation writes the acceleration in terms of the endpoint
Jacobi propagator \(H^{a'}{}_a\) and the second endpoint variations
\(L^{a'}{}_{bc},J^{a'}{}_{bc},I^{a'}{}_{bc}\) \citep{vines}:
\begin{equation}
\begin{aligned}{}
0={}&H^{a'}{}_a\ddot\xi^a
   +L^{a'}{}_{bc}k^bk^c
   +2J^{a'}{}_{bc}k^bv^c
   +I^{a'}{}_{bc}v^bv^c .
 \label{eq:vines-exact}
\end{aligned}
\end{equation}
Transporting the output index back to \(x\) identifies the three
contracted coefficients with Kim's tangent-bundle equation:
\begin{equation}
\begin{aligned}{}
(\mathsf P L)^a{}_{bc}k^bk^c
   &=(\nabla_cX^a{}_d+\widetilde X^a{}_{ec}X^e{}_d)k^ck^d,\\
 (\mathsf P J)^a{}_{bc}k^bv^c
   &=(\nabla_cY^a{}_d+\widetilde X^a{}_{ec}Y^e{}_d)k^cv^d,\\
 (\mathsf P I)^a{}_{bc}v^bv^c
   &=\left(\frac{\partial Y^a{}_d}{\partial\xi^c}
                +\widetilde Y^a{}_{ec}Y^e{}_d\right)v^cv^d.
 \label{eq:vines-kim-coefficients}
\end{aligned}
\end{equation}
The two-index matrices propagate endpoint variations; the three-index
tilded tensors account for the changing parallel transport. The connector
variations above define these objects and establish the identities. In
particular, the mixed identity uses the transposed slots in
\eqref{eq:connector-torsion-identity}. Substitution gives
\eqref{eq:connector-propagator-gde}, with no force expansion required.

The dictionary relates two descriptions of the same geodesics and
separation. Vines varies the two endpoints, while Kim varies the base
point and its tangent vector. Their exact equations therefore give the
same relative acceleration, and either can generate expansions beyond
cubic degree.

\begin{table}[!tbp]\centering\small
\begin{tabular}{p{0.19\textwidth}p{0.72\textwidth}}
\toprule
Object & Meaning and relation to the endpoint notation\\
\midrule
\(A(\tau),B(\tau)\) & Connector Jacobi solutions with position and
velocity initial data; \(X=A(1)\), \(Y=B(1)\).\\
\(X,Y\) & Transported endpoint propagators \(\mathsf P K\) and
\(\mathsf P H\) of Vines.\\
\(\widetilde X,\widetilde Y\) & Variations of the connector's parallel
transport. Together with derivatives of \(X,Y\), they give
\(\mathsf P L,\mathsf P J,\mathsf P I\) in
\eqref{eq:vines-kim-coefficients}.\\
\(\mathsf V_{[r]}\) & Degree-\(r\) coefficient of \(Y^{-1}\).\\
\(Q_n,\widetilde Q_n\) & Curvature matrices with \(n-2\) separation
derivatives, defined in \eqref{eq:connector-Q}.\\
\(\mathcal G_n\) & Degree-\(n\) numerator of the acceleration before
multiplication by \(-Y^{-1}\), given in \eqref{eq:connector-Gn-explicit}.\\
\bottomrule
\end{tabular}
\caption{Connector notation used in the force derivation. All output
indices in the right column have been transported to the reference
point; the relations to Vines's objects therefore include
\(\mathsf P\).}
\label{tab:connector-dictionary}\end{table}

The matrix \(Y\) can become singular at a conjugate point of the
\emph{connector}. This is distinct from an observer-to-source critical
curve, where the forward angular map loses rank. A local logarithm can
remain regular across that critical curve. Global nonuniqueness of a
connector at a cut locus is a further branch issue; invertibility alone
does not select a global logarithm.

\subsection{From connector propagators to relative acceleration}
We expand the connector matrices, differentiate their endpoint dependence,
and invert \(Y\) to extract each homogeneous relative acceleration.
The origin of the curvature products can be seen before expanding.
Integrating the two Jacobi equations with their respective initial data gives
\begin{equation}
\begin{aligned}{}
A(\tau)&=I+\int_0^\tau(\tau-s)\mathcal Q(s)A(s)\,\dd s,\\
 B(\tau)&=\tau I+\int_0^\tau(\tau-s)\mathcal Q(s)B(s)\,\dd s.
 \label{eq:connector-integral-equations}
\end{aligned}
\end{equation}
Curvature first changes a connector variation; curvature farther along
that connector then acts on the changed variation. Iterating these
integrals produces ordered curvature products, even when the curvature
is covariantly constant. The order matters when the matrices at
different points do not commute. These products belong to the local
finite-separation relation. Coupling between already propagated angular
maps is a separate operation, as in
\eqref{eq:cubic-hierarchy-main}.

\subsubsection{Ordinary coefficients and noncommuting curvature products}\label{app:propagators}
We abbreviate the curvature matrices at successive separation degrees by
\begin{equation}
\begin{aligned}{}
(Q_n)^a{}_b&=(\nabla_\xi^{\,n-2}R)^a{}_{\xi\xi b},&
 (\widetilde Q_n)^a{}_{bc}&=(\nabla_\xi^{\,n-2}R)^a{}_{b\xi c},
 \qquad n\geq2 .
 \label{eq:connector-Q}
\end{aligned}
\end{equation}

Taylor expansion of curvature along a geodesic gives
\(\mathcal Q(\tau)=\sum_{r\geq0}\tau^rQ_{r+2}/r!\).
Expand \(A=\sum_j\tau^jA_j\) and \(B=\sum_j\tau^jB_j\).
These are ordinary coefficients in the connector parameter \(\tau\),
with no Taylor factorial removed. They therefore differ in normalisation
from the angular derivative tensors. Equating the coefficient of \(\tau^n\) in their Jacobi equations gives
\begin{equation}
\begin{aligned}{}
A_{n+2}&=\frac{1}{(n+1)(n+2)}
                   \sum_{r=0}^n\frac{Q_{r+2}A_{n-r}}{r!},&
 A_0&=I,\quad A_1=0,\\
 B_{n+2}&=\frac{1}{(n+1)(n+2)}
                   \sum_{r=0}^n\frac{Q_{r+2}B_{n-r}}{r!},&
 B_0&=0,\quad B_1=I .
 \label{eq:connector-prop-recursion}
\end{aligned}
\end{equation}

For example, \(A_2=Q_2/2\), \(A_3=Q_3/6\), and
\(A_4=(Q_4+Q_2^2)/24\). At the next degree the terms are
\begin{equation}
 A_5=\frac1{20}\left[Q_2\frac{Q_3}{6}
                  +Q_3\frac{Q_2}{2}+\frac{Q_5}{6}\right].
\end{equation}
The coefficients of \(Q_3Q_2\) and \(Q_2Q_3\) differ because the
ordered products record which curvature acts first. The endpoint series are
\begin{equation}
\begin{aligned}{}
X={}&I+\tfrac12Q_2+\tfrac16Q_3+
               \tfrac1{24}(Q_4+Q_2Q_2)
    +\tfrac1{120}(Q_5+3Q_3Q_2+Q_2Q_3)+O(\xi^6),\\
 Y={}&I+\tfrac16Q_2+\tfrac1{12}Q_3+
               \tfrac1{120}(3Q_4+Q_2Q_2)
    +\tfrac1{360}(2Q_5+2Q_3Q_2+Q_2Q_3)+O(\xi^6) .
 \label{eq:connector-propagators}
\end{aligned}
\end{equation}

The separation degree of \(A_j\) is \(j\); that of \(B_j\) is
\(j-1\). For instance \(B_1=I\) is degree zero. This one-unit shift
must be retained when translating connector powers into force degrees.

For the transport-variation tensors, the subscript \([d]\) denotes
homogeneous degree \(d\) in the separation. Inserting the connector
series into \eqref{s:transport-integrals}, each term integrates
\(\tau^{r+j}\) to \(1/(r+j+1)\), giving
\begin{equation}
\begin{aligned}{}
(\widetilde X_{[d]})^a{}_{bc}
 &=\sum_{\substack{r+j=d-1\\r,j\geq0}}
 \frac{(\widetilde Q_{r+2})^a{}_{be}(A_j)^e{}_c}{r!(r+j+1)},\\
 (\widetilde Y_{[d]})^a{}_{bc}
 &=\sum_{\substack{r+j=d\\r\geq0,\ j\geq1}}
 \frac{(\widetilde Q_{r+2})^a{}_{be}(B_j)^e{}_c}{r!(r+j+1)} .
 \label{eq:connector-transport-coefficients}
\end{aligned}
\end{equation}
\begin{equation}
\begin{aligned}{}
\widetilde X^a{}_{bc}={}&(\widetilde Q_2)^a{}_{bc}
 +\tfrac12(\widetilde Q_3)^a{}_{bc}
 +\tfrac16\left[(\widetilde Q_4)^a{}_{bc}
 +(\widetilde Q_2)^a{}_{bd}(Q_2)^d{}_c\right]\\
 &+\tfrac1{24}\left[(\widetilde Q_5)^a{}_{bc}
 +3(\widetilde Q_3)^a{}_{bd}(Q_2)^d{}_c
 +(\widetilde Q_2)^a{}_{bd}(Q_3)^d{}_c\right]+O(\xi^5),\\
 \widetilde Y^a{}_{bc}={}&\tfrac12(\widetilde Q_2)^a{}_{bc}
 +\tfrac13(\widetilde Q_3)^a{}_{bc}
 +\tfrac1{24}\left[3(\widetilde Q_4)^a{}_{bc}
 +(\widetilde Q_2)^a{}_{bd}(Q_2)^d{}_c\right]+O(\xi^4).
 \label{eq:connector-transport-propagators}
\end{aligned}
\end{equation}

Let \(\mathsf V_{[r]}\) denote the ordinary degree-\(r\) coefficient of
\(Y^{-1}\), so that \(Y^{-1}=\sum_{r\geq0}\mathsf V_{[r]}\). Matching degrees in \(YY^{-1}=I\) gives
\begin{equation}
\begin{aligned}{}
\mathsf V_{[0]}&=I,\qquad \mathsf V_{[r]}=-\sum_{s=1}^rY_{[s]}\mathsf V_{[r-s]},\\
 Y^{-1}&=I-\tfrac16Q_2-\tfrac1{12}Q_3
              -\tfrac1{40}Q_4+\tfrac7{360}Q_2Q_2+O(\xi^5) .
 \label{eq:connector-inverse-Y}
\end{aligned}
\end{equation}

The coefficient \(7/360=1/36-1/120\) combines the square of the
quadratic coefficient \(1/6\) with \(-1/120\) from subtracting
the quartic term of \(Y\).

\subsubsection{Homogeneous force extraction}\label{app:force-recipe}
At a fixed reference event we expand the acceleration in its separation
and rate arguments:
\begin{equation}
\begin{aligned}{}
\ddot\xi^a&=\calF^a(\xi,v)
       =-R^a{}_{k\xi k}+\sum_{n\geq2}\calF_n^a(\xi,v),\\
 \calF_n(h\xi,hv)&=h^n\calF_n(\xi,v).
 \label{eq:vines-force-taylor}
\end{aligned}
\end{equation}
Denote the right side of \eqref{eq:connector-propagator-gde} by \(\mathcal G\).
Its degree-\(n\) coefficient has \(kk\), \(kv\) and \(vv\)
sectors. Each explicit \(v\) contributes one degree. In the last sector
the vertical derivative also removes one separation, explaining the
shifted subscripts on \(Y\):
\begin{equation}
\begin{aligned}{}
\mathcal G_n^a={}&
 \left[\nabla_c(X_{[n]})^a{}_d+
   \sum_{r+s=n}(\widetilde X_{[r]})^a{}_{ec}(X_{[s]})^e{}_d\right]k^ck^d\\
 &+2\left[\nabla_c(Y_{[n-1]})^a{}_d+
   \sum_{r+s=n-1}(\widetilde X_{[r]})^a{}_{ec}(Y_{[s]})^e{}_d\right]k^cv^d\\
 &+\left[\frac{\partial(Y_{[n-1]})^a{}_d}{\partial\xi^c}+
   \sum_{r+s=n-2}(\widetilde Y_{[r]})^a{}_{ec}(Y_{[s]})^e{}_d\right]v^cv^d .
 \label{eq:connector-Gn-explicit}
\end{aligned}
\end{equation}

Here coefficients of negative degree vanish, \(X_{[0]}=Y_{[0]}=I\),
\(X_{[1]}=Y_{[1]}=0\), and the degree-zero transport-variation tensors vanish.
Equating degrees in \(-Y\sum_n\mathcal F_n=\sum_n\mathcal G_n\)
gives
\begin{equation}
\begin{aligned}{}
\calF_n^a&=-\mathcal G_n^a-
       \sum_{r=2}^{n-1}(Y_{[r]})^a{}_b\calF_{n-r}^b .
 \label{eq:connector-force-recursion}
\end{aligned}
\end{equation}

There are two distinct recursions: this one produces the local force in
arbitrary \((\xi,v)\); the later angular recursion inserts propagated
maps into those slots.

\subsubsection{Linear and quadratic forces, step by step}
For compact intermediate algebra set
\(T_j(a,b)^c=(\widetilde Q_j)^c{}_{de}a^db^e\).
For example \(T_2(a,b)=R(a,\xi,b)\) and
\(Q_2b=R(\xi,\xi,b)\). Then
\begin{align}
 \mathcal G_1&=T_2(k,k),\\
 \mathcal G_2&=\tfrac12(\nabla_kQ_2)k
              +\tfrac12T_3(k,k)+2T_2(v,k).
\end{align}
No inverse-\(Y\) contribution occurs until degree three. Thus
\(\mathcal F_1=-R(k,\xi,k)\), and \(-\mathcal G_2\) gives the three
quadratic terms of \eqref{eq:F2}, with the sign of the second term
following from \(R(\xi,\xi,k)=-R(\xi,k,\xi)\). The first two terms
have units \(\nabla R\,\xi^2k^2\sim\mathrm{length}^{-1}\), and the
last has units \(R\,v\xi k\sim\mathrm{length}^{-1}\).

\subsubsection{Cubic force and its mixed derivative}
The uncombined degree-three source is
\begin{align}
 \mathcal G_{3,0}={}&\tfrac16(\nabla_kQ_3)k
  +\tfrac16T_4(k,k)+\tfrac16T_2(k,Q_2k)+\tfrac12T_2(Q_2k,k),\\
 \mathcal G_{3,1}={}&\tfrac13(\nabla_kQ_2)v+T_3(v,k),\\
 \mathcal G_{3,2}={}&\tfrac16(\partial_vQ_2)v+\tfrac12T_2(v,v).
 \label{s:G3}
\end{align}
The second subscript counts powers of \(v\). Since
\((\partial_vQ_2)v=R(v,\xi,v)+R(\xi,v,v)=R(v,\xi,v)\),
the last line is \(2R(v,\xi,v)/3\). Solving the inverse factor gives
\begin{equation}
 \mathcal F_3=-\mathcal G_3-\tfrac16Q_2\mathcal F_1.
\end{equation}
Substituting these sources and reversing the antisymmetric last two
curvature slots gives \eqref{eq:F3}. The three local curvature-product
coefficients are fixed by \(\widetilde X X\) and by the inverse of
\(Y\). All have the same separation degree as \(\nabla^2R\,\xi^3\),
even though their curvature-factor counts differ.

In particular the mixed source is \((\nabla_kQ_2)v/3\), since it comes
from the explicitly horizontal term \(2(\nabla_kY)v\). Replacing it
by \((\nabla_vQ_2)k/3\), as in the cubic line of Kim's published equation
(32), gives the acceleration difference \emph{printed minus corrected},
\begin{equation}
\begin{aligned}{}
\delta\calF_3^a&=\tfrac13\left((\nabla_vR)^a{}_{\xi k\xi}
                                     -(\nabla_kR)^a{}_{\xi v\xi}\right)\\
 &=\tfrac13(\nabla_\xi R)^a{}_{\xi k v}.
 \label{eq:connector-Kim-delta}
\end{aligned}
\end{equation}

The differential Bianchi identity proves the second equality. For a
two-dimensional curved spatial factor choose orthonormal components
\(k=(1,0)\), \(\xi=(0,y)\), \(v=(0,w)\). The discrepancy is
\begin{equation}
 \delta\mathcal F_3=\frac13(\nabla_2K)wy^2(1,0).
\end{equation}
A transverse curvature gradient makes this difference nonzero.
Adding flat time and a further flat spatial factor gives a null-ray
optical example. Both the varying-curvature null test and the
four-dimensional check recover a cubic remainder when the printed mixed
term is used.

\subsubsection{The quartic force}\label{app:quartic}
At quartic degree the recurrence \eqref{eq:connector-force-recursion} gives
\begin{equation}
 \mathcal F_4=-\mathcal G_4-\tfrac16Q_2\mathcal F_2
                              -\tfrac1{12}Q_3\mathcal F_1 .
\end{equation}
The two inverse-propagator terms affect the zero- and one-\(v\)
sectors. The two-\(v\) sector follows directly from
\(\mathcal G_{4,2}=(\partial_vQ_3)v/12+T_3(v,v)/3\).
Here the vertical derivative changes the derivative direction as well
as the explicit curvature slots:
\begin{equation}
 (\partial_vQ_3)v=(\nabla_vR)_{\xi\xi v}
                 +(\nabla_\xi R)_{v\xi v},
\end{equation}
where the remaining contraction vanishes by antisymmetry. We retain the
derivative order when expanding the zero-, one- and two-\(v\) sectors
below. The first two contain curvature products as well as derivatives;
the two-\(v\) sector contains one derivative of curvature. All seventeen
contractions are retained:\begin{equation}
\begin{aligned}{}
\calF_4^a&=\calF_{4,0}^a+\calF_{4,1}^a+\calF_{4,2}^a .
 \label{eq:connector-F4}
\end{aligned}
\end{equation}
\begin{equation}
\begin{aligned}{}
\calF_{4,0}^a={}&
 -\tfrac1{24}(\nabla_k\nabla_\xi\nabla_\xi R)^a{}_{\xi\xi k}
 -\tfrac1{24}(\nabla_\xi\nabla_\xi\nabla_\xi R)^a{}_{k\xi k}\\
 &-\tfrac1{24}(\nabla_kR)^a{}_{\xi\xi b}R^b{}_{\xi\xi k}
 +\tfrac1{24}R^a{}_{\xi\xi b}(\nabla_kR)^b{}_{\xi\xi k}\\
 &-\tfrac1{24}R^a{}_{k\xi b}(\nabla_\xi R)^b{}_{\xi\xi k}
 -\tfrac18(\nabla_\xi R)^a{}_{k\xi b}R^b{}_{\xi\xi k}\\
 &-\tfrac16R^a{}_{b\xi k}(\nabla_\xi R)^b{}_{\xi\xi k}
 -\tfrac14(\nabla_\xi R)^a{}_{b\xi k}R^b{}_{\xi\xi k}\\
 &+\tfrac1{12}R^a{}_{\xi\xi b}(\nabla_\xi R)^b{}_{k\xi k}
 +\tfrac1{12}(\nabla_\xi R)^a{}_{\xi\xi b}R^b{}_{k\xi k} .
 \label{eq:connector-F40}
\end{aligned}
\end{equation}
\begin{equation}
\begin{aligned}{}
\calF_{4,1}^a={}&
 -\tfrac16(\nabla_k\nabla_\xi R)^a{}_{\xi\xi v}
 -\tfrac13(\nabla_\xi\nabla_\xi R)^a{}_{v\xi k}\\
 &-\tfrac13R^a{}_{v\xi b}R^b{}_{\xi\xi k}
 -\tfrac13R^a{}_{b\xi k}R^b{}_{\xi\xi v}
 +\tfrac13R^a{}_{\xi\xi b}R^b{}_{v\xi k} .
 \label{eq:connector-F41}
\end{aligned}
\end{equation}
\begin{equation}
\begin{aligned}{}
\calF_{4,2}^a={}&-\tfrac1{12}(\nabla_vR)^a{}_{\xi\xi v}
                    -\tfrac5{12}(\nabla_\xi R)^a{}_{v\xi v} .
 \label{eq:connector-F42}
\end{aligned}
\end{equation}

The seventeen contractions agree with expansion of the uncombined
propagator source.

\subsubsection{The single-curvature force at arbitrary degree}\label{app:curvature-orders-details}
The single-curvature sector gives all terms with one Riemann tensor
and any required ordered derivatives. For \(n\geq3\), its complete
form is
\begin{equation}
\begin{aligned}{}
\left.\calF_n^a\right|_R={}&
 -\frac{1}{n!}(\nabla_\xi^{\,n-1}R)^a{}_{k\xi k}
 +\frac{1}{n!}(\nabla_k\nabla_\xi^{\,n-2}R)^a{}_{\xi k\xi}\\
 &-\frac{2}{(n-1)!}(\nabla_\xi^{\,n-2}R)^a{}_{v\xi k}
 +\frac{2(n-2)}{n!}(\nabla_k\nabla_\xi^{\,n-3}R)^a{}_{\xi v\xi}\\
 &-\frac{(n-2)(n+1)}{n!}(\nabla_\xi^{\,n-3}R)^a{}_{v\xi v}\\
 &-\frac{n-2}{n!}\sum_{j=0}^{n-4}
 (\nabla_\xi^{\,j}\nabla_v\nabla_\xi^{\,n-4-j}R)^a{}_{\xi\xi v} .
 \label{eq:one-curvature-all}
\end{aligned}
\end{equation}
The last sum is empty at \(n=3\); degree two is \eqref{eq:F2}.
Contraction vectors are held fixed and derivative order is preserved.

We obtain the coefficients by selecting one curvature tensor from the
connector recurrence. With no second curvature factor available, it
must act on the flat-space propagator. The required terms are therefore
\begin{equation}
\begin{aligned}{}
\left.X_{[d]}\right|_R&=\frac{Q_d}{d!},\qquad
 \left.Y_{[d]}\right|_R=\frac{d-1}{(d+1)!}Q_d,
 &\qquad d\geq2,\\
 \left.(\widetilde X_{[d]})^a{}_{bc}\right|_R
      &=\frac1{d!}(\nabla_\xi^{\,d-1}R)^a{}_{b\xi c},\\
 \left.(\widetilde Y_{[d]})^a{}_{bc}\right|_R
      &=\frac{d}{(d+1)!}(\nabla_\xi^{\,d-1}R)^a{}_{b\xi c},
 &\qquad d\geq1 .
 \label{eq:connector-one-curvature-propagators}
\end{aligned}
\end{equation}

Inverse-\(Y\) corrections multiply an already curved force, so they
contain at least two curvature factors. Consequently
\(\mathcal F_n|_R=-\mathcal G_n|_R\). The zero-\(v\) sector comes from the horizontal derivative of \(X\)
and its transport variation. Its terms are
\(-[(\nabla_kQ_n)k+(\nabla_\xi^{n-1}R)_{k\xi k}]/n!\).

For one \(v\), the horizontal derivative of \(Y\) and the transport
variation supply the two contributions. In particular,
\(Y_{[n-1]}|_R=(n-2)Q_{n-1}/n!\) yields
the along-ray derivative coefficient \(2(n-2)/n!\), while
\(\widetilde X_{[n-1]}\) yields \(2/(n-1)!\).

For two \(v\), varying the initial tangent differentiates both the
explicit separation slots and their derivative directions. The vertical
derivative contributes
\begin{equation}
\begin{aligned}{}
v^c\frac{\partial}{\partial\xi^c}
       (\nabla_\xi^{\,n-3}R)^a{}_{\xi\xi v}
 ={}&(\nabla_\xi^{\,n-3}R)^a{}_{v\xi v}\\
 &+\sum_{j=0}^{n-4}
    (\nabla_\xi^{\,j}\nabla_v\nabla_\xi^{\,n-4-j}R)^a{}_{\xi\xi v} .
 \label{eq:connector-one-curvature-vertical}
\end{aligned}
\end{equation}

Its prefactor is \((n-2)/n!\). The transport term contributes another
\((n-2)/(n-1)!\) to the first contraction. Their sum is
\((n-2)(n+1)/n!\). Restoring the signs and reversing the antisymmetric slots gives
\eqref{eq:one-curvature-all}. This proves all derivative placements in the last line without commuting
covariant derivatives. At \(n=3\) the sum is empty and the result
contains the non-product terms of \eqref{eq:F3}; at \(n=4\) it contains
all single-curvature terms in the displayed quartic force.

\subsubsection{Curvature-factor counting and the first cubic-curvature
sector}\label{app:curvature-grading}\label{app:curvature-flrw}
Every term is quadratic in the reference or relative tangent before
collecting the angular degrees. With \(r\) factors of \(v\), there
are \(2-r\) factors of \(k\), \(n-r\) factors of \(\xi\),
and \(p\) curvature tensors with \(d\) curvature derivatives.
Dimension matching to an acceleration gives
\begin{equation}
 \mathrm{length}^{n-r-2p-d}=\mathrm{length}^{-1},
 \qquad d+2p=n+1-r.
\end{equation}
Together with \(r\leq2\), this provides a useful selection rule. It
does not determine coefficients or contractions. Counting can equivalently
be implemented by multiplying each curvature tensor or curvature jet
\(\nabla^dR\) once by a formal variable \(t\), irrespective of how
many derivative operators it contains:
\begin{equation}
\begin{aligned}{}
[t^p]\calF_n^a={}&-[t^p]\mathcal G_n^a\\
 &-\sum_{r=2}^{n-1}\sum_{q=1}^{p-1}
       \bigl([t^q]Y_{[r]}\bigr)^a{}_b
       \bigl([t^{p-q}]\calF_{n-r}\bigr)^b .
 \label{eq:connector-curvature-count-recursion}
\end{aligned}
\end{equation}

The variable \(t\) counts curvature factors; the summation index \(r\)
in this recursion labels connector derivatives, rather than the number
of relative-direction factors in the preceding selection rule. Since a Ricci commutator
replaces an ordered derivative pair by curvature products, the division
into sectors labelled by \(p\) depends on the derivative ordering.

At \(n=5,p=3\), the selection rule requires \(d=r=0\). Set
\(Q^a{}_b=R^a{}_{\xi\xi b}\) and retain only undifferentiated factors in
the propagators. The degree-five source is
\begin{equation}
\begin{aligned}{}
\left.\mathcal G_5^a\right|_{R^3}
 &=\tfrac1{120}R^a{}_{k\xi(Q^2k)}
   +\tfrac1{12}R^a{}_{(Qk)\xi(Qk)}
   +\tfrac1{24}R^a{}_{(Q^2k)\xi k},\\
 \left.\calF_5^a\right|_{R^3}
 &=-\left.\mathcal G_5^a\right|_{R^3}
   -\tfrac16Q^a{}_b\left.\calF_3^b\right|_{R^2}
   -\tfrac1{120}(Q^2)^a{}_b\calF_1^b .
 \label{eq:connector-quintic-curvature-derivation}
\end{aligned}
\end{equation}

The inverse-factor contribution combines
\(-1/36+1/120=-7/360\). The complete result in this one sector is
\begin{equation}
\begin{aligned}{}
\left.\calF_5^a\right|_{R^3}={}&
 -\tfrac1{120}R^a{}_{k\xi(Q^2k)}
 -\tfrac1{12}R^a{}_{(Qk)\xi(Qk)}
 -\tfrac1{24}R^a{}_{(Q^2k)\xi k}\\
 &+\tfrac1{36}Q^a{}_bR^b{}_{k\xi(Qk)}
 +\tfrac1{12}Q^a{}_bR^b{}_{(Qk)\xi k}
 -\tfrac7{360}(Q^2)^a{}_bR^b{}_{k\xi k} .
 \label{eq:connector-curvature-cube}
\end{aligned}
\end{equation}

Selecting the \(R^3\) sector does not assume constant curvature:
it extracts one part of the general quintic force. The remaining one-
and two-curvature sectors are generated by the same recurrence, with
the one-curvature result displayed above. The exact constant-curvature
connector check tests the \(R^3\) expression; validation of the full
varying-curvature quintic force would also require the other sectors.

\subsection{Corrections to the published Vines--Kim expansions and independent tests}
\label{arch-evidence}\label{arch-version-comparison}
The exact formulations of \citet{vines} and \citet{kim} agree through
the dictionary \(X=\mathsf P K,\ Y=\mathsf P H\), derived in
section~\ref{app:vines-kim-dictionary}. Some terms in their printed
expansions differ from the relative acceleration obtained from those
identities. Table~\ref{tab:version-comparison} separates these expansion
errors from agreement of the exact equations. All contractions below
use our curvature sign and ordered covariant derivatives.

For comparison with the published sources, Vines's exact equation (45)
and combined propagator expansion (83) are equations (5.6) and (C13)
in the arXiv version. Kim's equations have the same numbers (31)--(33)
in both versions.

\begin{table}\centering\small
\begin{tabular}{p{0.22\textwidth}p{0.35\textwidth}p{0.34\textwidth}}\toprule
Published equation&Comparison in our convention&Evidence and consequence\\\midrule
Vines (45); Kim (31)&Exact equations related by output-index transport&Identical contracted
endpoint and transport coefficients; section~\ref{app:vines-kim-dictionary}\\[5pt]
Kim (32), cubic line&\((\nabla_kQ_2)v/3\) required in \(\mathcal G_3\), versus printed
\((\nabla_vQ_2)k/3\)&Horizontal derivative of \(Y\); nonzero Bianchi difference and independent
relative-acceleration test below\\[5pt]
Kim (33b), mixed slots&\(\partial_cX^a{}_d+\widetilde
Y^a{}_{ec}X^e{}_d=\nabla_dY^a{}_c+\widetilde X^a{}_{ed}Y^e{}_c\)&Differentiated exponential
map, equation~\eqref{eq:connector-torsion-identity}; the transposition is explicit\\[5pt]
Vines (83), quartic terms&Four coefficient differences listed in
table~\ref{tab:vines-quartic}&Cubic extraction unaffected; quartic construction uses the exact
recurrence\\\bottomrule
\end{tabular}
\caption{Exact-equation agreement and differences in the published expansions, in the stated
derivative convention.}
\label{tab:version-comparison}
\end{table}

The mixed cubic discrepancy is given in
\eqref{eq:connector-Kim-delta}. Its two separation factors and one
relative-direction factor give an \(h^3\) error. To test the force
independently, we integrated the coordinate exponential map and required
both endpoint curves to be geodesics, obtaining their relative
acceleration without using the covariant force expansion. The test
used a four-dimensional diagonal metric whose unequal components depend
on all four coordinates, so that the curvature varies. The reference
tangent was null and the neighbouring tangent was unrestricted.
Scaling each of three separation--rate pairs by a common \(h\) gave
fitted remainder powers of \(3.977\)--\(4.013\) after the corrected
cubic force and \(5.001\)--\(5.013\) after the quartic force.
Replacing only the mixed cubic term by Kim's printed slot order gave
powers of \(2.985\)--\(3.108\), recovering the predicted cubic error.
The smallest quartic remainders were \(3.4\)--\(5.2\times10^{-10}\)
in the test units. Tightening the connector integration changed the
exact accelerations by less than \(10^{-16}\), well below these
remainders.
At quartic degree, \citet{kim} already identified the first three
differences in table~\ref{tab:vines-quartic}. The fourth reverses
the sign of the displayed derivative-of-curvature times curvature term.
The exact connector recurrence fixes all four without commuting derivatives.

\begin{table}\centering\small
\begin{tabular}{p{0.54\textwidth}rr}\toprule
Contraction in \(\ddot\xi^a\)&Corrected (used here)&Printed Vines (83)\\\midrule
\(R^a{}_{b\xi k}(\nabla_\xi R)^b{}_{\xi\xi k}\)&\(-4/24\)&\(+1/24\)\\[5pt]
\(R^a{}_{v\xi b}R^b{}_{\xi\xi k}\)&\(-8/24\)&\(+8/24\)\\[5pt]
\(R^a{}_{b\xi k}R^b{}_{\xi\xi v}\)&\(-8/24\)&\(+8/24\)\\[5pt]
\((\nabla_\xi R)^a{}_{b\xi k}R^b{}_{\xi\xi k}\)&\(-6/24\)&\(+6/24\)\\\bottomrule
\end{tabular}
\caption{The corrected column gives the coefficients obtained from the exact recurrence and
used in this paper. The right column gives the incorrect coefficients obtained from Vines's
printed equation (83), corresponding to (C13) in the preprint.
Kim identifies the first three in the correction note accompanying
equation (32), footnote 4 in the published paper. The fourth changes
the sign of the displayed \(6R^\alpha{}_{\delta\xi\gamma;\xi}
R^\delta{}_{\xi\beta\xi}\) contribution. Derivatives have not been commuted.}
\label{tab:vines-quartic}
\end{table}

We also tested the four entries in table~\ref{tab:vines-quartic}
individually in the variable-curvature metric. With the corrected
coefficients the remainder scales as \(h^5\). Restoring any one
printed coefficient gives an \(h^4\) remainder. Across the three
separation--rate pairs the fitted powers for the individual reversions
range from \(3.94\) to \(4.14\).

%% file: appendices/b_curvature_and_optics.tex
\section{Optical interpretation and limiting cases}
\label{app:optical-physics}\label{app:cosmology}\label{app:cosmology-v12}\label{sec:flrw}
The endpoint derivatives describe changes in chosen coordinates. To
interpret them as image distortions, we first convert coordinate
increments into measured angles and physical source separations. We
then distinguish endpoint antisymmetry from optical twist, relate our
quadratic map to Paper~0, and separate the Ricci and Weyl contributions
to propagation. The final two subsections give the leading-gradient potential limit,
its thin-lens normalisation and the rotation produced by ordered
scalar deflections.
These conversions and limits allow the propagated amplitudes to be
compared with the usual optical fields.

\subsection{Physical optical fields from the roulette
polynomial}\label{app:physical-optical-fields}\label{sec:effective-optics}
Image stretching depends on how the endpoint changes between
neighbouring observed directions. Differentiating the roulette
polynomial first gives this relation in the chosen coordinates. We
normalise the Jacobian by the background source angular-diameter
distance \(\overline D_s\), a physical distance distinct from the
comoving or flat source-plane distance \(\chi_S\) used earlier.
Using \(\partial_B\rho=n_B\) and
\(\partial_B\vartheta=t_B/\rho\) gives
\begin{equation}
\begin{aligned}{}
\mathcal A^{\rm coord}_{AB}(\rho,\vartheta)
 &=\frac1{\overline D_s}\sum_{m=1}^{m_{\max}}\frac{\rho^{m-1}}{m!}
 \left[mP_A^{(m)}(\vartheta)n_B
                  +\mathrm d_\vartheta P_A^{(m)}(\vartheta)t_B\right].
 \label{eq:connector-roulette-jacobian-spins}
\end{aligned}
\end{equation}
At \(\rho=0\), the Cartesian polynomial supplies the regular limit.

Away from the coordinate origin, equal coordinate increments need not represent
equal measured angles. The unit-sky metric is
\(\dd\rho^2+\sin^2\rho\,\dd\vartheta^2\). The angular conversion matrix
\(L_o\) converts coordinate increments to components on an orthonormal
angular basis. At the source, let \(y\) collect the endpoint coordinates, and let \(L_s\) convert
an increment \(\dd y\) to a physical transverse separation at the
source. For a flat source plane described by orthonormal length coordinates,
\(L_s=I\); at the centre of the normal angular coordinates, \(L_o=I\) as well.
For general coordinates both conversions are needed, giving
\begin{equation}
\begin{aligned}{}
\mathcal A^{\rm phys}
  &=\overline D_s^{-1}L_s
        \frac{\partial y}{\partial\boldsymbol\zeta}L_o^{-1} .
 \label{eq:connector-physical-chart-jacobian}
\end{aligned}
\end{equation}

The conversion is pointwise. For a spacetime logarithm, \(y\) retains
all four components before projection, and \(L_s\) can include
\(\dd\exp_x\) and the source-surface intersection. Applying these
ruler and angular conversions gives the physical convergence, shear and
rotation from the coordinate derivatives of the endpoint map.

After this conversion, write \(\mathcal A=\mathcal A^{\rm phys}\).
For an equal-affine infinitesimal separation on the Sachs screen,
\(\mathcal A_{AB}=\mathcal J_{AB}/\overline D_s\), where
\(\mathcal J_{AB}\) is the Jacobi map of \eqref{eq:sachs-jacobi}.
For a different source surface, we use the endpoint and ruler conversion
in \eqref{eq:connector-physical-chart-jacobian}.
The trace, symmetric trace-free part and antisymmetric part of
\(\mathcal A\) define
\begin{equation}
\begin{aligned}{}
\mathcal A_{AB}&=\begin{pmatrix}
 1-\kappa_{\rm eff}-\gamma_1&-\gamma_2+\varpi\\
 -\gamma_2-\varpi&1-\kappa_{\rm eff}+\gamma_1
 \end{pmatrix},\\
 \kappa_{\rm eff}&=1-\tfrac12\operatorname{tr}\mathcal A,\qquad
 \gamma_1=\tfrac12(\mathcal A_{22}-\mathcal A_{11}),\\
 \gamma_2&=-\tfrac12(\mathcal A_{12}+\mathcal A_{21}),\qquad
 \varpi=\tfrac12(\mathcal A_{12}-\mathcal A_{21}).
 \label{eq:connector-effective-distortion}
\end{aligned}
\end{equation}

With the chosen screen convention the exact determinant is
\begin{equation}
 \det\mathcal A=(1-\kappa_{\rm eff})^2-\gamma_1^2-\gamma_2^2+\varpi^2.
\end{equation}
The physical angular-diameter distance is
\(D_A=\overline D_s\sqrt{|\det\mathcal A|}\), while unsigned
magnification relative to the same background is \(1/|\det\mathcal A|\).
The linearised identity \(D_A/\overline D_s=1-\kappa_{\rm eff}\)
is a perturbation about the background positive-parity branch; it should
not be applied across a caustic to finite strong distortions.
\subsection{Endpoint antisymmetry and optical twist}\label{app:optical-twist}
In a parallel screen, the Jacobi matrix obeys \(\ddot{\mathcal J}=\mathcal R\mathcal J\), as in
\eqref{eq:sachs-jacobi}.
The real tidal matrix is symmetric. Define its Jacobi Wronskian
\(W_{\mathcal J}=\mathcal J^{\mathsf T}\dot{\mathcal J}-\dot{\mathcal J}^{\mathsf T}\mathcal
J\). Differentiation gives
\begin{equation}
 \dot W_{\mathcal J}=\mathcal J^{\mathsf T}\ddot{\mathcal J}-\ddot{\mathcal J}^{\mathsf
T}\mathcal J=\mathcal J^{\mathsf T}\mathcal R\mathcal J-\mathcal J^{\mathsf T}\mathcal
R^{\mathsf T}\mathcal J=0.
\end{equation}
The vertex data \(\mathcal J(0)=0\), \(\dot{\mathcal J}(0)=-E_o I\) give \(W_{\mathcal J}(0)=0\), so it remains zero. Where \(\mathcal J\) is invertible,
left multiplication by \(\mathcal J^{-\mathsf T}\) and right multiplication by
\(\mathcal J^{-1}\) prove
\begin{equation}
 \dot{\mathcal J} \mathcal J^{-1}=(\dot{\mathcal J} \mathcal J^{-1})^{\mathsf T}.
\end{equation}
The optical deformation is twist free, even though ordered products
can make \(\mathcal J\) itself nonsymmetric. This resolves the apparent conflict
between a barred linear endpoint amplitude and a twist-free null bundle.
Neither equation requires defining an inverse at the caustic itself.

\label{app:two-planes}
The endpoint antisymmetry follows from the commutator in
\eqref{eq:two-plane}; appendix~\ref{app:ordered-deflections} gives its
continuous counterpart. This rotation arises from ordered linear
propagation between deflectors, independently of the local cubic
curvature-product force. Section~\ref{sec:weak-two-lens} demonstrates
it in a globally single-image model.

\subsection{Relation to the quadratic Hessian map}\label{app:paper-zero}
The Hessian-map construction of Paper~0
already retains the along-ray and relative-direction terms in the
quadratic deviation equation. Its expansion is
\(\xi_A=\mathcal J_{AB}\zeta^B+\mathcal H_{ABC}\zeta^B\zeta^C\).
For the same screen map, observed-angle coordinates and screen orientation, our derivative tensors obey
\begin{equation}
 \mathcal M_{AB}=\mathcal J_{AB},\qquad
 \mathcal M_{ABC}=2\mathcal H_{ABC}.
 \label{eq:paper-zero-normalisation}
\end{equation}
The factor of two is the Taylor factorial, not an additional physical
contribution. Paper~0's two spin-one parts and spin-three part correspond
to the unbarred and barred degree-two sectors.

At cubic degree the quadratic force acts on the already distorted beam
through the lower-degree insertions in \eqref{eq:cubic-hierarchy-main},
while the complete cubic force acts on the linear beam. The full spacetime launch data and
source-surface composition are retained as in appendix~\ref{app:maps}.
These operations fix the map being compared. The leading scalar,
transverse-gradient limit below recovers the familiar flexion and the
potential roulettes of Papers~I and II.

\subsection{The four-dimensional Ricci--Weyl
decomposition}\label{app:ricci-weyl-explicit}\label{app:ricci-weyl-discussion}
The linear decomposition is \eqref{eq:ricci-weyl}. Its null projection
separates matter focusing from the trace-free Weyl tidal field without
using field equations. In Einstein gravity the Ricci contraction becomes
\(8\pi G T_{kk}\); its sign requires an energy condition.

To separate the same contributions in every nonlinear force, we need
the Ricci-built tensor in arbitrary slots. Its contraction rule is
\begin{equation}
\begin{aligned}{}
(R^{\rm Ric})^a{}_{pqr}={}&
 \tfrac12\bigl(q^aR_{pr}-r^aR_{pq}
                    -(p\cdot q)R^a{}_r+(p\cdot r)R^a{}_q\bigr)\\
 &-\tfrac R6\bigl(q^a(p\cdot r)-r^a(p\cdot q)\bigr).
 \label{eq:connector-ricci-contraction-rule}
\end{aligned}
\end{equation}

Set \((p,q,r)=(k,\xi,k)\), multiply by minus one, and use \(k^2=0\).
The complete linear force is
\begin{equation}
\begin{aligned}{}
\ddot\xi^a\big|_{\rm linear}={}&-C^a{}_{k\xi k}
 -\tfrac12R_{kk}\xi^a+\tfrac12 k^aR_{k\xi}
 \\
 &+\tfrac12(k\cdot\xi)R^a{}_k
 -\tfrac R6(k\cdot\xi)k^a .\label{eq:connector-linear-ricci-weyl}
\end{aligned}
\end{equation}

Metric compatibility extends this contraction rule to differentiated
forces: \(R_{ab}\) and \(R\) are replaced by their ordered
derivatives. Here a differentiated \(R\) without tensor slots is
the Ricci scalar; with curvature slots it is the corresponding tensor.
Applying the rule to each curvature factor in \eqref{eq:F2} and
\eqref{eq:F3}, while preserving derivative and matrix order, separates
the nonlinear force into Ricci and Weyl contributions.

The local products contain pure Weyl, both mixed orderings, and pure
Ricci terms. The pure Weyl contribution is
\begin{equation}
\begin{aligned}{}
\calF_{3,CC}^a=\left.\calF_3^a\right|_{C^2}={}&
 \tfrac16C^a{}_{k\xi b}C^b{}_{\xi k\xi}
 +\tfrac12C^a{}_{b\xi k}C^b{}_{\xi k\xi}\\
 &-\tfrac16C^a{}_{\xi b\xi}C^b{}_{k\xi k} .
 \label{eq:connector-weyl-square}
\end{aligned}
\end{equation}
All retain the longitudinal contractions of the propagated separation;
the decomposition does not close the force on the screen alone.

\subsection{Electric and magnetic Weyl projection}\label{app:weyl-projection}
We now project the observer-dependent electric and magnetic Weyl fields
onto the null screen tidal matrix. Resolve \(k=E(u-n)\), with
\(u^2=-1,n^2=1,u\cdot n=0\), and define
\begin{equation}
\begin{aligned}{}
E_{ij}&=C_{i0j0},&
 H_{ij}&=\tfrac12\epsilon_i{}^{kl}C_{klj0},&
 \epsilon_{AB}&=\epsilon_{ABi}n^i,\qquad\epsilon_{12}=+1 .
 \label{eq:connector-electric-magnetic-definitions}
\end{aligned}
\end{equation}

The scalar \(E\) is photon energy, while \(E_{ij}\) is the electric
Weyl tensor. The inversion of these definitions gives
\begin{equation}
 C_{i0j0}=E_{ij},\qquad
 C_{ijk0}=\epsilon_{ij}{}^mH_{mk},\qquad
 C_{ijkl}=-\epsilon_{ij}{}^m\epsilon_{kl}{}^nE_{mn}.
\end{equation}
These relations satisfy all Weyl traces. Choose an oriented screen with
\(\epsilon_{AB}=\epsilon_{ABi}n^i\). Then
\begin{align}
 E^{-2}C_{AkBk}
 &=C_{A0B0}-C_{A0Bn}-C_{AnB0}+C_{AnBn},\\
 C_{AnBn}&=E_{AB}+\delta_{AB}E_{nn},\\
 -C_{A0Bn}-C_{AnB0}&=2\epsilon_{(A}{}^CH_{B)C}.
\end{align}
Therefore
\begin{equation}
\begin{aligned}{}
C_{AkBk}&=E^2\left(2E_{AB}+\delta_{AB}E_{nn}
                         +2\epsilon_{(A}{}^C H_{B)C}\right).
 \label{eq:connector-weyl-electric-magnetic}
\end{aligned}
\end{equation}

Holding \(u,n\) and the screen orientation fixed while replacing
\(E(u-n)\) by \(E(u+n)\) reverses the magnetic cross term.
Reversing the entire tangent, \(k\mapsto-k\), leaves the two-\(k\)
optical contraction unchanged. The magnetic sign also changes if the orientation or its defining
dualisation is changed. The formula is trace free because
\(\delta^{AB}E_{AB}=-E_{nn}\) and the trace of the symmetric magnetic cross term
vanishes.

A barred endpoint coefficient alone does not identify magnetic Weyl
curvature: the ordered electric-shear contribution is derived in
appendix~\ref{app:ordered-deflections}.

\subsection{The leading-gradient limit and potential roulettes}\label{app:born-roulettes}
To recover the potential description, we restrict to scalar
perturbations at first metric order, small angular offsets and the
leading transverse gradients. We use conformal Newtonian coordinates
on a spatially flat FLRW background, with scalar metric potentials
\(\Phi,\Psi\). Here \(\chi\) is comoving radial distance, increasing
from zero at the observer to the fixed source coordinate \(\chi_s\),
and \(\bar\nabla_A\) differentiates with respect to the comoving
Cartesian transverse coordinates. Primes denote \(\dd/\dd\chi\).
The reduced ray equation is
\begin{equation}
 x_A''(\chi)=-\bar\nabla_A(\Phi+\Psi),\quad
 x_A(0)=0,\quad x_A'(0)=\theta_A.
\end{equation}
Two integrations give the coordinate source angle
\begin{equation}
 \beta_A=\theta_A-\frac1{\chi_s}\int_0^{\chi_s}
               (\chi_s-\chi)\bar\nabla_A(\Phi+\Psi)(\chi,x(\chi))\,\dd\chi.
\end{equation}
Define \(\delta\mathcal A^{\rm Born}_{AB_1\cdots B_m}\) as the angular
derivatives of the lensing correction \(\beta_A-\theta_A\), excluding
the identity at \(m=1\). This is a comoving coordinate source-angle
map at fixed \(\chi_s\); a physical map at fixed observed redshift
also requires the observer and source-endpoint conversions. In the Born
approximation we evaluate the field along the unperturbed background ray,
\(x_A=\chi\theta_A\).
Each angular derivative gives a factor \(\chi\bar\nabla_B\), so
\begin{equation}
\begin{aligned}{}
\delta\mathcal A^{\rm Born}_{AB_1\cdots B_m}
 &=-\frac1{\chi_s}\int_0^{\chi_s}(\chi_s-\chi)\chi^m\\
 &\quad\times\bar\nabla_A\bar\nabla_{B_1}\cdots\bar\nabla_{B_m}
           (\Phi+\Psi)\,\dd\chi,\qquad m\geq1 .
 \label{eq:connector-flrw-born-roulettes}
\end{aligned}
\end{equation}

At \(m=2\) this gives
\(\mathcal H_{ABC}=(\chi_s/2)\delta\mathcal A^{\rm Born}_{ABC}\)
for the same comoving screen convention as Paper~0's leading flexion
limit. At every degree the commuting derivatives make the tensor fully
symmetric, so the barred amplitudes vanish. Higher angular derivatives
resolve finer variation across the image while the metric calculation
remains first order. Physical ruler and endpoint conversions, as in
appendix~\ref{app:physical-optical-fields}, are separate contributions
to an observed map.

Each potential roulette is a line-of-sight integral of the transverse
derivatives that generate its spin. Under the scalar, Born and
leading-transverse-gradient assumptions of
\eqref{eq:connector-flrw-born-roulettes}, spin projection gives
\begin{equation}
\begin{aligned}{}
\alpha_s^m[\delta\mathcal A^{\rm Born}]+i\beta_s^m[\delta\mathcal A^{\rm Born}]
 &=-\frac{\binom{m+1}{p}}{2^{m+\delta_{s0}}\chi_s}
 \int_0^{\chi_s}(\chi_s-\chi)\chi^m\\
 &\quad\times(\bar\nabla_\perp^2)^p
       (\bar\nabla_1+i\bar\nabla_2)^s(\Phi+\Psi)\,\dd\chi,\\
 p&=\frac{m+1-s}{2}.
 \label{eq:practical-born-amplitudes}
\end{aligned}
\end{equation}
The derivatives act on comoving Cartesian transverse coordinates along
the background ray; \(p\) is a nonnegative integer for each allowed spin.
The powers of \(\chi\) convert spatial derivatives into angular
derivatives. No assumption \(\Phi=\Psi\) is needed.

At \(m=1\), the result is
\(\alpha_0^1[\delta\mathcal A^{\rm Born}]=-\kappa_{\rm lens}\)
and
\(\alpha_2^1[\delta\mathcal A^{\rm Born}]+i\beta_2^1[\delta\mathcal A^{\rm Born}]=-(\gamma_1+i\gamma_2)\).
The full coordinate source-angle map also includes the identity,
adding \(\delta_{m1}\delta_{s0}\) to the unbarred cosine amplitude.
Equation~\eqref{eq:practical-curvature-map} gives the propagation
prescription when the scalar, Born and leading-gradient restrictions
are relaxed.

\subsubsection{Thin-lens maps and amplitude normalisation}\label{app:amplitude-dictionary}
For a specified single-plane potential, the normalisation follows
directly from the reduced lens equation. The lensing potential \(\psi\)
of Papers~I and II is dimensionless;
\(\chi_L^2\psi\) has units of length squared. With \(\nabla_A\)
denoting differentiation with respect to lens-plane length coordinates,
\begin{equation}
 U_A(R)=R_A-\chi_L^2\nabla_A\psi(R),\qquad
 u_A(x,R)=U_A(R+x)-U_A(R).
\end{equation}
Direct Taylor expansion gives
\begin{equation}
 u_A=x_A-\chi_L^2\sum_{m\geq1}\frac{1}{m!}
              \nabla_A\nabla_{B_1}\cdots\nabla_{B_m}\psi\,
              x^{B_1}\cdots x^{B_m}.\label{eq:thin}
\end{equation}
With \(x^A=\chi_L\zeta^A\) and \(\xi=\chi_Su/\chi_L\), define
the lensing correction at each degree by
\begin{align}
 \delta\mathcal A_{AB_1\ldots B_m}
     &=-\chi_L^{m+1}\nabla_A\nabla_{B_1}\cdots\nabla_{B_m}\psi,\\
 \mathcal A_{AB}&=\delta_{AB}+\delta\mathcal A_{AB},\\
 \mathcal A_{AB_1\ldots B_m}&=\delta\mathcal A_{AB_1\ldots B_m}
       \quad(m\geq2),\\
 \mathcal M_{AB_1\ldots B_m}&=\chi_S\mathcal A_{AB_1\ldots B_m}
       \quad(m\geq1).
\end{align}
Thus \(\mathcal A\) is the full dimensionless amplification tensor,
consistent with Paper~II's equations~(104) and (106), while
\(\delta\mathcal A\) excludes the identity at linear degree.
\(\mathcal M\) has units of length. All three tensors use the same
\(1/m!\) Taylor convention. At amplitude level, the conversion is
\(\alpha_s^m[\delta\mathcal A]=\alpha_s^m/\chi_S-
\delta_{m1}\delta_{s0}\) and
\(\beta_s^m[\delta\mathcal A]=\beta_s^m/\chi_S\).

For a potential, write
\(n\cdot\nabla=(e^{-i\vartheta}\nabla_+
                 +e^{i\vartheta}\nabla_-)/2\), with
\(\nabla_\pm=\nabla_X\pm i\nabla_Y\).
Expanding its \((m+1)\)-st power, pairing complex-conjugate frequencies,
and using \(\nabla_+\nabla_-=\nabla_\perp^2\) gives
\begin{equation}
\begin{aligned}{}
\alpha_s^m[\delta\mathcal A]+i\beta_s^m[\delta\mathcal A]
 &=-\frac{\chi_L^{m+1}}{2^{m+\delta_{s0}}}
   \binom{m+1}{(m+1-s)/2}\\
 &\quad\times(\nabla_X^2+\nabla_Y^2)^{(m+1-s)/2}
   (\nabla_X+i\nabla_Y)^s\psi .
 \label{eq:connector-complex-potential}
\end{aligned}
\end{equation}

The factor for \(s=0\) differs by two because its frequency is unpaired.
This agrees with the recursion normalisation of \citet{normann}.
To compare with \eqref{eq:practical-born-amplitudes}, we use the
projected potential of section~\ref{sec:weak-flexion-limit}.
Angular differentiation gives \(\partial/\partial\theta^A=\chi_L\nabla_A\)
in the thin-lens coordinates and
\(\partial/\partial\theta^A=\chi\bar\nabla_A\) inside the
line-of-sight integral. Since \(2p+s=m+1\), differentiating that
potential gives the weight \((\chi_s-\chi)\chi^m/\chi_s\) in
\eqref{eq:practical-born-amplitudes}. The two expressions then give
the same amplitudes in this limit.

\subsection{Ordered deflections and image rotation}\label{app:ordered-deflections}
Successive deflections give an instructive departure from the potential
map. A symmetric tidal matrix changes the beam shape; a later,
misaligned tidal field acts on that changed shape. Their ordered product
can have an antisymmetric part, as in the two-plane example.
Let \(T_{AB}=\bar\nabla_A\bar\nabla_B(\Phi+\Psi)\) on the background path.
Here \(D\) is the conformal propagation matrix, with background
\(D=\chi I\); division by \(\chi_s\) gives the reduced coordinate
source-angle Jacobian \(\mathcal A\) at fixed \(\chi_s\). Primes
below denote \(\dd/\dd\chi\), and the propagation weight is
\(W=\chi(\chi_s-\chi)/\chi_s\).
The first iteration of \(D''=-TD\) gives
\(D_1(\chi)=-\int_0^\chi(\chi-\chi')\chi'T(\chi')\dd\chi'\).
Inserting this result into \(D''=-TD\) and integrating twice gives
\begin{equation}
\begin{aligned}{}
\Delta\mathcal A^{\rm LL}_{AB}
 &=\frac1{\chi_s}\int_0^{\chi_s}\dd\chi\int_0^\chi\dd\chi'\,
   (\chi_s-\chi)(\chi-\chi')\chi'
                \bigl[T(\chi)T(\chi')\bigr]_{AB},\\
 \Delta\mathcal A^{\rm LL}_{[AB]}
 &=\frac1{2\chi_s}\int_0^{\chi_s}\dd\chi\int_0^\chi\dd\chi'\,
   (\chi_s-\chi)(\chi-\chi')\chi'
                \bigl[T(\chi),T(\chi')\bigr]_{AB} .
 \label{eq:connector-flrw-lens-lens}
\end{aligned}
\end{equation}

The later tidal matrix multiplies on the left. Its commutator with the
earlier matrix vanishes for aligned shear axes; their isotropic parts
commute for any orientation. Misaligned electric Weyl tidal fields
therefore suffice to generate rotation
\citep{penmao,pratten2016}. This is a barred linear roulette.
Spatial variation of the antisymmetric map contributes to higher barred
degrees, as in \eqref{eq:curl-barred}.

There is also a separate correction from evaluating the tidal field on
the displaced path. The first path displacement is
\begin{equation}
 \delta x_A^{(1)}(\chi)=-\int_0^\chi(\chi-\chi')
                     \bar\nabla_A(\Phi+\Psi)(\chi')\,\dd\chi'.
\end{equation}
Taylor-expanding the tidal field about the background path adds
\begin{equation}
 \Delta\mathcal A^{\rm path}_{AB}=-\int_0^{\chi_s}W\,
                    \delta x^{(1)C}\bar\nabla_C T_{AB}\,\dd\chi.
\end{equation}
It is symmetric because \(T_{AB}=T_{BA}\). Thus, within this reduced
equation and to its second metric order,
\begin{equation}
 \mathcal A=I-\int WT+\Delta\mathcal A^{\rm LL}+\Delta\mathcal A^{\rm path}+O(\epsilon^3).
\end{equation}
These are the second-metric-order terms generated by iterating the
reduced scalar ray equation. A complete relativistic prediction at that
order also requires the second-order metric and the redshift, observer
and source contributions.

%% file: appendices/c_calculating_maps.tex
\section{Calculating roulette maps}
\label{app:maps}
The initial direction, propagation and source intersection define one
map of the observer angles. To calculate its coefficients, we first
supply the launch data, then insert the lower-degree maps into the
inhomogeneous Jacobi equations and propagate each new degree. The
endpoint chain rules convert this separation to the specified source
surface. We retain all spacetime components until that conversion;
the final projection gives the roulette amplitudes. The curvature-source
matrix is interpreted for a round linear input beam. Appendix~\ref{app:amplitude-dictionary}
gives the potential-lens normalisation.

\subsection{Exact sky coordinates and launch coefficients}\label{app:sky-data}
The angular hierarchy needs initial data at every retained degree.
These follow by differentiating the exact sky direction
\eqref{eq:sky}, so they are fixed by the observer's angular coordinates.

The tangent is exactly null and has the same positive observed energy
\(-u_o\cdot k_o=E_o\) for every direction. Since all rays start at one
event, every initial position map vanishes. Expand sine and cosine:
\begin{equation}
 \cos\rho=\sum_{r\geq0}\frac{(-1)^r\rho^{2r}}{(2r)!},
 \qquad\frac{\sin\rho}{\rho}
       =\sum_{r\geq0}\frac{(-1)^r\rho^{2r}}{(2r+1)!}.
\end{equation}
Equating homogeneous angular polynomials gives
\begin{align}
 \dot{\mathcal M}^{a}{}_{B_1\ldots B_{2r}}\big|_o
 &=(-1)^{r+1}E_on_o^a\delta_{(B_1B_2}\cdots\delta_{B_{2r-1}B_{2r})},\quad r\geq1,\label{eq:initial-even}\\
 \dot{\mathcal M}^{a}{}_{B_1\ldots B_{2r+1}}\big|_o
 &=(-1)^{r+1}E_oe^a{}_{(B_1}\delta_{B_2B_3}\cdots
                       \delta_{B_{2r}B_{2r+1})},\quad r\geq0.\label{eq:initial-all}
\end{align}
Unit-weight symmetrisation is essential. Contracting all indices with
\(\zeta^A\) recovers \(\rho^{2r}\) or
\(\zeta_A\rho^{2r}\), with no additional combinatorial factor.
The even derivatives initially point along \(n_o\), while the odd
derivatives have screen components. These terms arise from the unit
sphere of observed directions before any curvature is encountered.

\subsection{The general angular recurrence relations}
\label{app:hierarchy-details}\label{app:polarisation}
At each angular degree, the curvature forces act on the separation and
relative direction already calculated at lower degrees. Their contribution
is found by substituting those solutions into the known forces and taking
the required angular derivative. All expansions below are at fixed
affine parameter.

For a fixed angular offset \(\boldsymbol\zeta\), write
\begin{equation}
 \xi^a(\lambda,h\boldsymbol\zeta)=\sum_{j\geq1}\frac{h^j}{j!}\xi_{(j)}^a,
 \qquad
 \dot\xi^a(\lambda,h\boldsymbol\zeta)=\sum_{j\geq1}\frac{h^j}{j!}\dot\xi_{(j)}^a .
 \label{eq:connector-angular-series}
\end{equation}
Substitution into the
nonlinear deviation equation gives the recurrence in one expression,
\begin{equation}
\begin{aligned}
 \ddot\xi_{(m)}^a+R^a{}_{k b k}\xi_{(m)}^b
 &=S_m^a(\lambda;\boldsymbol\zeta)\\
 &=m!\,[h^m]\sum_{n=2}^{m}\mathcal F_n^a
   \left(\sum_{j=1}^{m-1}\frac{h^j}{j!}\xi_{(j)},
         \sum_{j=1}^{m-1}\frac{h^j}{j!}\dot\xi_{(j)}\right).
 \label{eq:all-orders}
\end{aligned}
\end{equation}
Here \([h^m]\) selects the coefficient of \(h^m\), and the
factor \(m!\) converts it into an angular derivative. At \(m=1\)
the source vanishes. The source starts at force degree two, so a factor
of angular degree \(m\) cannot enter the degree-\(m\) source.

For example, the relative-direction term in \(\mathcal F_2\) contains
\(-2R^a{}_{bck}\dot\xi^b\xi^c\). Its cubic contribution is
\begin{equation}
 3!\,[h^3]\left(-2R^a{}_{bck}\dot\xi^b\xi^c\right)
 =-6R^a{}_{bck}
    \left(\dot\xi_{(1)}^b\xi_{(2)}^c
                      +\dot\xi_{(2)}^b\xi_{(1)}^c\right).
 \label{eq:connector-mixed-degree-example}
\end{equation}
The first product multiplies the linear separation rate by the
quadratic separation, and the second multiplies the quadratic rate by
the linear separation. The same substitution applies to every force slot.

Ordinary derivatives of the force polynomials keep these insertions
explicit without introducing new multilinear functions. Through cubic
degree,
\begin{equation}
\begin{aligned}
 S_2^a&=2\mathcal F_2^a(\xi_{(1)},\dot\xi_{(1)}),\\
 S_3^a&=3\left.\frac{\dd}{\dd t}\mathcal F_2^a
       (\xi_{(1)}+t\xi_{(2)},\dot\xi_{(1)}+t\dot\xi_{(2)})\right|_{t=0}
       +6\mathcal F_3^a(\xi_{(1)},\dot\xi_{(1)}) .
 \label{eq:connector-low-degree-sources}
\end{aligned}
\end{equation}
At quartic degree the quadratic force receives degrees one and three,
or two and two, while the cubic force receives degrees one, one and two.
Thus
\begin{equation}
\begin{aligned}
 \ddot\xi_{(4)}^a+R^a{}_{k b k}\xi_{(4)}^b=S_4^a
 ={}&4\left.\frac{\dd}{\dd t}\mathcal F_2^a
       (\xi_{(1)}+t\xi_{(3)},\dot\xi_{(1)}+t\dot\xi_{(3)})\right|_{t=0}\\
 &+6\mathcal F_2^a(\xi_{(2)},\dot\xi_{(2)})\\
 &+12\left.\frac{\dd}{\dd t}\mathcal F_3^a
       (\xi_{(1)}+t\xi_{(2)},\dot\xi_{(1)}+t\dot\xi_{(2)})\right|_{t=0}\\
 &+24\mathcal F_4^a(\xi_{(1)},\dot\xi_{(1)}) .
 \label{eq:connector-quartic-recursion}
\end{aligned}
\end{equation}
The parameter \(t\) only selects the displayed insertion into a
force polynomial. The factors are respectively \(4!/(1!3!)\),
\(4!/(2!2!)\), \(4!/(1!1!2!)\) and \(4!\), with the
derivatives counting the possible slots.

To calculate all directions together, retain the angular indices.
For each degree \(m\), use the lower-degree Taylor polynomial
in the following derivative,
\begin{equation}
\begin{aligned}
 S^a{}_{B_1\cdots B_m}(\lambda)
 &=\left.\partial_{B_1}\cdots\partial_{B_m}
        \sum_{n=2}^{m}\mathcal F_n^a(\xi,\dot\xi)\right|_{\boldsymbol\zeta=0},\\
 \xi^a(\lambda,\boldsymbol\zeta)
 &=\sum_{j=1}^{m-1}\frac1{j!}
       \mathcal M^a{}_{C_1\cdots C_j}(\lambda)
                       \zeta^{C_1}\cdots\zeta^{C_j}+O(\rho^m),
 \qquad \partial_B=\frac{\partial}{\partial\zeta^B} .
 \label{eq:amplitude-force-tensor}
\end{aligned}
\end{equation}
The rate uses the same polynomial with \(\dot{\mathcal M}\).
All lower-degree maps are evaluated where the curvature acts, at
\(\lambda\). The resulting tensors are symmetric in their angular
input indices and obey
\begin{equation}
 \ddot{\mathcal M}^a{}_{B_1\cdots B_m}
  +R^a{}_{k c k}\mathcal M^c{}_{B_1\cdots B_m}
 =S^a{}_{B_1\cdots B_m},\qquad S^a{}_B=0 .
 \label{eq:hierarchy}
\end{equation}
Equation~\eqref{eq:connector-hierarchy-green} propagates these tensor
sources, and section~\ref{sec:amplitude-integrals} projects their
screen maps into roulette amplitudes. Curvature products do not add
angular slots. A complete degree-five map still requires the full
\(\mathcal F_5\), not only its \(R^3\) sector.

\subsection{Propagation along the ray}\label{app:ray-propagation}
The angular map tensors obey the inhomogeneous Jacobi equation with
source \(S^a{}_{B_1\cdots B_m}\) defined in
\eqref{eq:amplitude-force-tensor}. To construct its Green function,
let \(\mathcal J\) and \(\mathcal K\) solve
\(\ddot{\mathsf X}^a{}_{b_o}+R^a{}_{k c k}\mathsf X^c{}_{b_o}=0\)
for \(\mathsf X=\mathcal J,\mathcal K\)
along the reference ray, with observer data
\[
 \mathcal J(0)=0,\qquad \dot{\mathcal J}(0)=-E_o I,
 \qquad \mathcal K(0)=I,\qquad \dot{\mathcal K}(0)=0 .
\]
These are four-dimensional propagators from the observer tangent space.
The Jacobi normalisation agrees with \eqref{eq:sachs-jacobi}; at
\(E_o=1\), both observer data agree with Paper~II \citep{clarksonII}.
The screen map is the transverse projection of \(\mathcal J^a{}_{b_o}\)
given in section~\ref{sec:amplitude-integrals}. Write \(\dagger\) for the metric adjoint between
the tangent spaces, using the Lorentz metric in a parallel frame.
On a Sachs screen it reduces to the ordinary transpose.

The tidal operator is self-adjoint. Its conserved Jacobi Wronskians
therefore give
\[
 \mathcal K\mathcal J^\dagger-\mathcal J\mathcal K^\dagger=0,
 \qquad
 \dot{\mathcal K}\mathcal J^\dagger
       -\dot{\mathcal J}\mathcal K^\dagger=E_o I .
\]
Consequently
\[
 \mathcal G(\lambda,\lambda')
 =\frac1{E_o}\left[
       \mathcal K(\lambda)\mathcal J(\lambda')^\dagger
       -\mathcal J(\lambda)\mathcal K(\lambda')^\dagger\right]
\]
solves the homogeneous equation in its first argument, with
\(\mathcal G(\lambda',\lambda')=0\) and
\(\left.D_\lambda\mathcal G(\lambda,\lambda')\right|_{\lambda=\lambda'}=I\).
Where \(\mathcal J(\lambda')\) is invertible, the first Wronskian
identity also gives
\[
 \mathcal G(\lambda,\lambda')
 =\frac1{E_o}\left[\mathcal K(\lambda)
  -\mathcal J(\lambda)\mathcal J(\lambda')^{-1}
                         \mathcal K(\lambda')\right]
                         \mathcal J(\lambda')^\dagger .
\]
At \(E_o=1\), its transverse restriction is the form in Paper~II,
equations~(9) and~(15). The inverse-free expression remains regular
at conjugate points. In flat spacetime \(\mathcal J=-E_o\lambda I\)
and \(\mathcal K=I\), giving
\(\mathcal G(\lambda,\lambda')=(\lambda-\lambda')I\).
In general \(\mathcal G(\lambda,0)=-\mathcal J(\lambda)/E_o\).
Variation of constants now gives the full map tensor and its final
screen projection,
\begin{equation}
\begin{aligned}
 \mathcal M^a{}_{B_1\cdots B_m}(\lambda)
 ={}&\mathcal G^a{}_{b_o}(\lambda,0)
                 \dot{\mathcal M}^{b_o}{}_{B_1\cdots B_m}(0)\\
 &+\int_0^\lambda\mathcal G^a{}_b(\lambda,\lambda')
                 S^b{}_{B_1\cdots B_m}(\lambda')\,\dd\lambda',\\
 \mathcal M_{AB_1\cdots B_m}(\lambda)
 ={}&e_{Aa}(\lambda)\mathcal M^a{}_{B_1\cdots B_m}(\lambda),
 \qquad
 S_m^a(\lambda;\boldsymbol\zeta)=S^a{}_{B_1\cdots B_m}(\lambda)\zeta^{B_1}\cdots\zeta^{B_m} .
 \label{eq:connector-hierarchy-green}
\end{aligned}
\end{equation}

The source \(S_m\) uses the lower-degree beam at \(\lambda'\), where
the curvature acts. The Green function then transports the resulting
correction to the endpoint. Replacing those intermediate maps by their
endpoint values would lose this ordering along the ray.
A truncated cubic ODE integrated without
angular truncation contains some fourth- and higher-degree feedback, but
misses the direct \(\mathcal F_4\) source and its successors.

\subsection{Endpoint composition through cubic degree}
\label{app:endpoint-composition}
A neighbouring ray generally reaches the emitting surface at a different
affine parameter. The change in arrival parameter combines with the ray
derivatives at every angular degree, as the flat-plane example shows.

We first convert the propagated logarithmic separation to the coordinate
ray family through \(x'=\exp_x\xi\). Beyond linear degree, the angular
derivatives of these coordinates differ from those of \(\xi\).
Let \(F(x)=0\) define the source surface, assume \(F_{,a}k^a\ne0\)
at the reference intersection, and set
\(X^a(\boldsymbol\zeta)=x^a(\lambda_s(\boldsymbol\zeta),\boldsymbol\zeta)\).
In that coordinate system, denote fixed-\(\lambda\) derivatives by
\(x_B,x_{BC},\ldots\), and evaluate all following quantities at the
central endpoint. The first two chain rules are
\begin{align}
 X_B^a&=x_B^a+k^a\lambda_B,\\
 X_{BC}^a&=x_{BC}^a+k_B^a\lambda_C+k_C^a\lambda_B
       +\partial_\lambda k^a\lambda_B\lambda_C+k^a\lambda_{BC}.
\end{align}
Here \(\partial_\lambda\) differentiates coordinate components at fixed
angular offset, so \(\partial_\lambda k^a=-\Gamma^a{}_{bc}k^bk^c\).
The covariant geodesic acceleration remains zero. Differentiating
\(F(X)=0\) gives
\begin{align}
 \lambda_B&=-\frac{F_a x_B^a}{F_a k^a},\\
 \lambda_{BC}&=-\frac{F_a[x_{BC}^a+k_B^a\lambda_C+k_C^a\lambda_B
                          +\partial_\lambda k^a\lambda_B\lambda_C]
                       +F_{ab}X_B^aX_C^b}{F_a k^a}.
\end{align}
The denominator is the transversality condition.

At cubic degree, let \(\widehat X^a{}_{BCD}\) denote the
part already known from lower endpoint derivatives:
\begin{align}
 \widehat X^a{}_{BCD}={}&x^a_{BCD}+3k^a_{(BC}\lambda_{D)}
          +3\partial_\lambda k^a_{(B}\lambda_C\lambda_{D)}
          +\partial_\lambda^2 k^a\lambda_B\lambda_C\lambda_D\\
 &+3k^a_{(B}\lambda_{CD)}+3\partial_\lambda k^a\lambda_{(B}\lambda_{CD)},\\
 X^a_{BCD}&=\widehat X^a{}_{BCD}+k^a\lambda_{BCD},\\
 \lambda_{BCD}&=-\frac{F_a\widehat X^a{}_{BCD}
               +3F_{ab}X^a_{(BC}X^b_{D)}
               +F_{abc}X_B^aX_C^bX_D^c}{F_ak^a}.
\end{align}
The symmetrisations have unit weight. Further angular differentiation
of the ray solution and endpoint condition gives the higher-degree
chain rules.

Finally, the surface endpoint must be expressed in the chosen output
coordinates. For a scalar coordinate \(y(x)\),
\begin{align}
 y_B&=y_aX_B^a,\\
 y_{BC}&=y_aX_{BC}^a+y_{ab}X_B^aX_C^b,\\
 y_{BCD}&=y_aX_{BCD}^a
      +y_{ab}(X_{BC}^aX_D^b+X_{BD}^aX_C^b+X_{CD}^aX_B^b)
      +y_{abc}X_B^aX_C^bX_D^c.
\end{align}
These are ordinary chain rules in fully specified coordinates. They explain
why source-plane derivatives and equal-affine logarithmic-screen
derivatives cannot be identified. In a covariant implementation the
corresponding exponential/logarithm differentials supply the same
geometric conversion.

\subsection{Curvature sources and their angular patterns}
\label{arch-matrix-proof}
Figure~\ref{fig:roulette-source-flowers} arranges the angular patterns
of selected curvature sources by force degree and source family.
To obtain these assignments, we first separate a screen polynomial into
its symmetric and barred parts, then evaluate the radial projection of
each displayed source on a round linear beam. The enlarged flower curves
show individual basis polynomials with the stated display gains; their
crossings need not be singularities of a physical lens map.

Let a degree-\(m\) screen polynomial be \(P=P_n n+P_t t\), and set
\(H_m=P_n\), \(K_m=P_t\). Its completely symmetric tensor part is
fixed by its radial contraction. Equivalently, the homogeneous potential
\(\rho^{m+1}H_m(\vartheta)/(m+1)\) has gradient
\(\rho^m[H_m n+H'_m t/(m+1)]\). Subtracting this vector from \(\rho^mP\)
leaves a purely tangential polynomial. With \(\epsilon_{12}=+1\),
\begin{equation}
\begin{aligned}{}
P=H_m n+\left(\frac{H'_m}{m+1}-\bar\sigma_m\right)t,
 \qquad \bar\sigma_m=\frac{H'_m}{m+1}-K_m.
\end{aligned}
\end{equation}
The highest Fourier harmonics of \(H'_m/(m+1)\) and \(K_m\) cancel,
leaving barred spins at most \(m-1\), as in \eqref{eq:spins}.
A homogeneous potential has \(\bar\sigma_m=0\); a radial-zero polynomial
has only barred content. This decomposition is local to the specified coordinates.

On the round linear input \(\xi_{(1)}\propto e_A n^A\),
\(v_{(1)}\parallel\xi_{(1)}\), radial contraction of the displayed
along-ray family gives a factor \(R_{\xi\xi k\xi}\). The corresponding
relative-direction family gives \(R_{\xi v\xi k}\). The first curvature
pair is antisymmetric, so each vanishes, including after covariant
differentiation. No analogous identity forces the radial transverse-gradient
projection to vanish. This proves the two local source restrictions shown in
figure~\ref{fig:roulette-source-flowers}. It does not classify every allowed curvature-product
contraction solely from its grade.

Homogeneity gives \(P(\vartheta+\pi)=(-1)^mP(\vartheta)\), whereas
\(n,t\) both reverse. Thus \(H_m,K_m\) and \(\bar\sigma_m\) have
Fourier parity \((-1)^{m+1}\): \(m+s\) is odd in both sectors.
This argument is independent of \(d+2p=n+1-r\). The latter counts
curvature structures; the former counts angular harmonics.

The source assignments are made at direct degree \(m=n\). The actual
propagated amplitude also receives anisotropic transport, lower-degree
insertions, initial data and endpoint composition. The figure therefore organises the local
inputs to the hierarchy;
its source assignments do not give a universal selection rule for the
propagated image.

\subsection{Evaluating the amplitudes in practice}\label{app:practical-amplitudes}
Each angular pattern contains the propagated launch data and the
distortions accumulated along the reference ray. Combining the
Green-function solution with the screen projection, evaluated on the
unit angular direction \(n(\vartheta)\), gives the pattern with the
separate factor \(\rho^m\) removed. At an equal-affine logarithmic
endpoint,
\begin{equation}
\begin{aligned}{}
P_A^{(m)}(\vartheta)={}&e_{Aa}(\lambda_s)
 \Bigl[\mathcal G^a{}_b(\lambda_s,0)\dot\xi_{(m)}^b(0;\vartheta)\\
 &\quad+\int_0^{\lambda_s}\mathcal G^a{}_b(\lambda_s,\lambda)
                    S_m^b(\lambda;\vartheta)\,\dd\lambda\Bigr].
 \label{eq:practical-curvature-map}
\end{aligned}
\end{equation}
Here \(\mathcal G\) is the spacetime Green function in
\eqref{eq:amplitude-jacobi-green}, and
\(S_m^a(\lambda;\vartheta)=S^a{}_{B_1\cdots B_m}(\lambda)n^{B_1}\cdots n^{B_m}\)
contracts the tensor source of \eqref{eq:amplitude-force-tensor} with the
unit angular direction. Here
\(\xi_{(m)}^a(\lambda;\vartheta)
=\mathcal M^a{}_{B_1\cdots B_m}(\lambda)n^{B_1}\cdots n^{B_m}\);
the coefficients for a general offset \(\zeta^A=\rho n^A\) in
\eqref{eq:connector-angular-series} have an additional factor \(\rho^m\).
The initial derivatives are
given in appendix~\ref{app:sky-data}. The curvature and its ordered derivatives are evaluated along that ray.
The longitudinal components remain in the propagated maps. For a
source-surface intersection, the composition in
section~\ref{app:endpoint-composition} gives the endpoint pattern in place
of the equal-affine projection above.

Project with \eqref{eq:connector-Fourier-normalisation}, then add the
central endpoint and evaluate the source. For a supplied homogeneous
polynomial, equally spaced sampling with \(N\geq2m+3\) evaluates
these projections without aliasing.
The symmetric/barred split \eqref{eq:split} gives
\(H_m=n^AP_A^{(m)}\) and \(K_m=t^AP_A^{(m)}\), with
\(\mathrm d_\vartheta H_m=(m+1)K_m\) in the symmetric sector.
The unbarred frequencies end at \(m+1\); the independent barred
frequencies end at \(m-1\). With the sign \(\epsilon n=-t\),
the projections of \eqref{eq:fourier} are

\begin{equation}
\begin{aligned}{}
\alpha_s^m&=\frac{1}{\pi(1+\delta_{s0})}
               \int_0^{2\pi}H_m\cos(s\vartheta)\,\dd\vartheta,\\
 \beta_s^m&=\frac1\pi\int_0^{2\pi}H_m\sin(s\vartheta)\,\dd\vartheta
                            \qquad(s>0),\\
 \bar\alpha_s^m&=\frac{1}{\pi(1+\delta_{s0})}
       \int_0^{2\pi}\left(\frac{\mathrm d_\vartheta H_m}{m+1}-K_m\right)\cos(s\vartheta)\,\dd\vartheta,
 \label{eq:connector-Fourier-normalisation}
\end{aligned}
\end{equation}

The barred sine projection has the same normalisation as its unbarred
counterpart. Spin-zero sine coefficients vanish. These projections give the component counts,
spin selection and barred
sign used in Paper~II \citep{clarksonII}.
Here unqualified amplitudes describe the dimensional map.
Appendix~\ref{app:amplitude-dictionary} gives the potential-lens
normalisation and the derivative formula
\eqref{eq:connector-complex-potential}.

For the SIS example of section~\ref{sec:review}, the cubic amplitudes of
the full dimensional map in \eqref{eq:SIS-worked} are
\begin{equation}
\begin{aligned}{}
\frac{\alpha_0^3}{\chi_S}&=-\frac{3b\chi_L^2}{8R^3},&
 \frac{\alpha_2^3}{\chi_S}&=-\frac{3b\chi_L^2}{2R^3},&
 \frac{\alpha_4^3}{\chi_S}&=\frac{15b\chi_L^2}{8R^3},&
 \beta_s^3&=0 .
 \label{eq:SIS-cubic}
\end{aligned}
\end{equation}
Their radial combination is
\(n^AP_A^{(3)}=(b\chi_S\chi_L^2/R^3)
[-3/8-(3/2)\cos2\vartheta+(15/8)\cos4\vartheta]\).
The tangential component follows from \eqref{eq:potential-spins}; all
barred amplitudes vanish. Direct differentiation and Fourier projection
independently check these coefficients.

%% file: appendices/d_lens_models.tex
\section{Models, sources and error measures}\label{app:models}
The comparisons fix the lens, observer, source surface and source
profile before varying the retained propagation terms or angular degree.
Here we give the inputs and error windows used in the figures,
derive the thin-lens position bounds in
section~\ref{app:thin-lens-bounds}, and define the image and beam errors.

\subsection{The common colour source}\label{app:common-source}
The colour images use a smooth four-quadrant disc on a white background,
with centres and radii listed in table~\ref{tab:colour-source-inputs}.
\begin{table}[!htbp]\centering\small
\begin{tabular}{lll}\toprule
Model&Source centre&Radius\\\midrule
Cored elliptical thin lens&\((0.28,0.08)R_{\rm E}\)&\(0.12R_{\rm E}\)\\
With compact component&\((0.22,0.08)R_{\rm E}\)&\(0.12R_{\rm E}\)\\
Two planes&\((0.520278,-0.050570)\theta_0\)&\(0.12\theta_0\)\\
Cluster&\((0.075,0.015)R_{\rm E}\)&\(0.12R_{\rm E}\)\\
Kerr, separated images&\((-0.105,0.055)D\)&\(0.045D\)\\
Kerr, long arcs&\((-0.150,0.035)D\)&\(0.045D\)\\
Wave burst&\((3.5847145144,4.2362028712)\ell\)&\(0.5\ell\)\\\bottomrule
\end{tabular}
\caption{Positions and radii for the common four-quadrant colour source. The scalar-brightness
examples have the
separate profiles specified below.}
\label{tab:colour-source-inputs}\end{table}

\subsection{Static lens spacetimes}\label{app:static-lens-controls}
The comparisons in section~\ref{sec:spacetime} use full spacetime rays
and deviation hierarchies in the same metric. All retain the
future-directed convention for \(k^a\); the backwards calculation uses
the increasing computational distance \(\ell=-\lambda\).

\subsubsection{The flattened matter lens}
\label{app:elliptical-propagation}
With \(G=c=1\) and our \((-+++)\) signature, relativistic
Miyamoto--Nagai model~I \citep{vogt2005galaxies} is
\begin{equation}
\begin{aligned}{}
\dd s^2&=-\left(\frac{1-f}{1+f}\right)^2\dd t^2
          +(1+f)^4(\dd x^2+\dd y^2+\dd z^2),\\
 f(x,y,z)&=\frac{\mathsf M}
 {2\sqrt{x^2+y^2+\bigl(a+\sqrt{z^2+b^2}\bigr)^2}}.
 \label{eq:flattened-metric}
\end{aligned}
\end{equation}
The symmetry axis is \(z\); the observer at \((-D,0,0)\) looks along
\(x\). The source plane is \(x=D\), with
\(\beta^A=(y,z)/(2D)\). The source grid spans
\([-1.1,1.1]^2\) in \(\beta^A/\theta_E\), with 24 points per side.
Its 508 singly imaged, 60 triply imaged and eight quintuply imaged sources
give 728 images. The profiles are flattened but need not have elliptical
isodensity contours of one fixed axis ratio.

We use \((\mathsf M/D,a/D,b/D)=(0.008,0.030,0.045)\), giving
\(\max f=0.05333\).
The energy density is the Newtonian Miyamoto--Nagai density divided by
\((1+f)^5\); \(a+b>\mathsf M/2\) keeps the density and principal
pressures positive in this model. The proper mass column
\(\int\varrho(1+f)^2\dd x\) has a contour through \(y=0.075D\)
whose vertical intercept is \(z=0.043323D\), giving an intercept ratio
\(0.578\). This describes the projected matter shape, not the
lensing convergence.

The global angular coordinates specify directions
\((\cos|\boldsymbol\theta|,
\operatorname{sinc}|\boldsymbol\theta|\,\boldsymbol\theta)\)
in the observer's orthonormal frame, with \(E_o=1\).
Each image direction defines its own normal angular coordinates for the local
circle in \eqref{eq:beam-rms-error}. Neighbours are compared at the
reference ray's arrival parameter, followed by a four-dimensional
connecting-geodesic solve and parallel-screen projection.
The fixed plane determines the central dots; it is not substituted for
the local equal-affine endpoint of the colour test.

\subsubsection{The relativistic isothermal sphere}
\label{app:sis-propagation}
For the scale-free static perfect fluid, \(p=w_{\rm iso}\varrho\),
the metric and energy density in areal radius \(R_{\rm ar}\) are
\citep{remmen2021sis}
\begin{equation}
\begin{aligned}{}
\dd s^2&=-\left(\frac{R_{\rm ar}}{r_o}\right)^{4w_{\rm iso}/(1+w_{\rm iso})}\dd t^2
 +\frac{1+6w_{\rm iso}+w_{\rm iso}^2}{(1+w_{\rm iso})^2}\dd R_{\rm ar}^2
 +R_{\rm ar}^2\dd\Omega^2,\\
 \varrho&=\frac{w_{\rm iso}}{2\pi(1+6w_{\rm iso}+w_{\rm iso}^2)R_{\rm ar}^2}.
 \label{eq:relativistic-sis}
\end{aligned}
\end{equation}
We set
\begin{equation}
\begin{aligned}{}
R_{\rm ar}&=r_o(r/r_o)^{s_{\rm iso}},&
 s_{\rm iso}&=\frac{1+w_{\rm iso}}{\sqrt{1+6w_{\rm iso}+w_{\rm iso}^2}},&
 r^2&=\delta_{ij}x^ix^j .
 \label{eq:sis-isotropic-radius}
\end{aligned}
\end{equation}
so that the spatial metric is conformally flat.

For the isothermal reference family in section~\ref{sec:relevance}, the observer is at
\((-D,b_o,0)\), with \(r_o=\sqrt{D^2+b_o^2}\); the coordinate basis
is orthonormal there. The reference ray is launched backwards along the
local positive \(x\) direction, with \(E_o=1\), and the initial screen
axes point along the positive transverse Cartesian axes. They are parallel
transported to \(\lambda_s=-L=-2D\). The family fixes the observer offset through
\begin{equation}
\begin{aligned}{}
\theta_E&=2\pi w_{\rm iso},& b_o&=1.10D\theta_E .
 \label{eq:sis-propagation-family}
\end{aligned}
\end{equation}
The nominal distances are \(D_L=D_{LS}=D\), \(D_S=2D\).
Observer and source remain at finite distances within this solution;
an exterior match and an expanding background would define a different
lens calculation.

The Gaussian source used for the isothermal control on the logarithmic
screen has centre \((0,0.008)Lb_o/D\) and standard deviation \(\sigma=0.003Lb_o/D\):
\(I_s(\xi)=\exp[-|\xi-\xi^{\rm src}|^2/(2\sigma^2)]\).
\subsubsection{Schwarzschild geometry and source}
\label{app:schwarz-propagation}
In isotropic Cartesian coordinates, with \(G=c=1\),
\begin{equation}
\begin{aligned}{}
\dd s^2&=-A^2(r)\dd t^2+B^2(r)\delta_{ij}\dd x^i\dd x^j,\\
 A(r)&=\frac{1-\mathsf M/(2r)}{1+\mathsf M/(2r)},\\
 B(r)&=\left(1+\frac{\mathsf M}{2r}\right)^2,\\
 r^2&=\delta_{ij}x^ix^j .
 \label{eq:schwarzschild}
\end{aligned}
\end{equation}
For the source-plane and amplitude controls, the static observer lies
at \((x,y,w)=(-D,b_o,0)\), and the reference
ray is launched backwards along the local positive \(x\) direction.
We set \(E_o=1\), take \(\lambda_s=-L=-2D\), and use
\begin{equation}
\begin{aligned}{}
b_o&=1.2\sqrt{2\mathsf M D} .
 \label{eq:schwarz-family}
\end{aligned}
\end{equation}
The initial screen axes point along positive \(y\) and \(w\).
The offset \(b_o\) is not the conserved asymptotic impact parameter.
For the nominal distances \(D_L=D_{LS}=D\), \(D_S=2D\),
\(\theta_E=\sqrt{2\mathsf M/D}\); the small-deflection limit is
\begin{equation}
\begin{aligned}{}
\frac{\calM_{11}}{L}&\longrightarrow 1+\frac{2\mathsf M D}{b_o^2},&
 \frac{\calM_{22}}{L}&\longrightarrow 1-\frac{2\mathsf M D}{b_o^2} .
 \label{eq:schwarz-thin-limit}
\end{aligned}
\end{equation}
This family changes deflection while keeping the leading thin-lens
focusing fixed.

For the Schwarzschild comparison in table~\ref{tab:propagation-results},
the source is the sum of three Gaussians,
\begin{equation}
\begin{aligned}{}
I_s(U)&=\sum_{j=1}^3 A_j\exp\!\left[-\frac12
 \left(\frac{(U_1-c_{j1})^2}{4\sigma_j^2}
             +\frac{(U_2-c_{j2})^2}{\sigma_j^2}\right)\right].
 \label{eq:schwarz-source-profile}
\end{aligned}
\end{equation}
We use this profile separately on the fixed plane and on the logarithmic
screen, replacing \(U_A\) by \(\xi_A\) for the latter. The principal
widths \((2\sigma_j,\sigma_j)\) lie along components one and two.
The component inputs are
\begin{center}\begin{tabular}{rrr}\toprule
\(A_j\)&\(c_j/D\)&\(\sigma_j/D\)\\\midrule
\(1\)&\((0,0.005)\)&\(0.001\)\\
\(0.65\)&\((-0.001,0.0065)\)&\(0.00055\)\\
\(0.4\)&\((0.0006,0.0025)\)&\(0.00045\)\\\bottomrule
\end{tabular}\end{center}
Both brightness norms use the full angular rectangle
\([-0.004,0.002]\times[-0.008,0.046]\) rad, sampled at
\(241\times677\) points.

The compact fixed-plane test uses \(\mathsf M/D=0.07\),
\(b_o=6.41\mathsf M\). Its stationary source plane passes through
the reference endpoint at \(x=x_c\simeq0.6591D\), with coordinates
\(U_A=(y-y_c,w-w_c)\). The restricted hierarchies determine
transverse propagation; to compare their plane intersections we supply
the temporal and longitudinal coefficients of the full solution:
\begin{equation}
 \xi_T^a=\xi_{\rm full}^a
   +e_A^a(\xi_T^A-\xi_{\rm full}^A),
 \qquad T\in\{\mathrm{TG},\text{TG+C}\}.
 \label{eq:plane-completion}
\end{equation}
We make the same replacement for \(v^a\), then apply the exponential-map
conversion and angle-dependent arrival shift of
appendix~\ref{app:endpoint-composition} to all three maps. With these
common longitudinal data, the fixed-plane brightness errors are
\(33\%\), \(32\%\) and \(2.2\%\) for TG, TG+C and Full.
On the equal-affine logarithmic screen the corresponding errors are
\(19\%\), \(18\%\) and \(0.45\%\). The source profile is
specified separately in each surface's coordinates.

\subsubsection{Matched spherical dot fields}
\label{app:spherical-dot-inputs}
Figure~\ref{fig:spherical-dots} uses the isothermal metric
\eqref{eq:relativistic-sis}, transformed by
\eqref{eq:sis-isotropic-radius} with \(r_o=D\), and the isotropic
Schwarzschild metric \eqref{eq:schwarzschild}. For both fields the
observer is at \((-D,0,0)\), the source plane is \(x=D\), and
\(\beta^A=(y,z)/(2D)\). We set
\(w_{\rm iso}=0.01\) and \(\mathsf M/D=2\pi^2w_{\rm iso}^2
=0.00197392\), giving the common nominal angle
\(\theta_E=2\pi w_{\rm iso}\). This matching uses the leading
weak-deflection Einstein angles; the full-ray critical radii are
\(0.961809\theta_E\) and \(1.045646\theta_E\), respectively.

The source coordinates are all pairs drawn from 24 equally spaced
values in \([-1.1,1.1]\theta_E\). The global normal angular coordinates
and \(E_o=1\) are as for the flattened lens. We retain the direct
transmitted image branches, with increasing \(x\) along the backwards
ray. There are 348 doubly imaged and 228 singly imaged isothermal
sources, giving 924 dots, and two such Schwarzschild images for each
source, giving 1152. Schwarzschild winding images are outside this
comparison.

Each image direction supplies a separate reference ray. Its initial
screen axes point in the radial and tangential directions of the
observer's sky and are parallel transported. A local normal angular
circle has radius \(\rho=0.04\min(\theta_E,|\boldsymbol\theta|)\).
The smaller aperture for central images keeps neighbouring rays clear
of the singular direction in the isothermal model. All three
hierarchies use that same aperture and central ray. The endpoint is
the affine parameter at which the central ray reaches \(x=D\);
each neighbour is evaluated at this parameter, joined by a
four-dimensional geodesic logarithm, and projected on the parallel
screen. The local comparison therefore uses equal-affine separation,
although the displayed dot positions lie on the fixed source plane.
The Gaussian controls above instead fix \(\lambda_s=-2D\).

Rotational symmetry reduces the calculation to 102 distinct
isothermal image radii and 128 Schwarzschild radii. Coefficients are
propagated separately at every radius; no coefficient or beam error is
interpolated between them. Four angular directions determine each
homogeneous polynomial through cubic degree, and 24 equally spaced
directions test it against direct rays. The colours use the RMS
normalisation in appendix~\ref{app:accuracy-measures}.

\subsection{The exact plane-wave burst}\label{app:wave-model}
Use Brinkmann coordinates with \(u=(t-z)/\sqrt2\),
\(v=(t+z)/\sqrt2\) and \(\boldsymbol X=(x,y)\):
\begin{equation}
\begin{aligned}{}
\dd s^2&=-2\dd u\dd v+\dd\boldsymbol X^T\dd\boldsymbol X
       +\boldsymbol X^T H_{\rm gw}(u)\boldsymbol X\,\dd u^2,\\
 H_{\rm gw}(u)&=\frac{0.7}{\ell^2}
 \left[f_{-2.4,-1}(u/\ell)
       \begin{pmatrix}1&0\\0&-1\end{pmatrix}
 +0.8f_{-4,-2.6}(u/\ell)
       \begin{pmatrix}0&1\\1&0\end{pmatrix}\right],\\
 f_{a,b}(q)&=\begin{cases}
 \sin^4\!\left[\pi(q-a)/(b-a)\right],&a<q<b,\\
 0,&\hbox{otherwise}.
 \end{cases}
 \label{eq:wave-metric}
\end{aligned}
\end{equation}
The profile is \(C^3\), has zero trace and vanishes outside the burst.
Thus the Ricci tensor vanishes exactly; the independent curvature
components in our convention are \(R_{uiuj}=-(H_{\rm gw})_{ij}\).
Here \(H_{\rm gw}\) specifies the wave's tidal profile; it is
distinct from the magnetic Weyl tensor \(H_{ab}\) in
appendix~\ref{app:weyl-projection}. Its amplitude is kept exact, while
the image map is expanded in angular offset.

The observer is at the origin, with tetrad aligned to the flat
\((t,x,y,z)\) coordinates. The central past spatial direction and
source-ruler axes are
\begin{equation}
\begin{aligned}{}
\boldsymbol n&=(\sin\iota,0,\cos\iota),&\iota&=0.65,\\
 \boldsymbol s_1&=(\cos\iota,0,-\sin\iota),&
 \boldsymbol s_2&=(0,1,0).
\end{aligned}
\end{equation}
Past-directed launch vectors have unit observed energy and spatial
part \(\boldsymbol n\cos\rho+
(\boldsymbol s_1\zeta_1+\boldsymbol s_2\zeta_2)\sin\rho/\rho\).
Changing to a past-increasing parameter reverses both \(k\) and
\(v\) in the NLGDE and leaves the acceleration equation unchanged.
The source plane is \(F=\boldsymbol n\cdot\boldsymbol r-D=0\),
with \(D=8\ell\), and its coordinates are
\(U_A=\boldsymbol s_A\cdot\boldsymbol r\). The common source has
radius \(0.5\ell\), centred at the exact reference-ray endpoint
\(U_A/\ell=(3.5847145144,4.2362028712)\). All source events used
below have \(u_s<-4\ell\), in the flat region before the burst.
Comparing the reference ray's spatial directions in the flat endpoint
Cartesian frames gives a change of \(44.1^\circ\); its direction
at the source is \(34.6^\circ\) out of the \((x,z)\) plane.

The observer fixes the screen vectors, which are parallel transported
along the reference ray. Since
\(\dd u(e_1)=\sin\iota/\sqrt2\), a transverse derivative in
this oblique screen samples the changing wave phase. The split into
transverse and along-ray forcing uses that screen convention.
Adding a multiple of \(k\) to a screen vector would require corresponding
changes to the launch and endpoint conversions to describe the same
observed map; at the observer it would also break orthogonality to
the specified four-velocity.

For an exact reference map, solve only the two-dimensional transfer
matrix \(B\):
\begin{equation}
\begin{aligned}{}
B''(u)&=H_{\rm gw}(u)B(u),& B(0)&=0,&B'(0)&=I,\\
 \boldsymbol X(u)&=B(u)\boldsymbol p,&
 \boldsymbol p&=\boldsymbol K_\perp/K^u,\\
 v(u)&=\tfrac12\boldsymbol X(u)\cdot\boldsymbol X'(u).
 \label{eq:wave-transfer}
\end{aligned}
\end{equation}
A prime denotes \(\dd/\dd u\); the last relation follows by
integrating the null condition with \(v(0)=0\). In the earlier flat
region write \(\boldsymbol X=\boldsymbol a+u\boldsymbol q\),
where \(\boldsymbol a=[B(u_b)-u_bB'(u_b)]\boldsymbol p\),
\(\boldsymbol q=B'(u_b)\boldsymbol p\) and \(u_b=-4\ell\).
The physical endpoint is then given algebraically by
\begin{equation}
\begin{aligned}{}
u_s&=\frac{D-\boldsymbol n_\perp\cdot\boldsymbol a
             -n_z\boldsymbol a\cdot\boldsymbol q/(2\sqrt2)}
 {\boldsymbol n_\perp\cdot\boldsymbol q
             +n_z(\boldsymbol q\cdot\boldsymbol q/2-1)/\sqrt2}.
 \label{eq:wave-endpoint}
\end{aligned}
\end{equation}
This expression, the exact sky coordinates and \eqref{eq:wave-transfer}
define a second route to all angular coefficients, independent of the
covariant force implementation.

The NLGDE calculation evolves the full four-dimensional separation and
rate in a parallel tetrad. After reaching the flat source region,
we compose their coefficients with the plane intersection as in
appendix~\ref{app:endpoint-composition}. For TG and TG+C we replace
the transverse propagation and retain the full longitudinal coefficients,
so that the comparison isolates the screen force with a common endpoint
conversion. Their errors therefore refer to this specified transverse
benchmark. The image colours are source labels, with no frequency-shift
or flux factor applied.

The colour norm uses the full window
\(-0.35\leq\zeta_1,\zeta_2\leq0.35\) rad on a \(1001^2\) grid;
figure~\ref{fig:wave} shows a smaller rotated crop. The errors are
\(53.37\%\), \(54.15\%\) and \(0.152\%\) for TG, TG+C and
Full. On a \(501^2\) grid the latter two become \(54.14\%\) and
\(0.153\%\). Source emission stays in the flat region,
\(u_s\leq-9.97\ell\), and \(|F_aK^a|\geq0.532\) for the
unit-energy past-directed tangent, so the intersections are transverse.

\subsection{Sphere-product geometry and sources}
\label{app:sphere-geometry}\label{app:sphere-inputs}
The optical spacetime is the product of a unit sphere and flat time and
spatial directions,
\begin{equation}
\begin{aligned}{}
\dd s^2&=-\dd t^2+\dd\Omega_2^2+\dd w^2 .
 \label{eq:sphere-metric}
\end{aligned}
\end{equation}

The sphere radius is our length unit. A ray follows a great circle while
moving uniformly along the additional flat direction \(w\). We choose
\begin{equation}
\begin{aligned}{}
k_o^a&=u_o^a-\cos\iota\,e_x^a-\sin\iota\,e_w^a,&
 \iota&=0.7,&L\cos\iota&=3.1 .
 \label{eq:sphere}
\end{aligned}
\end{equation}
The vector \(e_x\) is tangent to the initial great circle, and the
final affine parameter is \(-L\).
The reference ray ends just before the sphere's antipode: the first
focus occurs at \(L\cos\iota=\pi\), slightly beyond the chosen
endpoint. In the transported screen the linear map is diagonal:
\begin{equation}
\begin{aligned}{}
\frac{\calM_{11}}{L}&=\frac{\sin(L\cos\iota)}{L\cos\iota},&
 \frac{\calM_{22}}{L}&=1 .
 \label{eq:sphere-J}
\end{aligned}
\end{equation}

The source coordinates use an oriented parallel screen. At the observer
take the sphere's embedding position \(p_o=(1,0,0)\), its initial
great-circle tangent \(e_x=(0,1,0)\), and \(e_y=(0,0,1)\).
With \(e_w\) along the additional flat direction, the observer screen is
\(e_1=e_y\), \(e_2=-\sin\iota\,e_x+\cos\iota\,e_w\).
We parallel transport it along the reference ray.

For source centre \(c\) and width \(\sigma\) in \(\xi/L\) units,
put \(q=(\xi/L-c)/\sigma\). The cubic image test uses
\begin{equation}
\begin{aligned}{}
I_s(q)&=0.08e^{-|q|^2/2}+\sum_{j=1}^3 A_j e^{-|q-d_j|^2/h_j},\\
 (d_1,d_2,d_3)&=((0.72,0.28),(-0.48,0.62),(-0.10,-0.67)),\\
 (A_1,A_2,A_3)&=(0.55,0.35,0.25),\qquad
 (h_1,h_2,h_3)=(0.12,0.10,0.08).
 \label{eq:marked-source}
\end{aligned}
\end{equation}
The errors use this scalar intensity before colour conversion. The
comparison region and source parameters are listed in
table~\ref{tab:sphere-source-inputs}. The norm grid has \(1101\times501\) points. Refining the
cubic test to \(3301\times1501\) leaves
its rounded errors at \(96\%\) and \(3.5\%\).
\begin{table}[!htbp]\centering\small
\begin{tabular}{llll}\toprule
Comparison&\(c\)&\(\sigma\)&Angular rectangle [rad]\\\midrule
Cubic propagation&\((0.001,-0.08)\)&\(0.004\)&\([-0.42,0.42]\times[-0.16,0.04]\)\\
\bottomrule
\end{tabular}
\caption{Sphere-product source and full error window for the cubic test. We
restrict the norm to \(\rho\leq0.42\).}
\label{tab:sphere-source-inputs}\end{table}

\subsection{Two-plane and cluster lenses}\label{app:thin-model-inputs}
\subsubsection{The weak two-plane lens}\label{app:weak-two-lens-inputs}
The reduced potentials used in figure~\ref{fig:weak-two-lens} are
\begin{equation}
\begin{aligned}{}
\phi_i(\theta)&=a\theta_0^2 f_i(\theta),&
 f_i^2&=1+\frac{\theta^A Q^{(i)}_{AB}\theta^B}{\theta_0^2},\\
 Q^{(i)}&=\mathsf R(\varphi_i)
       \operatorname{diag}(q,q^{-1})\mathsf R(\varphi_i)^T,\\
 a&=0.1,&q&=0.6,\qquad b_{12}=0.6,\qquad
 \varphi_1=0,\quad\varphi_2=\pi/4 .
 \label{eq:weak-potentials}
\end{aligned}
\end{equation}
Here \(\mathsf R\) is the planar rotation matrix and \(\theta_0\)
is the common angular core scale. Both lenses are centred at the origin.

\subsubsection{The stronger two-plane lens}\label{app:two-plane-strong-control}
The deflections in figures~\ref{fig:model-gallery}, \ref{fig:two-plane}
and \ref{fig:two-plane-coupling} follow from
\begin{equation}
\begin{aligned}{}
\phi_i(\theta_i)&=b_i\theta_0^2
       \sqrt{c_i^2+r_i^{\mathsf T}Q_i r_i},&
 r_i&=\frac{\theta_i-\theta_{c,i}}{\theta_0},\\
 Q_i&=\mathsf R(\varphi_i)\operatorname{diag}(q_i,q_i^{-1})
                  \mathsf R(\varphi_i)^{\mathsf T},&
 \theta_0&=20\,\mathrm{arcsec}.
 \label{eq:strong-two-plane-potentials}
\end{aligned}
\end{equation}
Table~\ref{tab:strong-two-plane-parameters} gives the two cored
potentials. Their surface convergence is positive everywhere:
\begin{equation}
\begin{aligned}{}
\kappa_i=\frac12\nabla^2\phi_i
 =\frac{b_i[c_i^2\operatorname{tr}Q_i+|r_i|^2]}
 {2(c_i^2+r_i^{\mathsf T}Q_i r_i)^{3/2}}.
 \label{eq:strong-two-plane-density}
\end{aligned}
\end{equation}
The surfaces in the figures display this projected density. The source
lies at \(z_s=2\). In a flat \(\Lambda\)CDM background with
\(\Omega_m=0.3\) and \(H_0=70\,\mathrm{km\,s^{-1}\,Mpc^{-1}}\),
the distance factor is
\(b_{12}=D_{12}D_s/(D_2D_{1s})=0.5961148542\).

\begin{table}[!htbp]\centering\small
\begin{tabular}{lrrrrrr}\toprule
Plane&\(z_i\)&\(b_i\)&\(c_i\)&\(q_i\)&\(\varphi_i\)&\(\theta_{c,i}/\theta_0\)\\\midrule
Foreground&\(0.30\)&\(1.00\)&\(0.12\)&\(0.76\)&\(18^\circ\)&\((-0.20,-0.08)\)\\
Background&\(0.60\)&\(0.70\)&\(0.09\)&\(0.78\)&\(-37^\circ\)&\((0.44,0.22)\)\\\bottomrule
\end{tabular}
\caption{Reduced potential parameters for the stronger two-plane lens.}
\label{tab:strong-two-plane-parameters}\end{table}

The expansion centre is image 6 in figure~\ref{fig:two-plane}, with
\begin{equation}
\begin{aligned}{}
\theta_*/\theta_0&=(-0.6352941176,0.6352941176),\\
 U^{\rm src}/\theta_0&=(0.5202775346,-0.0505703374).
 \label{eq:strong-two-plane-centres}
\end{aligned}
\end{equation}
The colour source of appendix~\ref{app:common-source} has
\(r_s=0.12\theta_0\). At the centre, the singular values of \(\mathcal A\)
are \(0.4056053\) and \(0.04203725\), giving \(|\mu|=58.65\).
The along-arc and across-arc unit vectors are
\((0.52994463,0.84803224)\) and
\((-0.84803224,0.52994463)\).

The full comparison window is
\(24\times8.8\,\mathrm{arcsec}\) and includes source-centre images
5, 6 and 7. We use a common \(1001\times401\) grid and remove eight
pixels at each edge from the colour norm. Doubling both grid intervals,
with the same angular boundary strip, changes the degree-nine TG and
TG+C errors from \((48.0045,0.9337)\%\) to
\((47.9979,0.9303)\%\).

The closed TG expression provides a check independent of its angular
truncation. With
\(\Delta\alpha_1=\alpha_1(\theta_*+\boldsymbol\zeta)-\alpha_1(\theta_*)\)
and \(\theta_{2*}=\theta_*-b_{12}\alpha_1(\theta_*)\), it is
\begin{equation}
\begin{aligned}{}
U_{\mathrm{TG},\infty}
 &=U^{\rm src}+\boldsymbol\zeta-\Delta\alpha_1
 -[\alpha_2(\theta_{2*}+\mathsf L_{12}\boldsymbol\zeta)-\alpha_2(\theta_{2*})]\\
 &\quad+b_{12}H_2(\Delta\alpha_1-H_1\boldsymbol\zeta).
 \label{eq:two-plane-tg-sum}
\end{aligned}
\end{equation}
The last term transports the first lens's nonlinear distortion with
the linear Hessian of the second. Its insertion into the nonlinear
part of the second deflection is omitted. Table~\ref{tab:two-plane-convergence}
shows the resulting error floor. The maps agree through quadratic
degree and at every tested degree in the coplanar limit. Direct
differentiation of both deflections reproduces the cubic difference
\eqref{eq:two-plane-cubic} at all seven image centres.

\begin{table}[!htbp]\centering\small
\begin{tabular}{lrr}\toprule
\(m_{\max}\)&TG&TG+C\\\midrule
\(1\)&\(64.53\)&\(64.53\)\\
\(2\)&\(64.40\)&\(64.40\)\\
\(3\)&\(48.40\)&\(40.33\)\\
\(5\)&\(47.70\)&\(7.39\)\\
\(7\)&\(47.75\)&\(7.10\)\\
\(9\)&\(48.00\)&\(0.934\)\\
\(11\)&\(47.87\)&\(0.829\)\\
\(15\)&\(47.87\)&\(0.112\)\\
Closed TG&\(47.87\)&---\\\bottomrule
\end{tabular}
\caption{Colour errors in per cent for the fixed window in
figure~\ref{fig:two-plane-coupling}. The unsmoothed source and
\(1001\times401\) grid are used throughout. TG+C converges towards
the direct map, whereas TG approaches a nonzero error even when its
degree truncation is removed.}
\label{tab:two-plane-convergence}\end{table}

The central direction was selected to give a large TG beam error while
keeping the degree-five TG+C beam error below \(2\%\) on a circle
of radius \(8.4\,\mathrm{arcsec}\). This selection used no source
brightness; the colour source and enlarged image window were applied
afterwards. The example demonstrates nonlinear coupling for this
alignment, rather than its frequency in a cluster population.

\subsubsection{The eleven-component cluster}\label{app:cluster-inputs}
The cluster potential sums three translated cored terms of
\eqref{eq:elliptical}, eight compact terms of \eqref{eq:compact},
and one external shear of \(0.07\) at \(-23^\circ\).
For an extended component centred at \(C_j\), we use
\(Q_j=\mathsf R(\varphi_j)\operatorname{diag}(q_j,q_j^{-1})
\mathsf R(\varphi_j)^T\). Table~\ref{tab:cluster-parameters}
gives all components; lengths are in units of the common \(R_{\rm E}\).

\begin{table}[!htbp]\centering\small
\begin{tabular}{rrrrrr}\toprule
\(b_j/R_{\rm E}\)&\(q_j\)&\(c_j/R_{\rm E}\)&\(\varphi_j\) [deg]&\(C_{j1}/R_{\rm
E}\)&\(C_{j2}/R_{\rm E}\)\\\midrule
\(0.63\)&\(0.78\)&\(0.050\)&\(25\)&\(-0.43\)&\(0.06\)\\
\(0.43\)&\(0.70\)&\(0.045\)&\(-32\)&\(0.43\)&\(0.23\)\\
\(0.26\)&\(0.88\)&\(0.035\)&\(68\)&\(0.12\)&\(-0.54\)\\\bottomrule
\end{tabular}\par\medskip
\begin{tabular}{rrrr}\toprule
\(R_{p1}/R_{\rm E}\)&\(R_{p2}/R_{\rm E}\)&\(b_p/R_{\rm E}\)&\(a_p/R_{\rm E}\)\\\midrule
\(-0.18\)&\(1.42\)&\(0.16\)&\(0.012\)\\
\(0.53\)&\(1.28\)&\(0.20\)&\(0.014\)\\
\(0.98\)&\(0.88\)&\(0.18\)&\(0.010\)\\
\(1.38\)&\(0.10\)&\(0.17\)&\(0.015\)\\
\(0.55\)&\(-1.16\)&\(0.16\)&\(0.012\)\\
\(0.90\)&\(-0.90\)&\(0.15\)&\(0.011\)\\
\(-1.27\)&\(-0.25\)&\(0.20\)&\(0.014\)\\
\(-0.83\)&\(0.99\)&\(0.17\)&\(0.012\)\\\bottomrule
\end{tabular}
\caption{Extended (above) and compact (below) components of the cluster
lens in figures~\ref{fig:model-gallery}, \ref{fig:cluster}--\ref{fig:cluster-covers}.}
\label{tab:cluster-parameters}\end{table}
The final cover starts with the whole square \([-1.8,1.8]^2R_{\rm E}\)
and splits each tile into four until its uniform degree-25 position bound
is below \(10^{-6}R_{\rm E}\). Its \(\mathcal E_C\) values use
\(3601^2\) samples across this full square. Refinement from \(1801^2\)
changes the degree-one and degree-five errors by less than \(0.2\%\)
and the much smaller degree-nine and degree-25 errors by about \(13\%\),
relative to their respective values. The position bound is independent
of this sampling.

The filled-source panel instead tiles the source plane with squares of
side \(0.20R_{\rm E}\), with vertices at
\((0.075,0.015)R_{\rm E}+0.20R_{\rm E}\mathbb Z^2\) and the four
colours repeating in a \(2\times2\) block. It does not enter the
compact-source error norm.
The reduced cluster map was also checked against
\texttt{lenstronomy} 1.14.2 \citep{lenstronomy2018,lenstronomy2021},
using the component parameters above.

\subsection{Error bounds and patch selection for thin lenses}
\label{app:thin-lens-bounds}
The analytic lens equations give bounds on the position error throughout
a patch, including regions with no source light. We use these bounds to
decide where additional expansion centres are needed.

\subsubsection{Uniform position bounds}\label{app:bounds}
About an expansion centre \(R\), define
\begin{equation}
\begin{aligned}{}
d_0&=c^2+R^TQR,& \ell&=\frac{2R^TQx}{d_0},&
 h&=\frac{x^TQx}{d_0},&\eta_1&=\sqrt h,\\
 p&=R-R_p,&d_p&=a_p^2+|p|^2,&
 \ell_p&=\frac{2p\cdot x}{d_p},&
 h_p&=\frac{|x|^2}{d_p},\quad\eta_p=\sqrt{h_p}.
\end{aligned}
\end{equation}
Here \(x\) is the displacement from the expansion centre. The two
component maps have generating functions
\begin{equation}
\begin{aligned}{}
(1+\ell t+ht^2)^{-1/2}&=\sum_{m\geq0}g_mt^m,&
 (1+\ell_pt+h_pt^2)^{-1}&=\sum_{m\geq0}c_mt^m,\\
 g_{-1}&=c_{-1}=0,&g_0&=c_0=1,\\
 g_m&=-\frac{2m-1}{2m}\ell g_{m-1}
       -\frac{m-1}{m}h g_{m-2},&
 c_m&=-\ell_pc_{m-1}-h_pc_{m-2}.
 \label{eq:component-recurrences}
\end{aligned}
\end{equation}
For \(m\geq1\), the homogeneous degree-\(m\) lensing corrections are
\(-R_{\rm E}[(QR)_Ag_m+(Qx)_Ag_{m-1}]/\sqrt{d_0}\) and
\(-b_p^2[p_Ac_m+x_Ac_{m-1}]/d_p\), respectively. The linear
identity and external-shear term \(-\Gamma_{AB}x^B\) are added
separately. The SIS follows by setting \(c=0,Q=I,\Gamma=0\).

Since \(\ell^2\leq4h\), the first generating function is the
Legendre generating function with argument
\(-\ell/(2\sqrt h)\in[-1,1]\), giving \(|g_m|\leq\eta_1^m\).
The corresponding Chebyshev generating function of the second kind,
with argument \(-\ell_p/(2\sqrt{h_p})\in[-1,1]\), gives
\(|c_m|\leq(m+1)\eta_p^m\). The zero-offset limits follow by
continuity. Summing these bounds beyond degree \(M\geq1\) gives
\begin{equation}
\begin{aligned}{}
B_{1,M}(x)&=\frac{R_{\rm E}}{\sqrt{d_0}}
 \frac{|QR|\eta_1^{M+1}+|Qx|\eta_1^M}{1-\eta_1},
 \qquad \eta_1<1,
 \label{eq:host-bound}
\end{aligned}
\end{equation}
\begin{equation}
\begin{aligned}{}
B_{p,M}(x)&=\frac{b_p^2}{d_p}
       \left[|p|T_{M+1}(\eta_p)+|x|T_M(\eta_p)\right],\\
 T_k(r)&=\frac{r^k[(k+1)-kr]}{(1-r)^2},\qquad r<1 .
 \label{eq:compact-bound}
\end{aligned}
\end{equation}
Their sum bounds the compound-map position error. To bound a whole
square tile, let \(w\) be its centre displacement from the expansion
centre, \(\delta\) its half-diagonal, and \(q_{\max}\) the largest
eigenvalue of \(Q\). Throughout the tile,
\begin{equation}
\begin{aligned}{}
\eta_1&\leq\frac{\sqrt{w^TQw}+\sqrt{q_{\max}}\,\delta}{\sqrt{d_0}},&
 \eta_p&\leq\frac{|w|+\delta}{\sqrt{d_p}},\\
 |Qx|&\leq|Qw|+q_{\max}\delta,&|x|&\leq|w|+\delta .
 \label{eq:tile}
\end{aligned}
\end{equation}
Substitution in the positive tail sums gives a uniform bound whenever
both radius bounds are below one. Bounds for additional components
add in the same way. The covering criterion applies in exact arithmetic;
floating-point errors are checked separately.

\subsubsection{Greedy selection for the compact component}\label{app:compact-greedy}
For section~\ref{sec:changing-centres}, the square window of side
\(0.72R_{\rm E}\), centred at \((0.17,1.10)R_{\rm E}\), starts
with \(41^2\) tiles. A tile is split until its half-diagonal is at
most \(0.06\min[\sqrt{d_0/q_{\max}},\sqrt{d_p}]\), evaluated at
its centre using the quantities in appendix~\ref{app:bounds}.
All degrees use this fixed tiling. The first expansion centre is
\((0.35974349895,1.05976680815)R_{\rm E}\).
Among tiles with no finite bound from an existing patch, we next choose
the tile centre maximising \(\min_j\eta(R_{\rm tile}-R_j;R_j)\).
Once all tiles have finite bounds, we choose the tile with the largest
remaining bound. Each tile keeps the patch giving its smallest bound;
selection stops when every tile meets the target.

\begin{figure}[!htb]
  \centering
\includegraphics[width=\textwidth,height=.72\textheight,keepaspectratio]{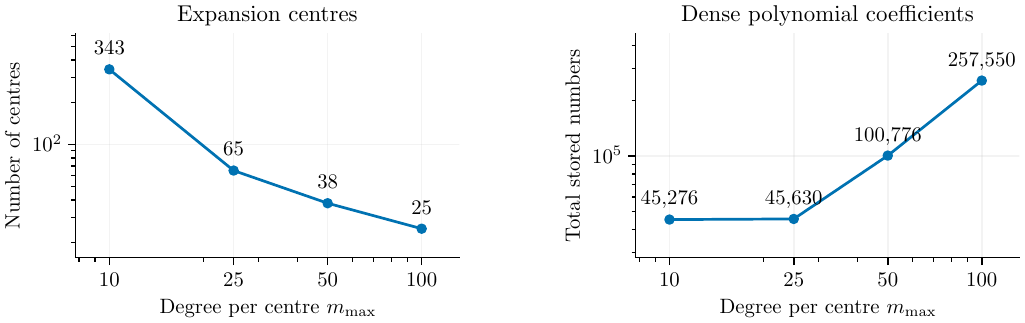}
  \caption{Angular degree and storage for the compact-component cover.
Both panels use the same uniform position-error target,
\(10^{-6}R_{\rm E}\). Increasing degree reduces the number of
expansion centres, but the total dense coefficient count eventually
rises. It includes both map components and the constant endpoint at
each centre. These counts do not measure runtime.}
  \label{fig:atlas-cost}
\end{figure}
\subsection{Kerr frames, shadows and comparison windows}\label{app:kerr-inputs}
For the static observer in section~\ref{sec:kerr},
\(u=(-g_{tt})^{-1/2}\partial_t\). We project the positive Cartesian
directions into its rest space and apply metric Gram--Schmidt in the
order \(\partial_x,\partial_y,\partial_w\), preserving each original
positive direction. The first spatial vector defines the sky origin;
the other two define its axes.

The shadow boundary is calculated from unstable spherical photon orbits
\citep{grenzebach2014,perlicktsupko2022}. We solve the radial double-root
conditions for their conserved quantities, retain the admissible
future-directed tangents at the finite observer, and transform them from
Boyer--Lindquist to Cartesian Kerr--Schild coordinates before projection
into this tetrad and the fixed sky coordinates. The zero-spin case uses its
separate spherical limit.

The computational mask is imposed separately from this physical
boundary. For all three spins, we retain only initial
directions whose straight tangent meets \(x=0\) at transverse radius
greater than \(0.085D\). The spin comparison covers
\([-0.36,0.075]\times[-0.25,0.22]\) rad and shows transmitted rays
on the monotonic integration branch. These transmitted rays pass at impact parameters of order
\(14\mathsf M\), with deflections of order \(0.3\) rad; the
near-photon-orbit and higher-order images are outside this calculation.

\begin{table}[!htbp]\centering\small
\begin{tabular}{lrrrr}\toprule
Source and parity&\(\zeta_{c1}\)&\(\zeta_{c2}\)&\(\zeta_1\) interval&\(\zeta_2\) interval\\\midrule
Long arcs, positive&\(-0.103406476\)&\(0.141768802\)&\([-0.280,0.030]\)&\([0.015,0.185]\)\\
Long arcs, negative&\(-0.196853616\)&\(-0.157754672\)&\([-0.320,-0.035]\)&\([-0.195,-0.025]\)\\
Separated, positive&\(-0.040627823\)&\(0.118179745\)&\([-0.140,0.030]\)&\([0.015,0.185]\)\\
Separated, negative&\(-0.241025254\)&\(-0.125823357\)&\([-0.317,-0.147]\)&\([-0.195,-0.025]\)\\\bottomrule
\end{tabular}
\caption{Kerr expansion centres and full comparison windows in radians
in the fixed observer coordinates.}
\label{tab:kerr-windows}\end{table}
The reconstruction norms use 1001 samples vertically and approximately
equal spacing horizontally in table~\ref{tab:kerr-windows}. Their common
validity mask is obtained from the \(641\times701\) reference grid,
eroded by eight cells; every common valid pixel enters the norm,
including false polynomial images. Halving the sampling density changes
the small seven-centre errors by at most \(5.1\%\) relative to their
original values.

To place seven centres, we sample each rectangle on a 321-row grid and
retain pixels with source envelope \(B>1/2\). Their polar origin is the
inward sky direction normal to the Kerr--Schild \(r=\mathrm{constant}\)
surface at the observer. Writing the outward unit normal as
\(s^a=\nabla^ar/\sqrt{\nabla_br\nabla^br}\), the future-directed
tangent \(u^a+s^a\) has this inward observed direction. Its sky
coordinates at \(a/\mathsf M=0.9\) are
\((-0.145301586,-0.019655999)\) rad. Seven equal azimuthal bins each
supply a centre at their median angle and radius. Every image pixel then
uses its nearest centre in the fixed coordinates and that centre's complete
degree-five polynomial, including its constant endpoint.

\subsection{Image and beam errors}\label{app:accuracy-measures}
A position error changes the image according to the source structure.
We measure it separately from the scalar brightness error and the
colour error. For a spacetime map of characteristic length \(L\)
and an image window \(\Omega\), we use
\begin{equation}
\begin{aligned}{}
\mathcal E_M(\rho)&=\max_\vartheta
 \frac{|\boldsymbol\xi^{[m_{\max}]}(\rho n)
              -\boldsymbol\xi^{\rm exact}(\rho n)|}{L},\\
 \mathcal E_I&=\left[
 \frac{\int_\Omega|I^{[m_{\max}]}-I^{\rm exact}|^2\dd^2\zeta}
      {\int_\Omega|I^{\rm exact}|^2\dd^2\zeta}\right]^{1/2},\\
 \mathcal E_C&=\left[
 \frac{\sum_{p,c}|I_{p,c}^{[m_{\max}]}-I_{p,c}^{\rm exact}|^2}
      {\sum_{p,c}|1-I_{p,c}^{\rm exact}|^2}\right]^{1/2}.
 \label{eq:errors}
\end{aligned}
\end{equation}
The scalar norm uses intensity before any plotting stretch. The colour
norm measures contrast against the white background, summing over RGB
channels \(c\) and all common valid pixels \(p\) in the stated
window, including false polynomial images. Thin-lens position errors
are quoted in \(R_{\rm E}\), and Kerr errors in \(D\).

For the dot fields, the directly traced beam size supplies the ruler:
\begin{equation}
\begin{aligned}{}
\mathcal E_{\rm beam}&=\left[
 \frac{\left\langle|\boldsymbol\xi^{[3]}-\boldsymbol\xi^{\rm ray}|^2
                     \right\rangle_\vartheta}
      {\left\langle|\boldsymbol\xi^{\rm ray}|^2\right\rangle_\vartheta}
 \right]^{1/2}.
 \label{eq:beam-rms-error}
\end{aligned}
\end{equation}
The average uses 24 equally spaced directions around each angular
circle. This measures errors in the neighbouring rays' separation,
independently of the central dot position or source brightness. For the
contribution maps in figure~\ref{fig:spherical-dots}, the numerator
instead uses TG+C minus TG, or Full minus TG+C; the direct-ray denominator
is unchanged.
In the Schwarzschild fixed-plane and wave comparisons, all three
benchmarks share the full temporal and longitudinal coefficients and
the same source-surface conversion, so only the transverse propagation differs.

%% file: appendices/e_wavelets.tex
\section{Wavelets and local roulette amplitudes}\label{app:wavelets}\label{sec:wavelets}
The relation between wavelets and roulettes depends on which field
the wavelets represent: source light, a lens potential or a reconstructed
map \citep{galan2021slitronomy,galan2022potential,leonard2014glimpse}.
For source coefficients \(c_\nu\), surface-brightness conservation gives
\(I(\theta)=\sum_\nu c_\nu w_\nu[U(\theta)]\): the wavelets specify
the source and the roulettes specify its map. Shapelet source models
enter in the same way \citep{birrer2015shapelets}.

For a single-plane potential \(\psi=\sum_\nu c_\nu w_\nu\), including
the coarse component, the relation is a derivative projection. Use
angular coordinates, \(U_A=\chi_S(\theta_A-\partial_A\psi)\), and
write \(\mathcal A_s^m=(\alpha_s^m+i\beta_s^m)/\chi_S
-\delta_{m1}\delta_{s0}\) for the lensing part. Equation~\eqref{eq:connector-complex-potential}
then gives
\begin{equation}
\begin{aligned}{}
\mathcal A_s^m(\theta_*)&=-\frac{1}{2^{m+\delta_{s0}}}
 \binom{m+1}{p}\sum_\nu c_\nu
 \left[(\partial_1+i\partial_2)^s(\partial_1^2+\partial_2^2)^p
 w_\nu\right]_{\theta_*},\\
 p&=\frac{m+1-s}{2},\qquad s=m+1,m-1,\ldots,0\ \hbox{or}\ 1.
 \label{eq:wavelet-roulette}
\end{aligned}
\end{equation}
The \(w_\nu\) are the synthesis functions of the reconstruction;
an infinite sum must converge with the required derivatives near the
ray. One wavelet generally contributes to several degrees and spins.
For a general vector map \(U_A=\sum_\nu c_{A\nu}w_\nu\), instead take
\(\calM_{AB_1\cdots B_m}=\sum_\nu c_{A\nu}
\partial_{B_1}\cdots\partial_{B_m}w_\nu\) and use
\eqref{eq:fourier}; this also includes the barred amplitudes.

Conversely, an analysis wavelet samples derivatives over an aperture.
For a real wavelet \(w\) with zero mean and first moments, define
\begin{equation}
\begin{aligned}{}
W_\psi(a,\theta_*)&=\int \dd^2u\,w(u)\psi(\theta_*+au),&
 M_{ns}&=\int \dd^2u\,w(u)|u|^n e^{-is\arg u} .
 \label{eq:wavelet-moments}
\end{aligned}
\end{equation}
Taylor expansion along each radius and angular projection give
\begin{equation}
\begin{aligned}{}
W_\psi(a,\theta_*)&=-\sum_{n=2}^{\infty}\frac{a^n}{n!}
 \sum_{\substack{0\leq s\leq n\\n-s\ {\rm even}}}
 \operatorname{Re}\!\left[\mathcal A_s^{n-1}(\theta_*)M_{ns}\right].
 \label{eq:roulette-wavelet}
\end{aligned}
\end{equation}
This equality requires that the Taylor series can be integrated over
the aperture, for example by uniform convergence on compact support.
A circular wavelet selects only spin zero. A finite set of roulettes
fixes only the corresponding local derivative combinations, so it
cannot determine a general finite-aperture coefficient. The NLGDE
specifies how those local derivatives arise from spacetime curvature;
the wavelet representation organises the chosen field by scale and
position.